\documentclass[a4paper,12pt]{article}
\pdfoutput=1
\usepackage{graphicx, rotating,amssymb,amsmath,enumitem}
\usepackage[normalem]{ulem}
\usepackage{cancel}

\ifx\pdfoutput\undefined
\usepackage[dvips,bookmarks]{hyperref}	% This is for arXiv.org
\else
\usepackage{hyperref}	% This is for pdftex
\fi
\hypersetup{colorlinks,bookmarksopen,bookmarksnumbered,citecolor=rossoc,
linkcolor=verdes,pdfstartview=FitH,urlcolor=rosso}
\def\myurl#1#2{\href{http://#1}{#2}}
\def\hhref#1{\href{http://arxiv.org/abs/#1}{#1}} % in bibliography
\usepackage{multicol}
\usepackage{color}
\definecolor{rosso}{cmyk}{0,1,1,0.4}
\definecolor{rossos}{cmyk}{0,1,1,0.55}
\definecolor{rossoc}{cmyk}{0,1,1,0.2}
\definecolor{blu}{cmyk}{1,1,0,0.3}
\definecolor{blus}{cmyk}{1,1,0,0.6}
\definecolor{bluc}{cmyk}{1,1,0,0.1}
\definecolor{verde}{cmyk}{0.92,0,0.59,0.25}
\definecolor{verdec}{cmyk}{0.92,0,0.59,0.15}
\definecolor{verdes}{cmyk}{0.92,0,0.59,0.4}

\font\tenrsfs=rsfs10 at 12pt
\font\sevenrsfs=rsfs7
\font\fiversfs=rsfs5
\newfam\rsfsfam
\textfont\rsfsfam=\tenrsfs
\scriptfont\rsfsfam=\sevenrsfs
\scriptscriptfont\rsfsfam=\fiversfs
\def\mathscr#1{{\fam\rsfsfam\relax#1}}

\newcommand{\fig}[1]{~\ref{fig:#1}}
\newcommand{\eq}[1]{~{\rm (\ref{eq:#1})}}

\newcommand{\GeV}{\,{\rm GeV}}
\newcommand{\TeV}{\,{\rm TeV}}

\def\circa#1{\,\raise.3ex\hbox{$#1$\kern-.75em\lower1ex\hbox{$\sim$}}\,}

\newcommand{\beq}{\begin{equation}}
\newcommand{\eeq}{\end{equation}}
\newcommand{\MeV}{\,{\rm MeV}}
\def\circa#1{\,\raise.3ex\hbox{$#1$\kern-.75em\lower1ex\hbox{$\sim$}}\,}
\makeatletter

\def\art{\@ifnextchar[{\eart}{\oart}}
\def\eart[#1]#2#3#4#5#6{{\rm #2}, {#3 #4} {\rm (#6) #5} [{\hhref{#1}}]}
\def\hepart[#1]#2{{\rm #2, \hhref{#1}}}
\newcommand{\oart}[5]{{\rm #1}, {#2 #3} {\rm (#5) #4}}

\newcounter{alphaequation}[equation]
\def\thealphaequation{\theequation\hbox to
0.6em{\hfil\alph{alphaequation}\hfil}}
\def\eqnsystem#1{
\def\@eqnnum{{\rm (\thealphaequation)}}
\def\@@eqncr{\let\@tempa\relax \ifcase\@eqcnt \def\@tempa{& & &} \or
  \def\@tempa{& &}\or \def\@tempa{&}\fi\@tempa
  \if@eqnsw\@eqnnum\refstepcounter{alphaequation}\fi
\global\@eqnswtrue\global\@eqcnt=0\cr}
\refstepcounter{equation} \let\@currentlabel\theequation \def\@tempb{#1}
\ifx\@tempb\empty\else\label{#1}\fi
\refstepcounter{alphaequation}
\let\@currentlabel\thealphaequation
\global\@eqnswtrue\global\@eqcnt=0 \tabskip\@centering\let\\=\@eqncr
$$\halign to \displaywidth\bgroup \@eqnsel\hskip\@centering
$\displaystyle\tabskip\z@{##}$&\global\@eqcnt\@ne
\hskip2\arraycolsep\hfil${##}$\hfil& \global\@eqcnt\tw@\hskip2\arraycolsep
$\displaystyle\tabskip\z@{##}$\hfil
\tabskip\@centering&\llap{##}\tabskip\z@\cr}
\def\endeqnsystem{\@@eqncr\egroup$$\global\@ignoretrue} \makeatother

\begin{document}
\begin{center}
{\footnotesize CERN-PH-TH/2010-057
\hfill
SACLAY--T10/025
\hfill
IFUP-TH/2010-44}
\end{center}
%\begin{flushright}
%{\footnotesize CERN-PH-TH/2010-057\\
%SACLAY--T10/025\\
%IFUP-TH/2010-44}
%\end{flushright}
\color{black}

\vspace{-0.5cm}

\begin{center}
{\Huge\bf PPPC 4 DM ID:\\[1mm] A Poor Particle Physicist Cookbook\\[3mm] for Dark Matter Indirect Detection}

\color{black}\vspace{0.5cm}

{
{\large\bf Marco Cirelli}$^{a,b}$,
{\large\bf  Gennaro Corcella}$^{c,d,e}$,
{\large\bf  Andi Hektor}$^f$,\\[2mm]
{\large\bf  Gert H\"utsi}$^g$,
{\large\bf  Mario Kadastik}$^f$,
{\large\bf  Paolo Panci}$^{a,h,i,j}$,\\[2mm]
{\large\bf  Martti Raidal}$^{f}$,
{\large\bf  Filippo Sala}$^{d,e}$,
{\large\bf Alessandro Strumia}$^{a,e,f,k}$
}
\end{center}

\centerline{\large\bf Abstract}
\begin{quote}
\color{black}\large

We provide ingredients and recipes for computing signals of TeV-scale Dark Matter annihilations and decays in the Galaxy and beyond. 
For each DM channel, we present the energy spectra of $e^\pm,\bar p, \bar d, \gamma,\stackrel{\mbox{{\tiny (--)}}}{\nu}_{e,\mu,\tau}$ at production, computed by high-statistics simulations. We estimate the Monte Carlo uncertainty by comparing the results yielded by the {\sc Pythia} and {\sc Herwig} event generators.
We then provide the propagation functions for charged particles in the Galaxy, for several DM distribution profiles and sets of propagation parameters. Propagation of $e^\pm$ is performed with an improved semi-analytic method that takes into account position-dependent energy losses in the Milky Way. 
Using such propagation functions, we compute the energy spectra of $e^\pm,\bar p $ and $\bar d$ at the location of the Earth.
We then present the gamma ray fluxes, both from prompt emission and from Inverse Compton scattering in the galactic halo. 
Finally, we provide the spectra of extragalactic gamma rays.
All results are \myurl{www.marcocirelli.net/PPPC4DMID.html}{available in numerical form} and ready to be consumed.

\end{quote}

\vfill

{\footnotesize\noindent
{\it $^a$ CERN Theory Division, CH-1211 Gen\`eve, Switzerland}\\
{\it $^b$ Institut de Physique Th\'eorique, CNRS, URA 2306 \& CEA/Saclay,
	F-91191 Gif-sur-Yvette, France}\\
{\it $^c$ Museo Storico della Fisica, Centro Studi e Ricerche E.~Fermi,
P. del Viminale 1, I-00185 Rome, Italy}\\
{\it $^d$ Scuola Normale Superiore, Piazza dei Cavalieri 7, I-56126 Pisa, Italy}\\
{\it $^e$ INFN, Sezione di Pisa, Largo Fibonacci 3, I-56127 Pisa, Italy}\\
{\it $^f$ National Institute of Chemical Physics and Biophysics, Ravala 10, 10143 Tallinn, Estonia}\\
{\it $^g$ Tartu Observatory, T\~oravere 61602, Estonia}\\
{\it $^h$ Dipartimento di Fisica, Universit\`a degli Studi dell'Aquila, 67010 Coppito (AQ)}\\
{\it $^i$ INFN, Laboratori Nazionali del Gran Sasso, 67010 Assergi (AQ), Italy}\\
{\it $^j$ Universit\'e Paris 7-Diderot, UFR de Physique, 10, rue A. Domon et L.Duquet, 75205 Paris, France}\\
{\it $^k$ Universit{\`a} degli Studi di Pisa, Dipartimento di Fisica,
Largo Fibonacci 3, I-56127 Pisa, Italy}}

\tableofcontents
\newpage

\section{Introduction}
\label{introduction}

Cosmology and astrophysics provide several convincing evidences of the existence of Dark Matter~\cite{JungmanReview, BertoneReview, EinastoReview}. 
The observation that some mass is missing to explain the internal dynamics of galaxy clusters and the rotations of galaxies dates back respectively to the '30s and the '70s~\cite{rotation}. The observations from weak lensing~\cite{lensing}, for instance in the spectacular case of the so-called `bullet cluster'~\cite{bullet}, provide evidence that there is mass where nothing is optically seen. More generally, global fits to a number of cosmological datasets (Cosmic Microwave Background, Large Scale Structure and also Type Ia Supernovae) allow to determine very precisely the amount of DM in the global energy-matter content of the Universe at $\Omega_{\rm DM} h^2 =0.1123 \pm 0.0035$~\cite{cosmoDM}\footnote{Here $\Omega_{\rm DM} = \rho_{\rm DM}/\rho_c$ is defined as usual as the energy density in Dark Matter with respect to the critical energy density of the Universe $\rho_c = 3 H_0^2/8\pi G_N$, where $H_0$ is the present Hubble parameter. $h$ is its reduced value $h = H_0 / 100\ {\rm km}\, {\rm s}^{-1} {\rm Mpc}^{-1}$.}.
 
All these signals pertain to the gravitational effects of Dark Matter at the cosmological and extragalactical scale. Searches for explicit manifestation of the DM particles that are supposed to constitute the halo of our own galaxy (and the large scale structures beyond it) have instead so far been giving negative results, but this might be on the point of changing. 

\medskip

Indirect searches for Dark Matter aim at detecting the signatures of the annihilations or decays of DM particles in the fluxes of cosmic rays, intended in a broad sense: charged particles (electrons and positrons, protons and antiprotons, deuterium and antideuterium), photons (gamma rays, X-rays, synchrotron radiation), neutrinos. Pioneering works have explored this as a promising avenue of discovery since the late-70's: gamma rays from annihilations were first considered in~\cite{Gunn,Stecker,Zeldovich} and then revisited in~\cite{Ellis}, antiprotons in~\cite{SilkSrednicki,SteckerRudazWalsh} and then more systematically in~\cite{Ellis,SteckerRudaz,SteckerTylka}, positrons in~\cite{SilkSrednicki,Ellis,SteckerRudaz,TurnerWilczek}, antideuterons have been first discussed in~\cite{pioneerDbar,Dbar2,followupDbar}, radio-waves from synchrotron radiation from DM in~\cite{BerezinskyGurevich,BerezinskyBottino,Gondolo,BertoneSiglSilk} and later in~\cite{AloisioBlasiOlinto} (which questions the approach in~\cite{Gondolo,BertoneSiglSilk}), extragalactic gamma rays have been first discussed in~\cite{BergstomEdsjoUllio}. Inverse Compton gamma rays from DM have been only relatively recently considered as a possible signal (see e.g.~\cite{BaltzWai,Cholis, Zhang}). In general, a key point of all these searches is to look for channels and ranges of energy where it is possible to beat the background from ordinary astrophysical processes. This is for instance the basic reason why searches for charged particles focus on fluxes of antiparticles (positrons, antiprotons, antideuterons), much less abundant in the Universe than the corresponding particles. 

A well spread theoretical prejudice wants the DM particles to be thermal relics from the Early Universe. They were as abundant as photons in the beginning, being freely created and destroyed in pairs when the temperature of the hot plasma was larger then their mass. Their relative number density started then being suppressed as annihilations proceeded but the temperature dropped below their mass, due to the cooling of the Universe. Finally the annihilation processes also froze out as the Universe expanded further. The remaining, diluted abundance of stable particles constitutes the DM today. As it turns out, particles with weak scale mass ($\sim 100\, {\rm GeV} - 1\, {\rm TeV}$) and weak interactions \cite{JungmanReview, BertoneReview} could play the above story remarkably well, and their final abundance would automatically (miracolously?) be the observed $\Omega_{\rm DM}$. While this is not of course the only possibility, the mechanism is appealing enough that a several-GeV-to-some-TeV scale DM particle with weak interactions (WIMP) is often considered as the most likely DM candidate.

In any case, this mass range (TeV-ish DM) has the best chances of being thoroughly explored in the near future by charged particle and photon observatories, also in combination with direct DM searches (aiming at detecting the nuclear recoil produced by a passing DM particle in ultra-low background underground detectors) and, possibily, production at LHC collider. It is therefore the focus of our attention.

Supposing (and hoping) therefore that anomalous features are detected in the fluxes of cosmic rays, it will be crucial to be able to `reverse engineer' them to determine which Dark Matter is at their origin. Moreover, it will be useful to be able to quickly compute which other associated signals are implied by a possible positive detection and have to be looked for in other channels. Only via a cross-correlation of multi-messenger signals a putative detection of DM will be confirmed or disproved. More generally, in order to compute the predicted signatures of a given model of Dark Matter, a number of particle physics and astrophysics ingredients are needed. These ingredients are what we aim to provide.

\bigskip

This work does not contain any new theoretical proposal nor any new study of (existing or foreseen) data. It contains instead all the phenomenological ingredients that allow to perform the analyses sketched above in the most general possible way. 

More precisely, the rest of this compilation is organized as follows.
In Section~\ref{DMdistribution} we start by recalling the most commonly used DM distribution profiles in the Milky Way, that we will adopt for the computation of all signals. In Section~\ref{primary} we discuss the production of the fluxes of Standard Model particles from DM annihilations (and decays): we compare the {\sc Pythia} and {\sc Herwig} Monte Carlos and quantify the uncertainties. Section~\ref{charged} deals with the propagation in the Galaxy and the resulting fluxes of charged cosmic rays from DM: electrons, positrons, antiprotons, antideuterons. Section~\ref{promptgamma} deals with the basics of prompt gamma rays from Dark Matter annihilations (or decays). Section~\ref{ICSgamma} discusses the `secondary' radiation from $e^\pm$ produced by DM annihilations or decays: Inverse Compton Scattering (ICS) $\gamma$-rays and synchrotron radiation. Section~\ref{extragalactic} presents the results on extragalactic gamma rays. One signature that we do not discuss here is that of neutrino fluxes from the annihilation/decays of DM accumulated in the center of the Sun: we refer the reader to~\cite{DMnu,wimpsim}, where they have been discussed in a spirit very similar to the one of the present work, and we intend to upgrade those former results in the light of the current developements in upcoming work.

Several of these parts contain innovations with respect to the existing literature. For instance, the comparison among Monte Carlo generators; the propagation halo functions for $e^\pm$, which allow to take into account,  in a semi-analytic way, point-dependent energy losses and therefore to compute much more precisely the fluxes of charged cosmic rays and (above all) ICS photons; the introduction of a formalism in terms of (other) halo functions to compute the flux of IC $\gamma$ rays; the study of the impact of different choices for the model of extragalactic background light on the predicted fluxes of extragalactic gamma rays...

\medskip

All our numerical results are available at the \myurl{www.marcocirelli.net/PPPC4DMID.html}{website} referenced in~\cite{website}.
So finally in Section~\ref{summary} we give a summary of these provided numerical ingredients and we list the entry points in the text for the main recipes.

Of course, many refined numerical tools which allow to (directly or indirectly) compute Dark Matter indirect detection signatures have been developed in the latest decades. Among them, GALPROP~\cite{galprop}, DarkSUSY~\cite{darksusy}, MicrOMEGAs~\cite{micromegas}, IsaTools~\cite{isatools}, WimpSim~\cite{wimpsim}...
Rather than focusing on a particular DM model, we try to be model-independent 
and parameterize the observables in terms of the DM mass, of the DM decay or annihilation rates and channels,
as well as in terms of a few uncertain astrophysical assumptions. 
We prefer, whenever possible, a semi-analytical treatment that allows us to keep track of the approximations and choices that we make along the way. Also, we aim at providing the reader with ready-to-use final products, as opposed to the generating code. We make an effort to extend our results to large, multi-TeV DM masses (recently of interest because of possible multi-TeV charged cosmic ray anomalies) and small, few-GeV DM masses (recently discussed because of hints from DM direct detection experiments), at the edge of the typical WIMP window. Above all, our aim is to provide a self-consistent, independently computed, comprehensive set of results for DM indirect detection. Whenever possible, we have compared with existing codes, finding good agreement or improvements.

\section{Dark Matter distribution in the Galaxy}
\label{DMdistribution}

For the galactic distribution of Dark Matter in the Milky Way we consider several possibilities. 
The Navarro, Frenk and White (NFW)~\cite{Navarro:1995iw} profile (peaked as $r^{-1}$ at the Galactic Center (GC)) is a traditional benchmark choice motivated by N-body simulations. The Einasto~\cite{Graham:2005xx, Navarro:2008kc} profile (not converging to a power law at the GC and somewhat more chubby than NFW at kpc scales) is emerging as a better fit to more recent numerical simulations; the shape parameter $\alpha$ varies from simulation to simulation, but 0.17 seem to emerge as a central, fiducial value, that we adopt. Cored profiles, such as the truncated Isothermal profile~\cite{Begeman,Bahcall:1980fb} or the Burkert profile~\cite{Burkert}, might be instead more motivated by the observations of galactic rotation curves, but seem to run into conflict with the results of numerical simulations. On the other hand, profiles steeper that NFW had been previously found by Moore and collaborators~\cite{Moore04}.

As long as a convergent determination of the actual DM profile is not reached, it is useful to have at disposal the whole range of these possible choices when computing Dark Matter signals in the Milky Way.
The functional forms of these profiles read:
\begin{equation}
\begin{array}{rrcl}
{\rm NFW:} & \rho_{\rm NFW}(r)  & = & \displaystyle \rho_{s}\frac{r_{s}}{r}\left(1+\frac{r}{r_{s}}\right)^{-2} \\[4mm]
{\rm Einasto:} & \rho_{\rm Ein}(r)  & = & \displaystyle \rho_{s}\exp\left\{-\frac{2}{\alpha}\left[\left(\frac{r}{r_{s}}\right)^{\alpha}-1\right]\right\} \\[4mm]
{\rm Isothermal:} &  \rho_{\rm Iso}(r) & = & \displaystyle \frac{\rho_{s}}{1+\left(r/r_{s}\right)^{2}} \\[4mm]
{\rm Burkert:} & \rho_{\rm Bur}(r) & = & \displaystyle  \frac{\rho_{s}}{(1+r/r_{s})(1+ (r/r_{s})^{2})} \\[4mm]
{\rm Moore:} &  \rho_{\rm Moo}(r)  & = & \displaystyle \rho_{s} \left(\frac{r_s}{r}\right)^{1.16} \left(1+\frac{r}{r_s}\right)^{-1.84} 
\end{array}
\label{eq:profiles}
\end{equation} 
Numerical DM simulations that try to include the effects of the existence of baryons have consistently found modified profiles that are steeper in the center with respect to the DM-only simulations~\cite{DMandBaryons}. Most recently, \cite{Tissera} has found such a trend re-simulating the haloes of~\cite{Graham:2005xx, Navarro:2008kc}: steeper Einasto profiles (smaller $\alpha$) are obtained when baryons are added. To account for this possibility we include a modified Einasto profile (that we denote as EinastoB, EiB in short in the following) with an $\alpha$ parameter of 0.11. All profiles assume spherical symmetry~\footnote{Numerical simulations show that in general halos can deviate from this simplest form, and the isodensity surfaces are often better approximated as triaxial ellipsoids instead (e.g. \cite{Jing:2002np}). For the case of the Milky Way, however, it is fair to say that at the moment we do not have good observational determinations of its shape, despite the efforts already made studying the stellar tidal streams, see~\cite{Law:2009yq}. Thus the assumption of spherical symmetry, in absence of better determinations, seems to be still well justified. Moreover, it is the current standard assumption in the literature and we therefore prefer to stick to it in order to allow comparisons. In the future, the proper motion measurements of a huge number of galactic stars by the planned GAIA space mission will most probably change the situation and give good constraints on the shape of our Galaxy's DM halo, e.g. \cite{Gnedin:2005pt}, making it worth to reconsider the assumption. For what concerns the impact of non-spherical halos on DM signals, charged particles signals are not expected to be affected, as they are sensistive to the local galactic environment. For an early analysis of DM gamma rays al large latitudes see~\cite{Calcaneo}.} and $r$ is the coordinate centered in the Galactic Center.

\begin{figure}[t]
\begin{minipage}{0.5\textwidth}
\centering
\includegraphics[width=\textwidth]{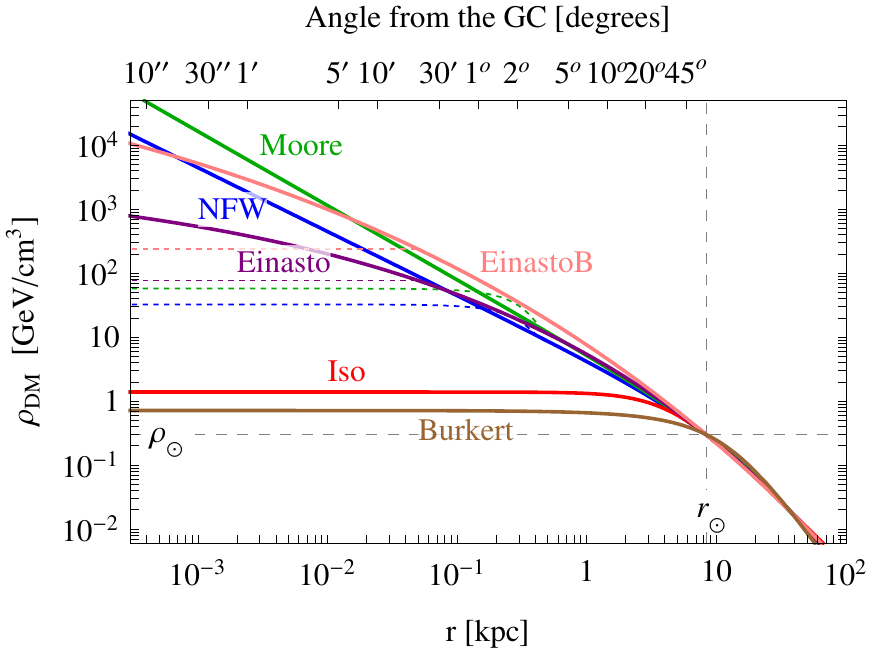}
\label{fig:DMprofiles}
\end{minipage}
\quad
\begin{minipage}{0.4\textwidth}
\centering
\footnotesize{  
\begin{tabular}{l|crc}
DM halo & $\alpha$ &  $r_{s}$ [kpc] & $\rho_{s}$ [GeV/cm$^{3}$]\\
  \hline \\[-1mm]
  NFW & $-$ & 24.42 & 0.184 \\
  Einasto & 0.17 & 28.44 & 0.033 \\
  EinastoB & 0.11 & 35.24 & 0.021 \\
  Isothermal & $-$ & 4.38 & 1.387 \\
  Burkert & $-$ & 12.67 & 0.712 \\
  Moore & $-$ & 30.28 & 0.105 
 \end{tabular}} %\caption{\em Parameters of the DM profiles.}
\label{tab:profiles}
\end{minipage}
\vspace{-5mm}
\caption{\em \small \label{fig:DMprofiles} {\bfseries DM profiles} and the corresponding parameters to be plugged in the functional forms of eq.~(\ref{eq:profiles}). The dashed lines represent the smoothed functions adopted for some of the computations in Sec.~\ref{propagationapprox}. Notice that we here provide 2 (3) decimal significant digits for the value of $r_s$ ($\rho_s$): this precision is sufficient for most computations, but more would be needed for specific cases, such as to precisely reproduce the $J$ factors (discussed in Sec.\ref{promptgamma}) for small angular regions around the Galactic Center.}
\end{figure}

Next, we need to determine the parameters $r_s$ (a typical scale radius) and $\rho_s$ (a typical scale density) that enter in each of these forms. Instead of taking them from the individual simulations, we fix them by imposing that the resulting profiles satisfy the findings of astrophysical observations of the Milky Way. Namely, we require:
\begin{itemize}
\item[-] The density of Dark Matter at the location of the Sun $r_\odot= 8.33$ kpc (as determined in~\cite{rSun}; see also~\cite{rSun2}~\footnote{The commonly adopted value used to be 8.5 kpc on the basis of~\cite{Kerr}.}) to be $\rho_\odot = 0.3$ GeV/cm$^{3}$. This is the canonical value routinely adopted in the literature (see e.g.~\cite{JungmanReview,BertoneReview,PDG}), with a typical associated error bar of $\pm 0.1$ GeV/cm$^{3}$ and a possible spread up to $0.2 \to 0.8$ GeV/cm$^{3}$ (sometimes refereed to as `a factor of 2'). Recent computations have found a higher central value and possibly a smaller associated error, still subject to debate~\cite{CatenaUllio,deBoer, Nesti,deBoer2}. 
\item[-] The total Dark Matter mass contained in 60 kpc (i.e.\ a bit larger than the distance to the Large Magellanic Cloud, 50 kpc) to be $M_{60}\equiv 4.7 \times \, 10^{11} M_\odot$. This number is based on the recent kinematical surveys of stars in SDSS~\cite{SDSSMWmass}. We adopt the upper edge of their 95\% C.L.  interval to conservatively take into account that previous studies had found somewhat larger values (see e.g.~\cite{SakamotoMWmass,PrzybillaMWmass}).
\end{itemize}
The parameters that we adopt and the profiles are thus given explicitly in fig.~\ref{fig:DMprofiles}. Notice that they do not differ much (at most 20\%) from the parameter often conventionally adopted in the literature (see e.g.~\cite{BertoneReview}), so that our results presented below can be quite safely adopted for those cases. 

As well known, the profiles differ most in the inner part of the galactic halo, close to the galactic center, while they are quite self-similar above a few kiloparsecs, and in particular around the location of the Earth. As a consequence, DM signals from the inner Galaxy (e.g.\ gamma ray fluxes from regions a few degrees around the GC) will be more sensitive to the choice of profile than DM signals that probe the local environment (e.g.\ the fluxes of high energy positrons, produced at most a few kpc away from the Earth) or that probe regions distant from the GC (e.g.\ gamma rays from high latitudes). 

\medskip

Notice that we will not consider throughout the paper (with the exception of Sec.~\ref{extragalactic}) the potential contribution from galactic DM substructures. It is well known that the Cold DM paradigm predicts DM halos to contain in a hierarchical fashion a copious number of subhalos, something which is clearly demonstrated by high resolution N-body simulations, e.g.~\cite{Ghigna:1999sn}. 
Most often this is taken into account via an effective overall boost factor that multiplies the fluxes (see~\cite{boost} for the earliest analyses). In reality, however, the phenomenological implications of substructure are more complicated than that.
Indeed, the intensity and morphology of the DM annihilation signal is highly sensitive to the way the substructure mass function and the subhalo concentration parameters are extrapolated down to several orders of magnitude below the actual resolution of the numerical simulations. As an effect of propagation and, as a consequence, of the different galactic volumes that contribute to the signal at Earth, the boost factor can be energy dependent and is in general different for different species (e.g. positrons vs antiprotons)~\cite{Lavalle}.  
Moreover, in models in which the DM annihilation rate is enhanced by the Sommerfeld effect (see~\cite{Sommerfeld,MDMastro, CKRS}, and then~\cite{AHFSW,Sommerfeld2}) there are claims that the contribution from substructures might outweigh the DM annihilation signal from the smooth main halo alone, e.g. \cite{Kuhlen:2009kx}. The gamma ray signal of DM annihilation from the Galaxy with a simply modeled population of subhalos is presented in~\cite{Hutsi:2010ai}. This is however still an active field of research and before the situation is somewhat better established we feel it too early to complicate the calculations with a highly uncertain additional component.

%%%%%%%%%%%%%%%%%%%%%%%%%%%%%%%%%%%%%%%%%%%%%%%%%%
%%%%%%%%%%%%%%%%%%%%%%%%%%%%%%%%%%%%%%%%%%%%%%%%%%

\section{Fluxes at production}
\label{primary}

We consider DM annihilations (parameterized by the DM DM cross section $\sigma v$)
and decays (described by the DM decay rate $\Gamma=1/\tau$) into the following primary channels:
\beq 
\begin{array}{ll}
e_L^+e_L^-,\ 
e_R^+e_R^-,\
\mu_L^+\mu_L^-,\ 
\mu_R^+\mu_R^-,\ 
\tau_L^+\tau_L^-,\ 
\tau_R^+\tau_R^-,\\[3mm]
q \bar q, \ 
c \bar c, \ 
b \bar b, \  
t \bar t, \ 
\gamma \gamma,\ 
g g,  \\[3mm]
W_L^+ W_L^-,\ 
W_T^+W_T^-,\ 
Z_LZ_L,\ 
Z_T Z_T, \\[3mm]
hh, \\[3mm]
\nu_e \bar\nu_e, \ 
\nu_\mu \bar\nu_\mu, \ 
\nu_\tau \bar\nu_\tau, \\[3mm] 
VV \to 4e, \
VV \to 4\mu, \
VV \to 4\tau,
\end{array}
\label{primarychannels}
\eeq
where  $q ={u,d,s}$ denotes a light quark and 
$h$ is the Standard Model (SM) Higgs boson, with its mass fixed at 125 GeV~\cite{higgsupdate04072012}.
The last three channels denote models in which the annihilation or 
decay first happens into some new (light) boson 
$V$ which then decays into a pair of leptons, 
along the lines of the models proposed in~\cite{AHFSW,PR}.

The particles produced in Dark Matter annihilation/decay
will be provided with  parton showers and  hadronization, in such a way to obtain
the fluxes of 
$e^\pm,\bar p, \bar d, \gamma, \stackrel{\mbox{{\tiny (--)}}}{\nu}_{e,\mu,\tau}$ 
at the production point. 
To this goal, we shall use the two most widely used 
Monte Carlo simulation programs:
{\sc Pythia}~\cite{pythia8} (version 8.135), already used in most DM studies 
carried out so far, and {\sc Herwig}~\cite{herwig6} (version 6.510).
In fact, the algorithms implemented in {\sc Herwig} and {\sc Pythia} are
quite different, in both parton showers and hadronization, which
makes compelling the employment of both codes for the 
sake of comparing and estimating the Monte Carlo 
uncertainty on a prediction. 

\bigskip

\noindent {\bf Polarizations and EW corrections.} Before moving forward, a brief discussion on polarizations and ElectroWeak corrections is in order.
In their current versions, {\sc Herwig} and {\sc Pythia} contain lepton- or 
$W$-pair production processes in resonance decays,
but, unlike the channels listed in eq.~(\ref{primarychannels}),
leptons and vector bosons are treated as unpolarized. Furthermore, in the terms which
will be clarified later, parton showers 
include gluon and photon radiation,
but not the emissions of $W$'s and $Z$'s.
However, as discussed in ~\cite{CCRSSU} (see also~\cite{EWbrem}) electroweak radiation effects
can be particularly relevant for the leptonic and $\gamma\gamma$
channels. In fact, the emission of $W$'s and $Z$'s 
leads to further hadrons in the final state, and therefore
it significantly modifies the flux of $\gamma$'s 
and $e^\pm$ at energies $E\ll M$, $M$ being the DM mass. Moreover,
$W/Z$ radiation leads to a 
$\bar p$ contribution, which is instead absent 
if weak corrections are neglected; this is also true for the 
the neutrino channels, that also yield
$e^\pm$'s, $\gamma$'s and $\bar p$'s.
In fact, $W/Z$ radiation results in contributions  
enhanced by one or more powers of $\ln (M/M_{W,Z})$, with $M\gg M_{W,Z}$,
which do not depend on the DM model.
Also, in the EW radiation processes, 
the polarization of the 
leptons ($L$eft- or $R$ight-handed fermion)
and of the massive vectors ($T$ransverse or $L$ongitudinal) plays a role.

Electroweak emissions can be added to 
the {\sc Pythia} event generator following the lines of \cite{CCRSSU}, 
wherein one accounts for 
the logarithmically-enhanced contributions due 
to $W/Z$ radiation,
at leading order in the electroweak coupling constant, as well 
as leptons/vector-bosons polarizations.
Of course, the corresponding unpolarized channels can 
be recovered by means of the following averages: 
$$e^+e^- =\frac{e_L^+e_L^- + e_R^+e_R^-}{2},\qquad W^+W^-=\frac{2 \, W_T^+ W_T^- + W_L^+W_L^-}{3},$$
and analogously for $\mu^+\mu^-$, $\tau^+\tau^-$ and $ZZ$.
As electroweak radiation has not been implemented yet in {\sc Herwig}, 
when comparing the two Monte Carlo codes and presenting
the primary fluxes in section 3.2, we will assume unpolarized
particles and turn $W/Z$ emissions off. Later on, we will 
employ only the {\sc Pythia} code for an extensive study of the particle 
fluxes, and therein the contribution of the large leading
electroweak logarithms will be taken into account. These effects of $W/Z$ emissions are also included
in the numerical results collected in the \myurl{www.marcocirelli.net/PPPC4DMID.html}{website}~\cite{website}.

\bigskip

Let us now go back to the discussion of the list of primary channels in eq.~(\ref{primarychannels}). Our approach is to consider all these channels on an equal footing, in a manner which is completely independent of the DM model.
In any given model,
annihilation or decay branching ratios into 
the specific channels will instead be dictated by 
the underlying theory. Some channels \
(such as $\gamma \gamma$, $\nu\bar\nu$, $gg$) 
are `unusual' as they are often suppressed in many models,  
but from a model-independent point 
of view they are as viable as any other, so that 
we shall include them and discuss them further below. 
Operationally, as discussed e.g.\ in~\cite{CKRS}, 
$s$-wave non-relativistic DM DM annihilation can 
be seen as equivalent to the decay of 
a $\mathscr{D}$ resonance with mass $M_{\mathscr{D}} = 2M_{\rm DM}$,
where $M_{\rm DM}$ is the DM particle mass.
Decays of $\mathscr{D}$ into any pair of Standard Model
particles can therefore be computed and implemented in  
Monte Carlo generators, along the lines which will be described 
hereafter.

\smallskip

The annihilation into SM Higgs, tau, photon and gluon pairs deserves further comments.
 
\noindent $\boldsymbol{hh.}$ Now that a particle consistent with the SM Higgs boson has been discovered~\cite{higgsupdate04072012}, we need more than ever to include the corresponding channel in the list of possible annihilations, and indeed we do so. On the other hand, the detailed properties of such a particle are obviously still under very active investigation, so that we have to make some guesses/assumptions. For its mass, we assume $m_h = 125$ GeV. For its branching ratios, we take those predicted by the Standard Model and embedded in the MonteCarlo codes. The values in {\sc Herwig} and {\sc Pythia} can differ by up to 25\% for a light higgs: {\sc Herwig} has a slightly smaller BR into $WW$ and $ZZ$ with respect to {\sc Pythia}, while it has has a slightly larger BR into $b \bar b$.
Such discrepancies are due to the different accuracy which is used to
compute the partial widths 
(see, e.g., \cite{daniela}). For example, in the decays
into $WW/ZZ$ {\sc Pythia} allows both vector bosons to be off-shell, 
whereas in {\sc Herwig} at least one is forced to be on-shell.
In the rate of $h\to b\bar b$ processes, {\sc Herwig} includes also the resummation
of mass logarithms $\sim\alpha_S^n(m_h^2)\ln^n(m_h/m_b)$, 
which are not resummed in {\sc Pythia}.
Hereafter, we shall stick to the default branching fractions for the two codes: more
accurate results can be obtained by means of the {\sc Hdecay} program \cite{hdecay}, 
whose use is nevertheless beyond the scopes of the present paper.
Above all, we stress that the mentioned branching ratios are obtained for the Standard Model Higgs
boson and that, Beyond the Standard Model, the Higgs decay fractions will clearly be different. Should the investigations at the LHC highlight a non-SM behavior of the $h$ particle, these assumptions will clearly have to be revised. 
For BSM scenarios both {\sc Herwig} and {\sc Pythia} are able to read external data files, provided e.g. by the {\sc Isajet/SuGra}~\cite{isasugra} package, containing masses, lifetimes and branching ratios of the particles predicted by the chosen new physics model, including non standard Higgs bosons. A study of Dark Matter annihilation and decay in scenarios such as the MSSM is therefore feasible along the same lines as our current study, but in this work we prefer not to make any hypotheses on possible Dark Matter models and we shall stick to decays only into Standard Model particles.

\noindent $\boldsymbol{\tau^+\tau^-.}$ As for $\tau$ leptons, {\sc Herwig} and {\sc Pythia} treat them as unpolarized
and implement the Standard Model three-body decay matrix elements. Alternatively, the two Monte Carlo codes can be interfaced with the {\sc Tauola} package~\cite{tauola}, which fully includes polarization effects and
implements several lepton and hadron decay modes, by means of hadronic
matrix elements. In the following, we shall nonetheless use the standard {\sc Herwig} and {\sc Pythia} routines even for the purpose of $\tau$ decays and subsequent showers and hadronization. In fact, this is a reasonable
approximation for the observables which we shall investigate, namely the
hadron/lepton/photon energy fraction in the Dark Matter rest frame and averaged over many, many events.
A remarkable impact of the inclusion of the $\tau$ polarization should instead
be expected if one looked at other quantities, such as angular
correlations between the $\tau$ decay products from the same event.
Furthermore, for the sake of consistency,
we prefer to use everywhere in our analysis the modelling of hadronization
contained in {\sc Herwig} and {\sc Pythia}, rather than the non-perturbative
matrix elements incorporated in the {\sc Tauola} package.

\noindent $\boldsymbol{\gamma\gamma.}$ We include $\gamma\gamma$ as a primary channel: Dark Matter, being 
dark, has no tree-level coupling to photons, but  
$\gamma\gamma$ production can occur at one loop. This is not to be confused with photons
emitted by charged particles or produced in
three-body annihilations or radiative hadron decays, such as $\pi^0\to\gamma\gamma$.
Photons in final-state showers  
or hadron decays are of course included in the fluxes 
yielded by {\sc Herwig} and {\sc Pythia} (see below
for more details).
Including instead DM annihilation into three-body 
final states would require a specific model 
of Dark Matter (see e.g.~\cite{IB,3body}), 
whereas in this work we shall stick to model-independent results.

\noindent $\boldsymbol{gg.}$ Neglecting the case of colored Dark Matter, the DM\ DM~$\to gg$ mode can also 
take place only at one loop. In the Monte Carlo codes which will be employed later on, we shall
implement the ${\mathscr{D}}\to gg$ decay in the same fashion as $h\to gg$, i.e.
with an effective ${\mathscr{D}}gg$ vertex, assuming that DM is color neutral.

\bigskip

In the following, we shall first review the basics of 
parton cascades (section 3.1), then we will compare
{\sc Herwig} and {\sc Pythia} fluxes for a few channels and
fixed $M_{\rm DM}$ (3.2)
and finally present the {\sc Pythia} fluxes for several
modes and values of the DM mass (3.3).

\subsection{Parton shower algorithms} 

In this subsection we review the basics of the 
parton shower algorithms in {\sc Herwig} and {\sc Pythia}, 
focussing in particular on the differences, 
with the aim of gaining some insight in the interpretation of
possible discrepancies in the predictions of the two codes.

\medskip

{\bf QCD (quark, gluon) final state radiation:} Showering algorithms 
rely on the universality of the elementary branching
probability in the soft or collinear approximation.
Referring first to quark/gluon final state radiation,
the probability to radiate a soft or collinear parton 
reads:
\begin{equation}
  \label{elem}
  d{\cal P}={{\alpha_S(k_T)}\over{2\pi}}{{dQ^2}\over{Q^2}}\ 
  P(z)\  dz\ 
 {{ \Delta_S(Q^2_{\mathrm{max}},Q^2)}\over {\Delta_S(Q^2,Q_0^2)}}.
\end{equation}
In (\ref{elem}), $P(z)$ is the Altarelli--Parisi splitting
function, $z$ is the energy fraction of the radiated parton with respect to the
emitter, $Q^2$ is the evolution variable of the shower.
In {\sc Herwig} \cite{herwig6}, $Q^2$ is an energy-weighted angle \footnote{
In the {\sc Herwig} showering frame, $Q^2\simeq E^2(1-\cos\theta)$, where $E$ is the
energy of the splitting parton and $\theta$ is the emission angle
\cite{marweb}.}, 
which corresponds
to angular ordering in the soft limit \cite{marweb}.
%In {\sc Pythia} \cite{pythia}, $Q^2$ is the virtuality of
%the radiating parton, with an option to veto branchings that do not
%fulfil the angular ordering prescription. 
%Moreover, the latest {\sc Pythia} version
%offers, as an alternative, the possibility to order 
%final-state showers according to the transverse momentum of the emitted
%parton with respect to the parent one \cite{pythiakt}.
%In either options, the {\sc Pythia} evolution variable is not completely
%equivalent to angular ordering in the soft limit:
%although in several cases the actual ordering does not make
%really big changes, when comparing with experimental observables sensitive to
%colour coherence, as done e.g.\ 
%in Ref.~\cite{cdf}, {\sc Herwig} agrees with the data  
%better than {\sc Pythia}.
%Transverse-momentum ordering, included in the latest {\sc Pythia}
%version, yields a better description of angular ordering. 
%Nevertheless,
%as discussed in \cite{bcd}, discrepancies with respect to {\sc Herwig} are still
%present when considering, e.g., the so-called non-global observables 
%\cite{ng}, sensitive to radiation in a limited part of phase space,
%such as the transverse energy flow in a rapidity gap.
%Throughout this paper, we shall stick to the
%{\sc Pythia} default shower modelling, i.e. ordering in virtuality and
%veto on possibly non-angular-ordered emissions.
In the traditional {\sc Pythia}~\cite{pythia} algorithm, implemented
in fortran language, $Q^2$ is the virtuality of
the radiating parton, with an option to veto branchings that do not
fulfil the angular ordering prescription. Moreover, the latest version
offers, as an alternative, the possibility to order final-state showers according to the transverse momentum ($k_T$) of the emitted
parton with respect to the parent one~\cite{pythiakt}.
$k_T$ is also the evolution variable in the novel object-oriented  {\sc Pythia} 8 generator~\cite{pythia8}, written in C++.
In either options, the {\sc Pythia} evolution variable is not completely
equivalent to angular ordering in the soft limit:
although in several cases the actual ordering does not make
really big changes, when comparing with experimental observables sensitive
to
colour coherence, as done e.g.\ in Ref.~\cite{cdf}, {\sc Herwig} agrees with the data better than {\sc Pythia}.
$k_T$-ordering yields a better inclusion of angular ordering, but nevertheless,
as discussed in~\cite{bcd}, discrepancies with respect to {\sc Herwig} are
still
present when considering, e.g., the so-called non-global observables~\cite{ng}, sensitive to the radiation in a limited part of the phase space, such as the transverse energy flow in a rapidity gap.
Throughout this paper, we shall use the
{\sc Pythia} 8 shower modelling, i.e. ordering in transverse momentum.

In (\ref{elem}) $\Delta_S(Q_1^2,Q_2^2)$ is the Sudakov form
factor, expressing the probability of evolution from $Q_1^2$ to $Q_2^2$ 
with no resolvable emission.
In diagrammatic terms, it 
sums up all virtual and unresolved real emissions to all orders.
In particular, the ratio of form factors
in (\ref{elem}) represents the probability that the considered emission
is the first, i.e. there is no emission
between $Q^2$ and $Q^2_{\mathrm{max}}$, where $Q^2_{\mathrm{max}}$ is set by
the hard-scattering process. $Q_0^2$ is instead the value of $Q^2$ at which
the shower evolution is terminated and hadronization begins.
The Sudakov form factor is given by the following equation:
\begin{equation}\label{sudakov}
\Delta_S(Q^2_{\mathrm{max}},Q^2)=\exp\left\{-\int_{Q^2}^{Q^2_\mathrm{max}}
{\frac{dk^2}{k^2}}
\int_{z_{\mathrm{min}}}^{z_{\mathrm{max}}}{dz\ \frac{\alpha_S(z,k^2)}{2\pi}\ 
P(z)}\right\}.
\end{equation}
In Eq.~(\ref{sudakov}), 
the lower integration limit is related to the evolution variable $Q$ 
and the shower cutoff $Q_0$: it is given
by $z_{\mathrm{min}}=Q_0^2/Q^2$ in {\sc Pythia}  
and $z_{\mathrm{min}}=Q_0/Q$ in {\sc Herwig}, consistently with virtuality/angular
ordering. The upper limit is instead $z_{\mathrm{max}}=1-z_{\mathrm{min}}$.
For any $Q$, such conditions
imply a larger $z$-evolution range in {\sc Pythia} with respect to {\sc Herwig}.

For multiparton radiation, 
iterating the branching probability (\ref{elem}) is equivalent to performing
the resummation of soft- and collinear-enhanced radiation.
As discussed, for example, in \cite{cmw} in the framework of the {\sc Herwig}
event generator, parton shower algorithms resum leading logarithms (LL) in the
Sudakov exponent, and include a class of subleading 
next-to-leading logarithms (NLL) as well.
The strong coupling constant in (\ref{elem}) is evaluated at the
transverse momentum ($k_T$) of the radiated quark/gluon with respect to the
parent parton: in this way, one includes in the showering
algorithm further subleading soft/collinear enhanced logarithms
\cite{cmw}. Moreover, in {\sc Herwig} the two-loop coupling constant in the
physical CMW scheme \cite{cmw} is used\footnote{The CMW scheme
consists in defining a Monte Carlo strong coupling 
$\alpha_S^{\mathrm{MC}}$ related to the usual $\overline{\mathrm{MS}}$ one
via $ \alpha_S^{\mathrm{MC}}=\alpha_S^{\overline{\mathrm{MS}}}[1+
K\alpha_S^{\overline{\mathrm{MS}}}/(2\pi)]$, where the explicit expression
for $K$ can be found in \cite{cmw}. By means of this rescaling,
the {\sc Herwig} Sudakov form factor includes threshold-enhanced corrections
in the NLL approximation.}; in {\sc Pythia} the $\beta$ function is
instead included in the one-loop approximation.
The actual value of $\alpha_S(M_Z)$ comes after fits to
LEP data, which yield $\alpha_S(M_Z)\simeq 0.116$ in {\sc Herwig}
and $\alpha_S(M_Z)\simeq 0.127$ in {\sc Pythia}.
Different values of $\alpha_S(M_Z)$ would lead
to different total cross sections for 
hard-scattering processes
mediated by the strong interaction, but
have very little impact on differential distributions,
such as the ones which we will investigate,
as long as tuned versions of {\sc Herwig} and {\sc Pythia}
are used (see e.g.\ the discussion in \cite{daniela}).

\smallskip

Also for the purpose of hadronization, the two programs implement very different
models, namely the cluster model \cite{cluster} ({\sc Herwig}), based on
colour preconfinement and closely related to
angular ordering, and the string model \cite{string}
({\sc Pythia}), both depending on a few non-perturbative parameters. 
These parameters, along with other quantities, such as
the shower cutoff or quark and gluon effective masses,
are fitted to experimental data, e.g., $e^+e^-$ or
$p\bar p$ data from LEP and Tevatron experiments
(see, e.g., Refs.~\cite{volker,cdf1}).
In principle, whenever one runs {\sc Herwig} or {\sc Pythia}
at much higher energies, as will be done in the
following, such fits may have to be reconsidered.
In fact, although the hadronization transition is
universal, when using a hadronization model along with
a perturbative calculation or a parton shower algorithm,
one assumes that the hadronization model even
accounts for the missing perturbative contributions, which are clearly
process-dependent.
Therefore, whenever one has data from other experiments, 
one should check that the best-fit parametrizations are still
able to reproduce the data.
In this paper, for the sake of a consistent comparison,
we shall use the default values of the parameters employed in the default versions of {\sc Herwig} and {\sc Pythia},
which  were fitted to LEP data (i.e. at the typical 100 GeV energy scale). In the future, data at higher energies (e.g. from the LHC) will be available and will possibly allow to retune the two Monte Carlo codes to better mimic high-mass Dark Matter annihilation/decay.

{\bf QED (photon) final state radiation: }
Extending the algorithm (\ref{elem}) to include photon
radiation off quarks and leptons, as well as
photon branching into quark or lepton pairs, is straightforward.
{\sc Pythia} (the version 8.135 that we use) include all such processes, {\it but} it does not include photon radiation from $W^+W^-$ final states. We have therefore to add this by hand as discussed in~\cite{CCRSSU} (see footnote 2 in v1 on the arXiv). 
%We refer to one of those authors (A.S.) for further information.
The latest version of {\sc Herwig}, on the other hand, contains 
%However, unlike {\sc Pythia}, which includes all such processes, 
%the latest fortran version of {\sc Herwig} contains 
$q\to q\gamma$ 
branchings, but does not implement
photon emissions off leptons as well as $\gamma\to f\bar f$ 
processes, which are instead present in {\sc Herwig}++ \cite{herpp},
the object-oriented version written in C++, along the
lines of \cite{hamilton}.
In the following we will nevertheless employ the latest fortran version of {\sc Herwig} and delay to future work the use of {\sc Herwig}++. We can indeed anticipate that, in the phase space regions in which
we will mostly be interested, the partial lack of QED radiation in fortran
{\sc Herwig} has a small effect in the comparison with {\sc Pythia}. Also, we should always
keep in mind that, although shower and hadronization are implemented in a different fashion, the two codes have been fitted to agree with LEP data (as discussed above), and therefore possible lacks are somehow compensated by means of suitable parameter tuning.\footnote{We also point out that, although in future perspectives the C++ generators will supersede the fortran ones,
fortran {\sc Herwig} and {\sc Pythia} will still be used as they provide parton shower and hadronization to the so-called `matrix-element
generators' like {\sc Alpgen}~\cite{alpgen} or {\sc Madgraph}~\cite{madgraph}.}

\subsection{Dark Matter fluxes comparing {\sc Herwig} and {\sc Pythia}}

We present the energy spectra of final-state 
particles from DM annihilation, yielded by {\sc Herwig} and {\sc Pythia}. 
For this purpose, as anticipated, we modified the two
codes in order to include the decay of a generic resonance
${\mathscr{D}}\to ab$, in such a way that we are allowed to fix
the mass $M_{\mathscr{D}}=2M$ and specify particles $a$ and $b$.
Moreover, we changed the hadron decay tables, to
make it sure that hadrons, such as kaons or pions, which
are often treated as stable when simulating collider phenomenology,
decay according to the branching ratios quoted in the PDG
\cite{PDG}. Our results will be expressed in terms of 
the energy fraction 
\begin{equation}
x=\frac{K}{M_{\rm DM}},
\end{equation}
where $K$ is the kinetic energy of the final-state stable 
hadrons/leptons/photons in the rest frame of ${\mathscr{D}}$.
We shall plot the particle multiplicity as a function of
the logarithmic energy fraction, i.e. $dN/d\log x$; 
our spectra will be normalized to the average multiplicity 
in the simulated high-statistics event sample.
Also, as pointed out before, this comparison will be carried
out for production of unpolarized particles and without including
any effect of final-state weak boson radiation.
\begin{figure}[t]
\begin{center}
$$\includegraphics[width=0.45\textwidth]{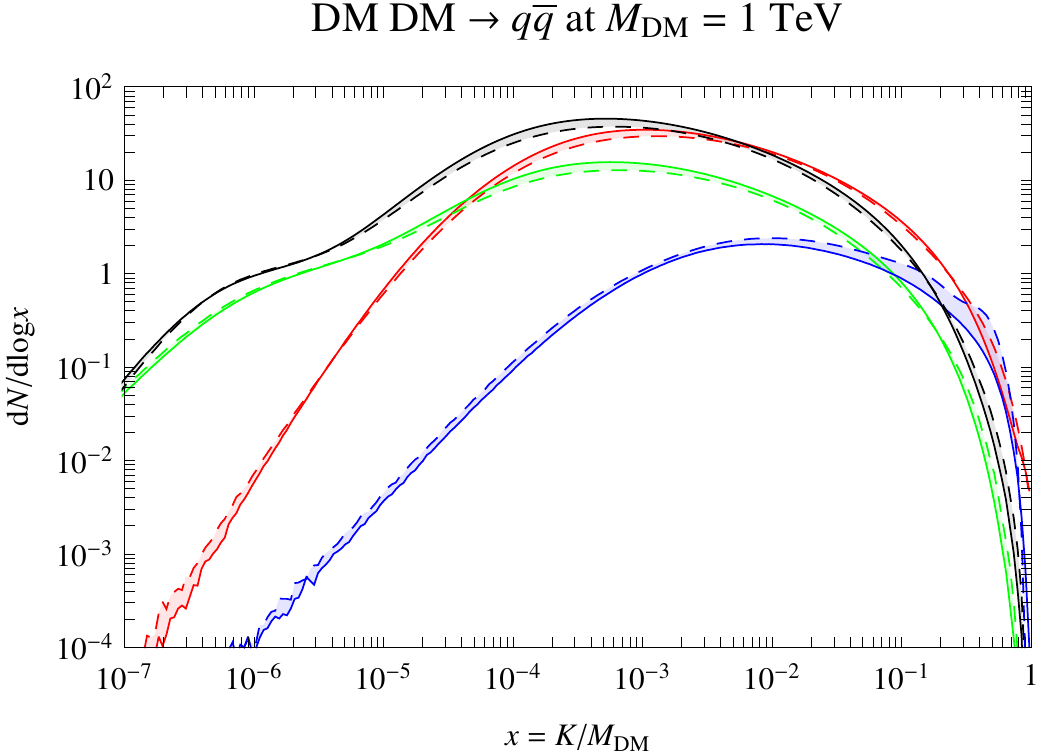}\qquad\includegraphics[width=0.45\textwidth]{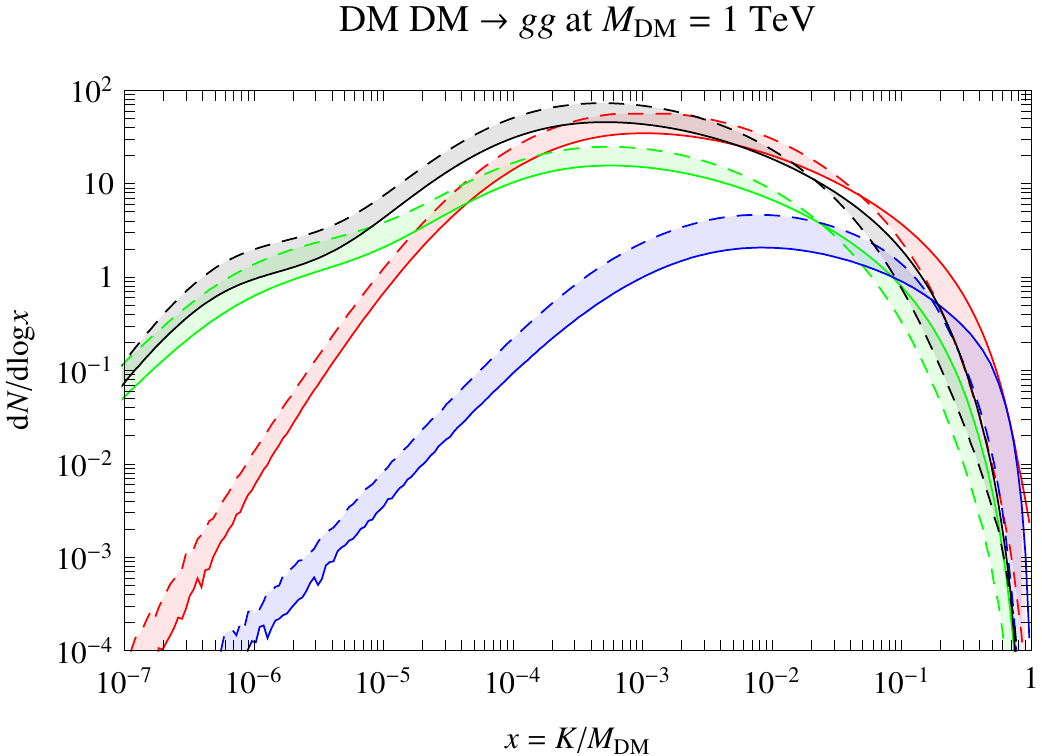}$$
$$\includegraphics[width=0.45\textwidth]{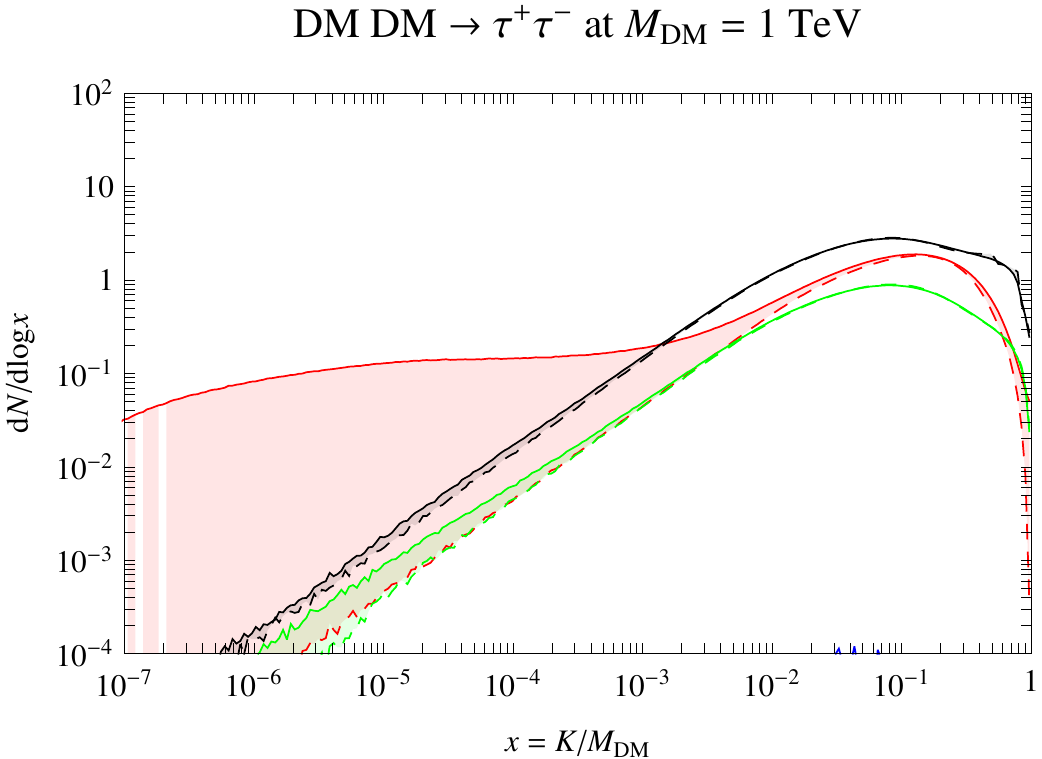}\qquad\includegraphics[width=0.45\textwidth]{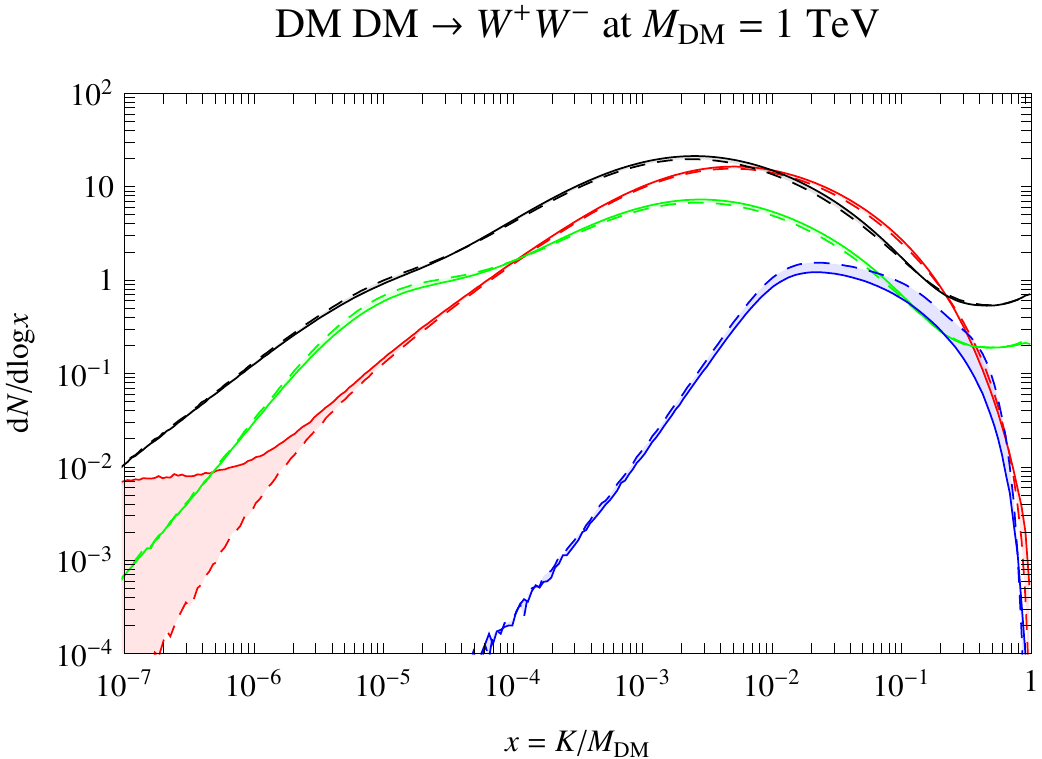}$$
\caption{\label{fig:PHW}\em \small {\bfseries Comparison between Monte Carlo results}: {\sc Pythia} is the continuous line, {\sc Herwig} is dashed.
Photons (red), $e^\pm$ (green), $\bar p$ (blue), $\nu = \nu_e +\nu_\mu+\nu_\tau$ (black).}
\end{center}
\end{figure}

An example of the comparison of the DM fluxes from {\sc Pythia} and {\sc Herwig} is presented
in Fig.~\ref{fig:PHW}, where we show the photon, electron, antiproton
and neutrino $dN/d\log x$ 
spectra for the channels DM DM~$\to q\bar q$, $gg$, $W^+W^-$ 
and $\tau^+\tau^-$. 
In Fig.~\ref{fig:PHW} we have set the DM mass to 
$M_{\rm DM}=1\TeV$, but we can anticipate that 
similar $dN/d\log x$ hold for all DM masses $M_{\rm DM}\gg M_Z,m_t$.
Astrophysical experiments are currently probing  
$K\circa{<} 100\GeV$, whose corresponding range of $x$ depends on the
chosen $M_{\rm DM}$;
in particular, the low-$x$ tails mostly determine the DM signals if $M_{\rm DM}$ is very large.
Overall, we note the following features:
\begin{itemize}
\item For the $q\bar q$ modes there 
is a reasonable agreement between {\sc Pythia} and {\sc Herwig},
for all final-state particles and through the whole $x$ spectrum,
including the low-energy tails.
In fact, although the centre-of-mass energy has been increased to 2 TeV, the
${\mathscr{D}}\to q\bar q$ is similar to $Z/\gamma^*\to q\bar q$ processes at LEP,
which were used when tuning the {\sc Herwig} and {\sc Pythia} user-defined parameters.
Nevertheless, we note some discrepancy, about 20\%, especially in the neutrino 
spectra, as {\sc Pythia} yields overall a higher multiplicity,
and in the $\bar p$ distribution, where {\sc Herwig} is above {\sc Pythia}
especially at large $x$.

\item Some discrepancy, up to a factor of 2, 
is instead found for the $gg$ mode (which is, however, 
presumably not the dominant one in DM phenomenology).
In fact, unlike the $q\bar q$ mode, the ${\mathscr{D}}\to gg$ channel does not
have a counterpart at LEP; the differences
in parton showers and hadronization in {\sc Herwig} and {\sc Pythia}, as well as
the fact that we are running the two codes at a much higher energy with
respect to LEP, may thus
be responsible for this discrepancy.
In detail, as far as the $\gamma$, $e^\pm$ and $\bar p$ spectra are concerned,
{\sc Herwig} is above {\sc Pythia} at small $x$ and below at large $x$; the {\sc Pythia}
neutrino multiplicity is instead above the {\sc Herwig} one in the whole
$x$ range, especially for $x>10^{-5}$.

\item 
Lepton modes (here exemplified by the $\tau^-\tau^+$ case) exhibit 
a significant disagreement, 
especially in the photon spectra, where {\sc Pythia} yields
a remarkably higher multiplicity with respect to {\sc Herwig}
for $x<10^{-2}$.
As we pointed out before, {\sc Pythia} includes $\ell\to\ell\gamma$, $\ell$ being 
a charged lepton, and $\gamma\to f\bar f$ processes, which are
instead absent in {\sc Herwig}, whose photons may come only from
$q\to q\gamma$ or radiative hadron decays.
The lack of this type of $\gamma$ radiation in {\sc Herwig}
might therefore explain the discrepancy in the
photon spectra in the $\tau^+ \tau^- $ channel.

\item 
The fourth panel shows an example of `composite' mode, i.e. the 
DM\ DM~$\to W^+W^-$ channel. This process, where a $W$ pair
is produced via annihilation of colourless particles, is alike
$e^+e^-\to W^+W^-$ at LEP 2, where {\sc Herwig} and {\sc Pythia} have been
carefully tested. In fact, the electron and positron spectra exhibit 
good agreement, and even the antiproton multiplicities are not too
far, although {\sc Herwig} yields a higher $dN/d\log x$ for $x>10^{-3}$.
More visible discrepancies are instead exhibited by the
neutrino spectra, as the {\sc Pythia} multiplicity is higher than
{\sc Herwig} throughout all $x$ range, and by low-$x$ photons,
where {\sc Pythia} is much above {\sc Herwig}.
The latest discrepancy can be traced back to the differences
in photon radiation off leptons commented on above.

\end{itemize}

Similar features 
are observed for other values of the DM mass and for other modes. 
In view of these considerations, mainly the fact that
{\sc Pythia} includes also $\ell\to\ell\gamma$ and $\gamma\to f\bar f$ 
processes and its spectra can be supplemented by large 
electroweak logarithms according to \cite{CCRSSU}, hereafter we shall
employ only {\sc Pythia} to present the spectra for several 
decay/annihilation modes and mass values.
The Monte Carlo uncertainty, gauged from the comparison
between {\sc Pythia} and {\sc Herwig}, can be estimated to be on average about
$\pm 20\%$, with the exception of the (probably negligible) gluon-gluon
channel and photon (neutrino) spectra in the lepton ($WW$) mode, which instead
exhibit a larger disagreement, especially for very small values of $x$.
As pointed out above, this latest discrepancy will likely get milder if
one implemented further QED-type branchings in {\sc Herwig} or possibly
used {\sc Herwig}++.

%%%%%%%%%%%%%%%%%%%%%%%%%%%%%%%%%%%%%%%%%%%%%%%%%%
%%%%%%%%%%%%%%%%%%%%%%%%%%%%%%%%%%%%%%%%%%%%%%%%%%

\subsection{Dark Matter fluxes: results}
\label{fluxesresults}

We therefore compute the fluxes~\footnote{We note that a similar work has been performed recently, limited to gamma rays, in~\cite{Cembranos:2010dm}.} of $e^\pm,\bar p, \bar d, \gamma,\stackrel{\mbox{{\tiny (--)}}}{\nu}_{e,\mu,\tau}$ in a large range of DM masses $M_{\rm DM} = 5\, {\rm GeV} \to 100\, {\rm TeV}$,
by using the {\sc Pythia} event generator, 
and provide them in numerical form on the~\myurl{www.marcocirelli.net/PPPC4DMID.html}{website}~\cite{website}, both in the form of {\sc Mathematica}$^{\tiny{\textregistered}}$ interpolating functions and numerical tables.\footnote{
Analogous results yielded by the {\sc Herwig} code can be obtained by
contacting the authors (G.C. and F.S.).}
Such computing-power demanding results have been obtained using the 
EU Baltic Grid facilities~\cite{BG}. 

\medskip

\begin{figure}[!p]
\begin{center}
\includegraphics[width=\textwidth]{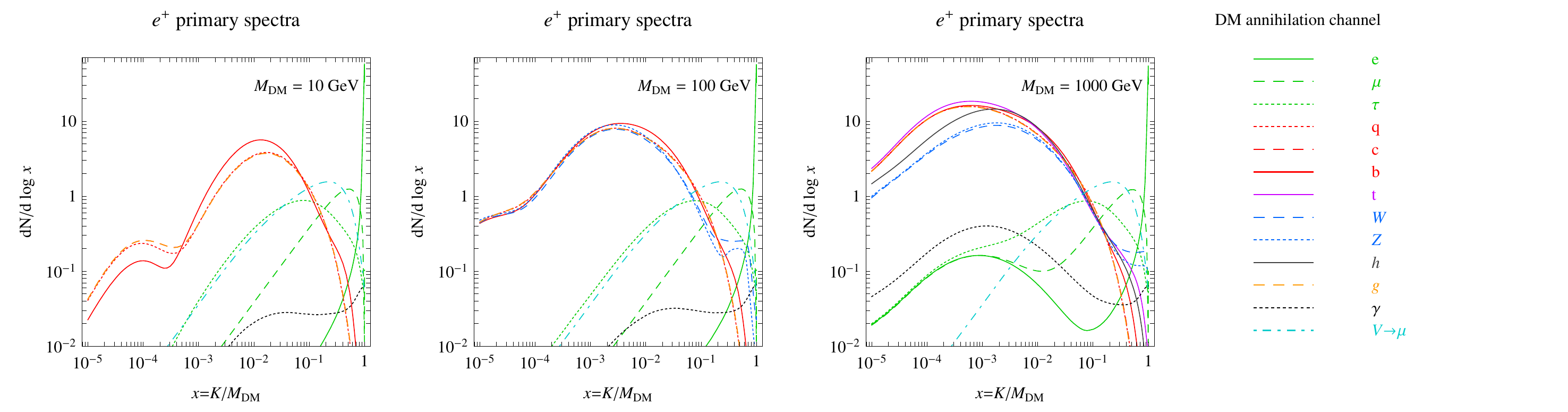}
\includegraphics[width=\textwidth]{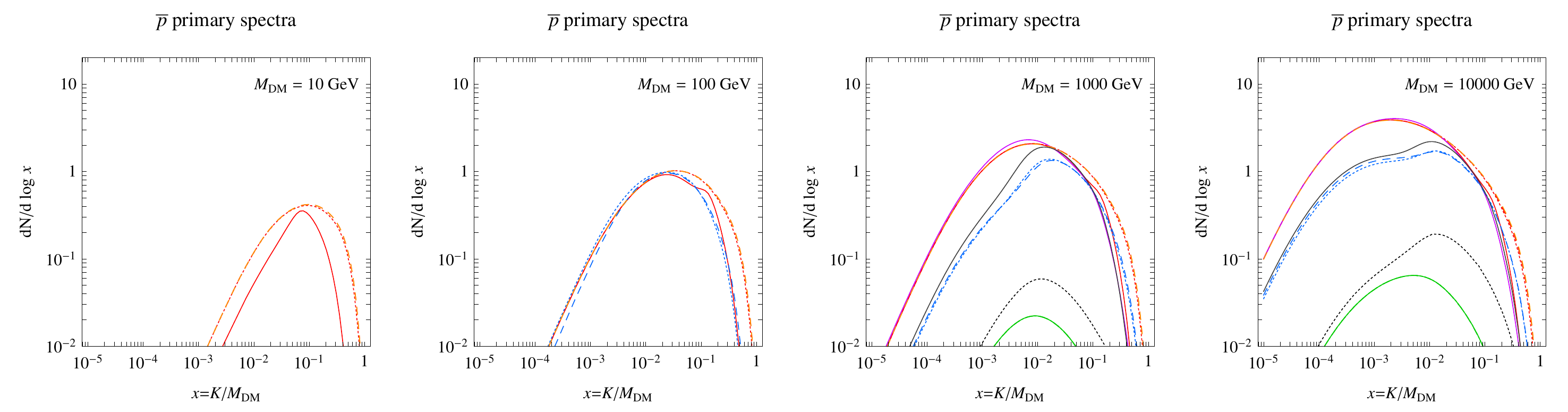}
\includegraphics[width=\textwidth]{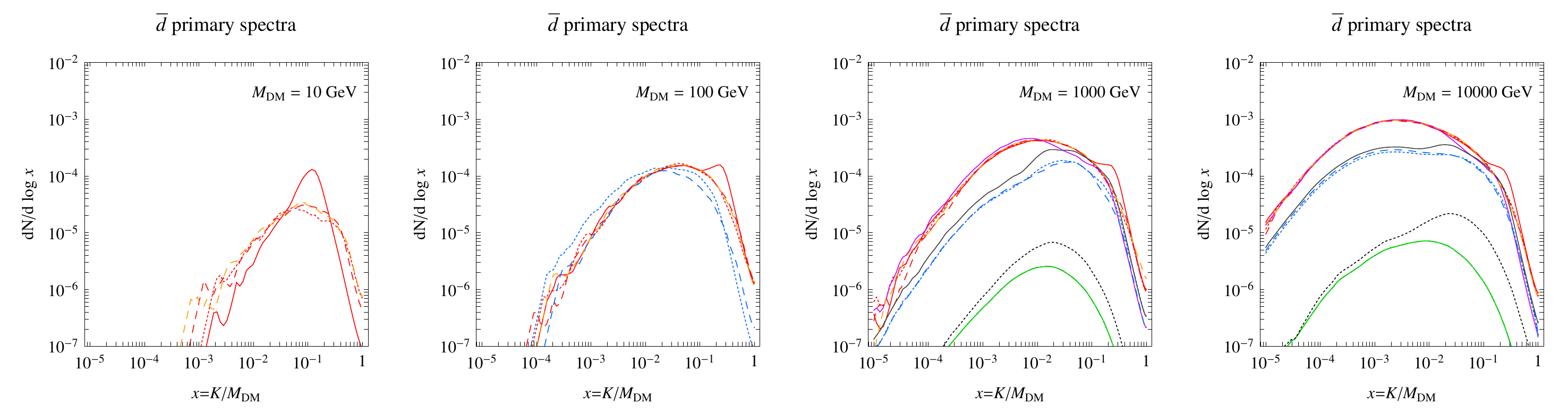}
\includegraphics[width=\textwidth]{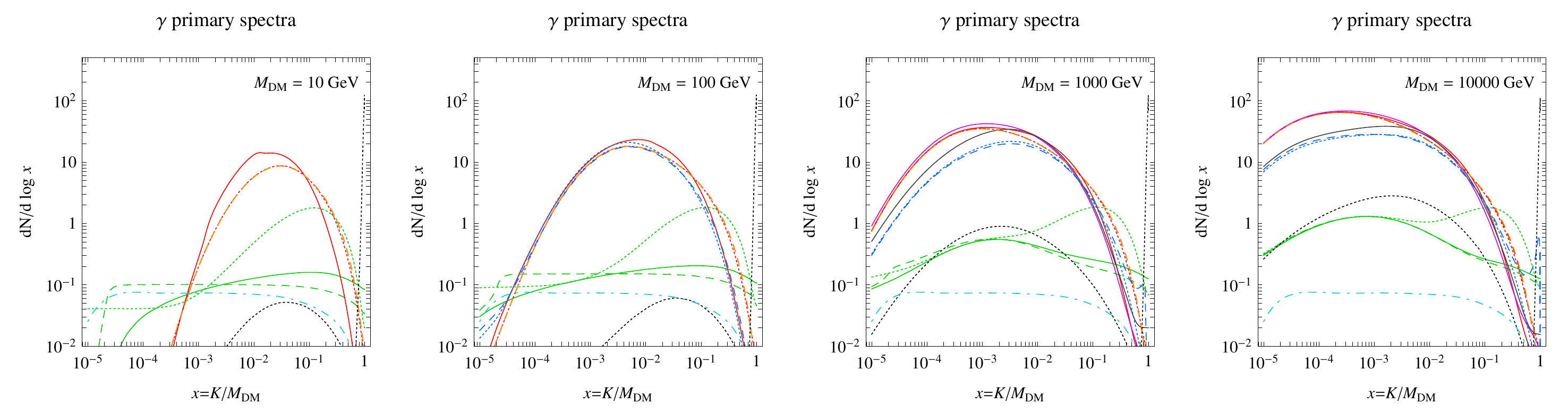}
\includegraphics[width=\textwidth]{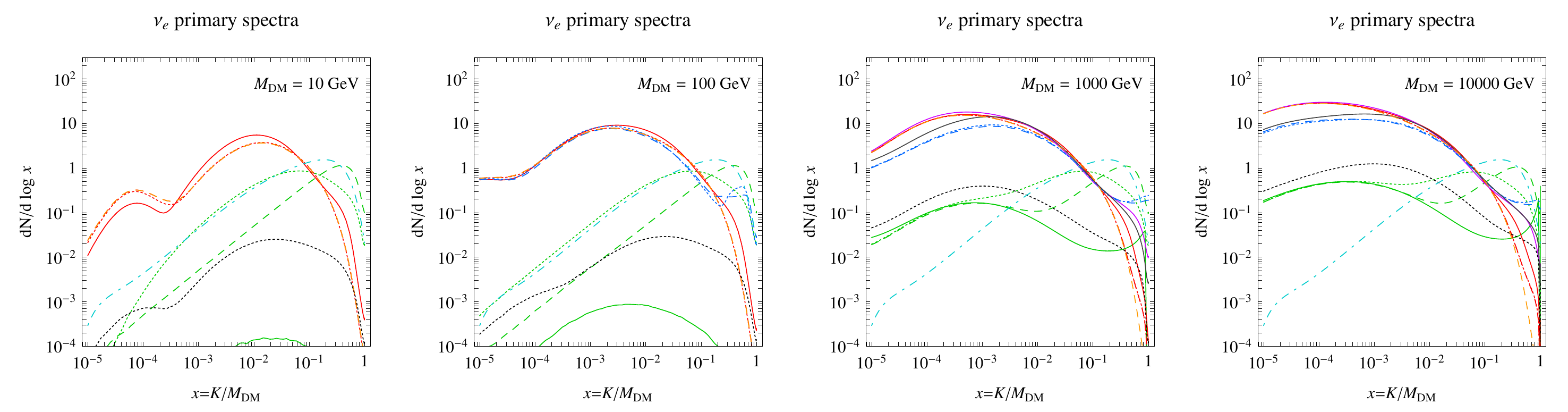}
\caption{\em \small \label{fig:primary} 
{\bfseries Primary fluxes} of $e^\pm$, $\bar p$, $\bar d$, $\gamma$ and $\nu_e$.}
\end{center}
\end{figure}

In fig.~\ref{fig:primary} we present some examples of the spectra produced by the annihilation of two DM particles with mass $M_{\rm DM}$ (normalized per annihilation), for four values of $M_{\rm DM}$. They correspond to the fluxes from the decay of a DM particle with mass $2 M_{\rm DM}$.

\medskip

\begin{figure}[!p]
\begin{center}
\includegraphics[width=\textwidth]{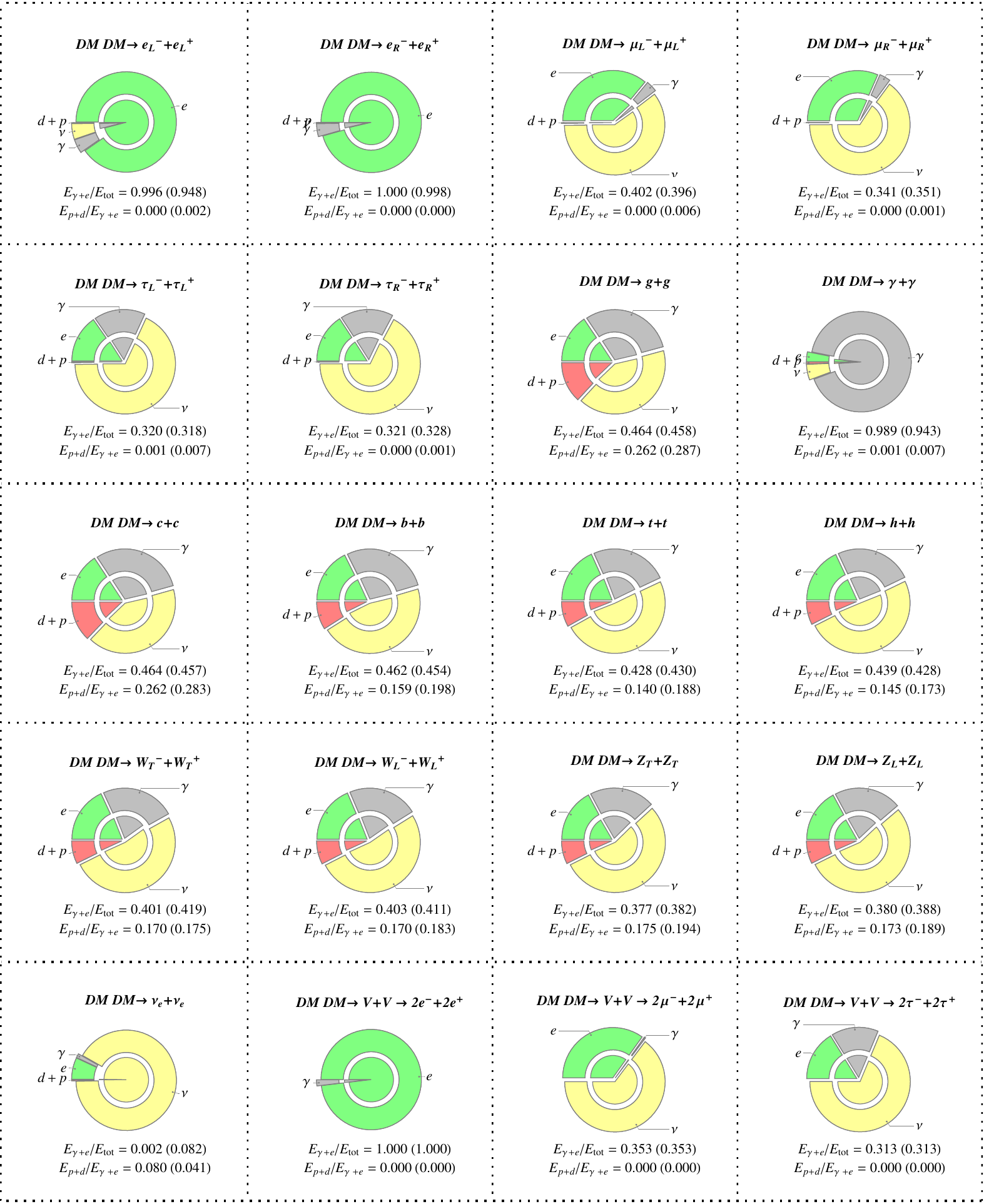}
\caption{\em \small \label{fig:energypie}
{\bfseries Energy distribution between the final states particles}: $e^\pm$, hadrons ($p+d$), $\gamma$ and $\nu$, for a set of characteristic annihilation channels. The inner (outer) pie refers to a DM mass of 200 GeV (5 TeV). For each pie chart, the first caption gives the energy fraction going into $\gamma$ and $e^\pm$ ($E_{\gamma+e}$) with respect to the total. The second caption gives the energy fraction into hadronic final states ($E_{p+d}$) with respect to $\gamma$ and $e^\pm$.}
\end{center}
\end{figure}

Some specifications on these fluxes are in order.
About all fluxes: The fluxes presented from here on and employed in the rest of the paper include EW corrections, as discussed above and in~\cite{CCRSSU}. However, in~\cite{website} we also provide, for comparison, all the spectra {\it before} EW corrections.\\
About $\gamma$ ray fluxes: We specify that of course the fluxes here include only the prompt emission and not the secondary radiation (e.g.\ due to Inverse Compton processes) that we discuss in Sec.~\ref{ICSgamma}.
Furthermore, we recall that by `prompt emission' we here mean all photons in final-state showers or hadron decays as given by {\sc Pythia}, including those from (IR-enhanced model-independent) QED and EW bremsstrahlung as discussed above and in~\cite{CCRSSU}. But further contributions to prompt emission can come from other three-body final states such as {\it internal} bremsstrahlung~\cite{IB,3body}: these can only be computed in the framework of a precise DM model because one needs to know the higher order QED annihilation/decay diagram. These are not included.\\
About fluxes of anti-deuterons: They are computed taking into account the jet structure of the annihilation products
scale with the cube of the uncertain coalescence parameter, here fixed to $p_0=160\MeV$; for details on the computation we refer the reader to~\cite{Dbar}.\\
About fluxes of neutrinos and anti-neutrinos: Those that we provide here are of course the neutrino spectra at production; the corresponding fluxes at detection are affected by oscillations (if travelling in vacuum, such as for neutrinos from DM annihilations/decays at the Galactic Center) and/or by interactions with matter (if e.g. from DM annihilations/decays in the center of the Sun). The fluxes at detection of neutrinos having traveled in vacuum from a distant astrophysical source can be obtained taking into account average oscillations with the formula
\beq \label{eq:Paveraged}
P(\nu_\ell \to \nu_{\ell '}) = P(\bar\nu_\ell \to \bar\nu_{\ell '})=
\sum_{i=1}^3 | V_{\ell i} V_{\ell' i}|^2\approx
\left(
\begin{array}{ccc}
0.6 & 0.2 & 0.2 \\
0.2 & 0.4 & 0.4 \\
0.2 & 0.4 & 0.4 
\end{array}
\right)\eeq
where $i$ runs over neutrino mass eigenstates, and the elements $|V_{\ell i}|$ of the neutrino mixing matrix can depend on its unknown CP-violating phases. The case of neutrinos from the center of the Sun is more complicated: they can be obtained along the lines of the analysis in~\cite{DMnu} and we leave to future work their detailed presentation. Finally, we recall that neutrinos detected after having crossed the Earth can experience additional oscillation effects.

\medskip

To conclude this quantitative presentation of the DM fluxes, in fig.~\ref{fig:energypie} we show some characteristic energy distributions between the final-states particles: $e^\pm$, hadrons ($p+d$), $\gamma$ and $\nu$. The inner pie refers to a DM mass of 200 GeV (a typical SuSy WIMP value) while the outer pie to 5 TeV (taken as a typical multi-TeV case). One can see that the portion of energy which goes into gamma rays and $e^\pm$ is often the most important one and always dominates over the energy fraction of the hadronic final states for all the channels. This is especially relevant in the context of extragalactic gamma rays signatures, where the energy fraction in $e^\pm$ is quickly converted to gamma rays due to Inverse Compton radiation.
For the channels involving $\mu^+\mu^-$ and $\tau^+\tau^-$ and of course for the $\nu \bar\nu$ channels, the portion of energy carried away by neutrinos becomes the dominant one. The fractions are rather independent of the mass of the DM particle, with some exceptions. For example, in the $\nu \nu$ channels, primary neutrinos start to radiate gamma rays and charged leptons due to radiative weak corrections when $M_{\rm DM}$ is above the electroweak scale (i.e. for the outer pie in the figure) and this increases the energy fraction of $\gamma$ and $e^\pm$.

%%%%%%%%%%%%%%%%%%%%%%%%%%%%%%%%%%%%%%%%%%%%%%%%%%%%%

\section{Charged Cosmic Rays}
\label{charged}

Having at disposal the energy spectra of charged particles per annihilation at production, as generated by MonteCarlos and as discussed in the previous section, we next need to consider where these fluxes of particles are produced in the galactic halo and how they propagate to the Earth. 

For simplicity, we present separately the propagation formalism for electrons or positrons, for antiprotons and for antideuterons. 
In this latter case, only a few trivial changes have to be implemented with respect to antiprotons, as we discuss below.

In general one ends up with a convenient form for the propagated fluxes in terms of a convolution of the spectra at production with a propagation function that encodes all the intervening astrophysics: see eq.~(\ref{eq:positronsflux}) for $e^\pm$ (or eq.~(\ref{eq:positronsfluxnox}) for the approximated treatment), eq.~(\ref{eq:fluxpbar}) for $\bar p$ and eq.~(\ref{eq:fluxdbar}) for $\bar d$.

\subsection{Propagation functions}
\label{chargedpropagation}

\subsubsection{Electrons or positrons: full formalism}
\label{positronpropagation}

The differential $e^\pm$ flux~\footnote{\label{footnotee+e-}Notice that with the notation $e^\pm$ we always refer to the independent fluxes of electrons $e^-$ or positrons $e^+$, which share the same formalism, and not to their sum (for which we use the notation $e^++e^-$ when needed) which of course differs by a trivial factor 2.} per unit of 
energy from DM annihilations or decays in any point in space $\vec x$ and time $t$ is given by $d\Phi_{e^\pm}/dE\, (t,\vec x,E) = v_{e^\pm} f/4\pi$
(units $1/\GeV\cdot{\rm cm}^2\cdot{\rm s}\cdot{\rm sr}$)
where $v_{e^\pm}$ is the velocity (essentially equal to $c$ in the regimes of our interest). 
The $e^\pm$ number density per unit energy, $f(t,\vec x,E)= dN_{e^\pm}/dE$,
obeys the diffusion-loss equation~\cite{SalatiCargese}:
\beq 
\label{eq:diffeq}
\frac{\partial f}{\partial t}-
\nabla\left( \mathcal{K}(E,\vec x) \nabla f \right) - \frac{\partial}{\partial E}\left( b(E,\vec x) f \right) = Q(E,\vec x)
\eeq
with diffusion coefficient function $\mathcal{K}(E,\vec x)$
and energy loss coefficient function $b(E,\vec x)$. %=E^2/(\GeV\cdot \tau_E)$ with $\tau_E = 10^{16}\,{\rm s}$.
They respectively describe transport through the turbulent magnetic fields and energy loss due to several processes, such as synchrotron radiation and Inverse Compton scattering (ICS) on CMB photons and on infrared or optical galactic starlight, as we discuss in more detail below. Notice that other terms would be present in a fully general diffusion-loss equation for Cosmic Rays, such as diffusive re-acceleration terms (describing the diffusion of CR particles in momentum space, due to their interactions on scattering centers that move in the Galaxy with an (Alfv\'en) velocity $V_a$) and convective terms. These are however negligible for $e^\pm$, see e.g.~\cite{SalatiCargese,Delahaye:2008ua}.
Eq.~(\ref{eq:diffeq}) is solved in a diffusive region with the shape of a solid flat cylinder that sandwiches the galactic plane, with height $2L$ in the $z$ direction and radius $R=20\,{\rm kpc}$ in the $r$ direction~\cite{DiffusionCylinder}. The location of the solar system corresponds to $\vec x  = (r_{\odot}, z_{\odot}) = (8.33\, {\rm kpc}, 0)$.
Boundary conditions are imposed such that the $e^\pm$ density $f$ vanishes on the surface of the cylinder, outside of which electrons and positrons freely propagate and escape.\footnote{See~\cite{perelstein} for the impact of not neglecting the propagation outside the cylinder.} 
Assuming that steady state conditions hold (as it is if one assumes that the typical time scales of the DM galactic collapse and of the variation of propagation conditions are much longer than the time scale of propagation itself, of the order of 1 Myr at 100 GeV energies~\cite{Blum}), the first term of eq.\eq{diffeq} vanishes and the dependence on time disappears.\footnote{A caveat on this point is that the time-independence of the diffusion process might not be justified in extreme environments such as the galactic central regions, where the propagation conditions may possibly change on a short enough time scale that they make this assumption invalid. E.g. recently the Fermi satellite has pointed out the existence of large gamma-ray structures (dubbed `Fermi bubbles') above and below the Galactic Center~\cite{fermibubbles}. A detailed modeling of the impact of these possible features on CR propagation is, for the moment, well beyond the scope of our analysis and probably of most DM related ones.}
Before illustrating the solution method, we briefly comment on the different pieces of the equation. 

\smallskip

\begin{figure}[t]
\begin{center}
\hspace{-1.3cm}
\includegraphics[width= 0.555 \textwidth]{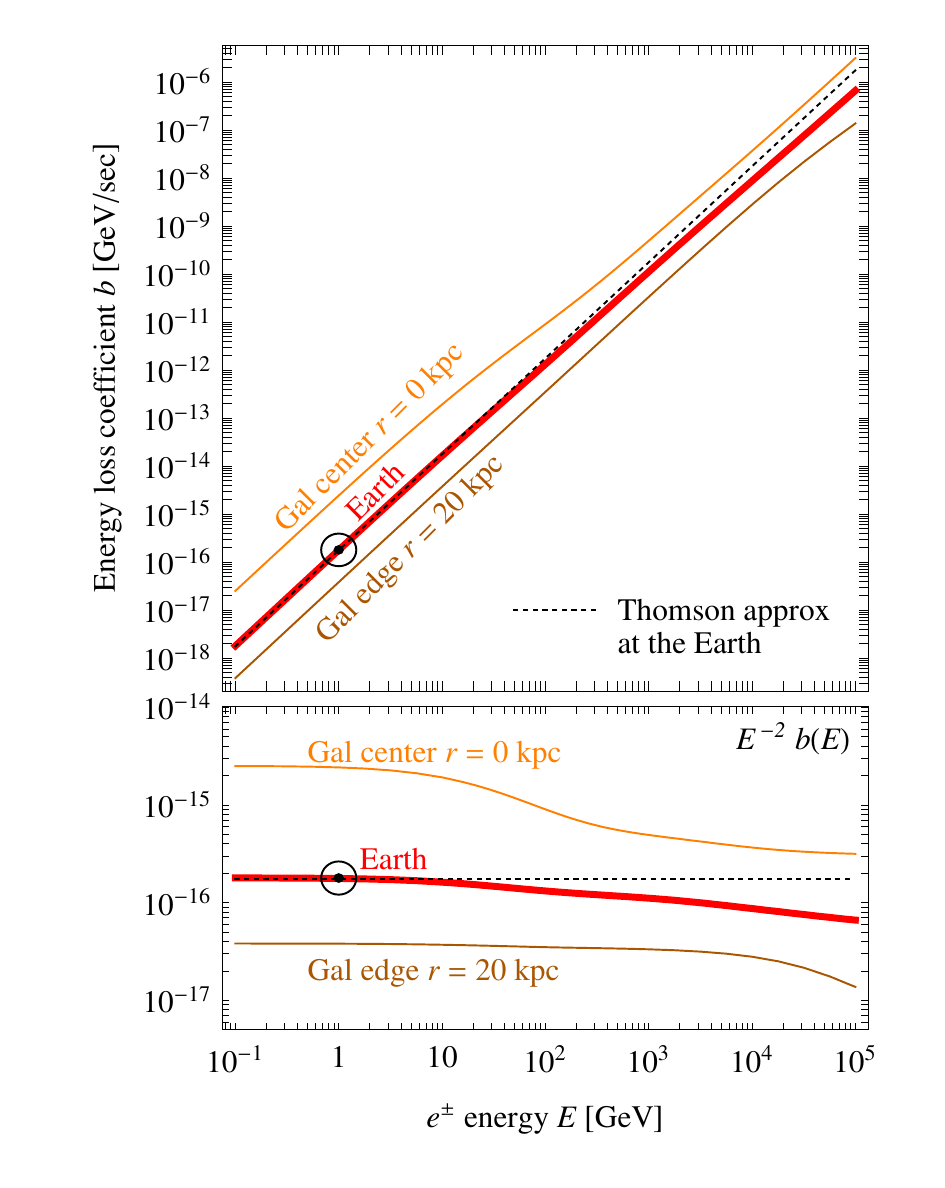}
\hspace{-0.95cm}
 \includegraphics[width= 0.555 \textwidth]{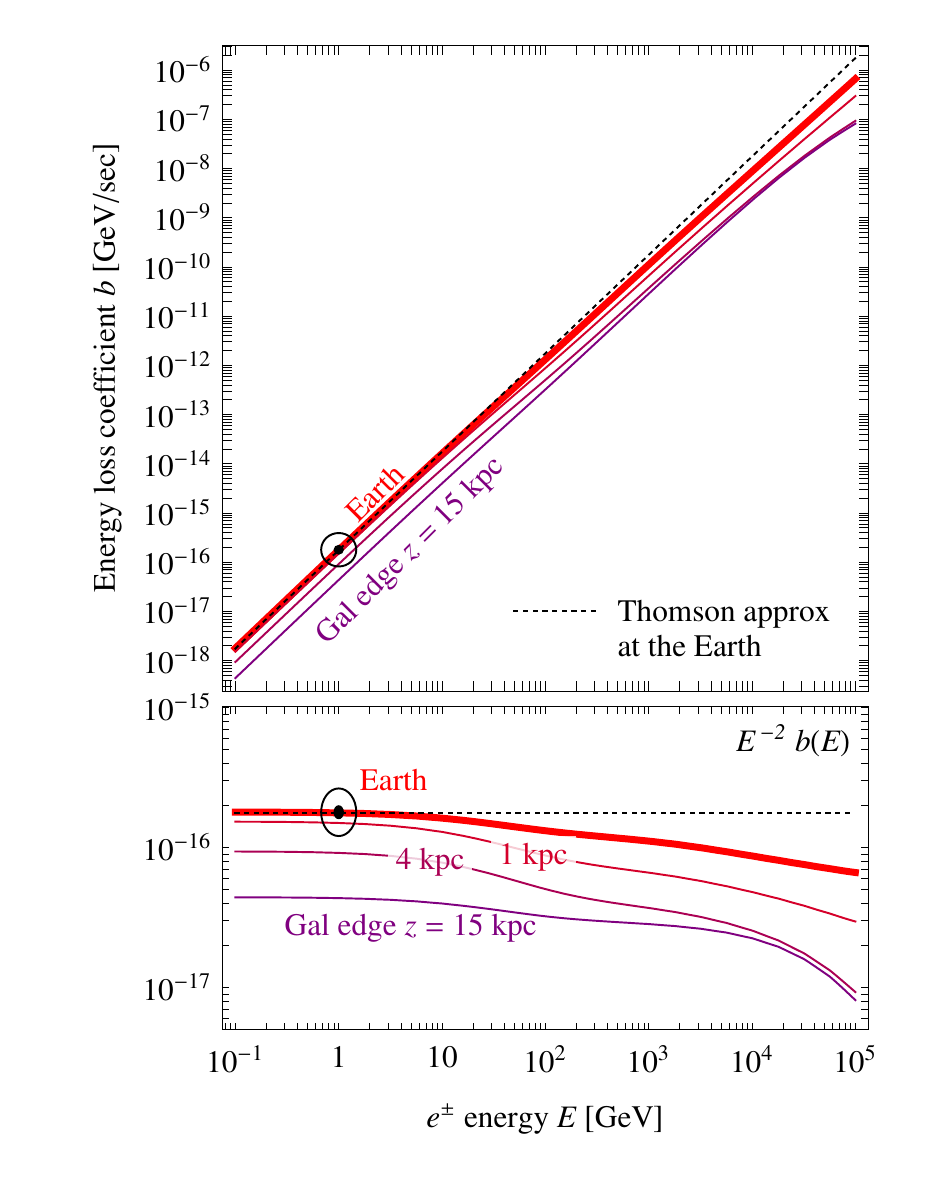}
\caption{\em \small \label{fig:b} {\bfseries Energy loss coefficient function} for electrons and positrons in the Milky Way. Left panel: at several locations along the galactic radial coordinate $r$, right panel: above (or below) the location of the Earth along the coordinate $z$. The dot points at the value of $\tau_\odot$ (see next subsection).}
\end{center}
\end{figure}

The {\em $e^\pm$ energy loss coefficient function} $b(E, \vec x)$  is in general dependent on the position $\vec x$, 
since the energy losses suffered by the $e^\pm$ are sensitive to the environment:
\beq\label{eq:b(E)}
 - \frac{dE}{dt}\equiv b(E,\vec x)= \frac{4\sigma_T }{3 m_e^2}  
 E^2  \tilde{u}
 \qquad , \qquad 
\tilde{u}= u_B(\vec x) + \sum _i u_{\gamma}(\vec x)  R_i^{\rm KN}(E),  \eeq
where $\sigma_T = 8\pi  r_e^2/3$, with $r_e = \alpha_{\rm em}/m_e$, is the Thompson cross section. The first addend in $\tilde{u}$ is associated with synchrotron losses, the second one with ICS losses~\footnote{So one can also define 
\beq
 \label{bIC}
b_{\rm syn}(E,\vec x)= \frac{4\sigma_T }{3 m_e^2}  E^2  u_B (\vec x) \quad {\rm and}\quad  b_{\rm IC}(E,\vec x)= \frac{4\sigma_T }{3 m_e^2} \sum_i E^2  u_{\gamma,i} (\vec x) R_i^{\rm KN}(E)
\eeq}.
$u_B = B^2/2$ is the energy density in galactic magnetic fields $B$ and $u_{\gamma,i} = \int dE\, n_i(E)$ is the energy density in light. Here $i$ runs over the three main components: CMB, star-light and dust-diffused InfraRed light. For the CMB, $n(E)$ is just the black body spectrum with $T=2.725$ K and one gets $u_{\gamma,{\rm CMB}}=0.260\, {\rm eV}/{\rm cm}^3$. For IR light and starlight, we extract the maps of their distribution and energy profile in the Galaxy from {\sc Galprop}~\cite{galprop} (an approximated formalism which employs a superposition of black body spectra also for IR and starlight has been discussed in~\cite{CP}). 
The $\vec x$ dependence in $b$ is due to the fact that the composition of the background light for ICS is different in different points of the halo (e.g.\ the center or the periphery of the Galaxy) and the value of the magnetic field also varies (much higher in the center than elsewhere). 

The dependence of $b$ on the energy $E$, in turn, is dictated by the energy dependence of the rates of the different loss processes. In particular, for IC scattering one has $b \propto E^2$ as long as the scattering happens in the Thomson regime, where the factor $R_i^{\rm KN}(E) =1$.\footnote{We recall that the Thomson regime in electron-photon Compton scattering is identified by the condition $\epsilon^\prime_{\rm max} = 2 \gamma \epsilon < m_e$, where $\epsilon$ denotes the energy of the impinging photon, $\epsilon^\prime$ the same quantity in the rest frame of the electron, $\gamma$ is the Lorentz factor of the electron and $m_e$ is the electron mass. When $e^\pm$ scatter on CMB photons ($\epsilon \simeq 2\ 10^{-4}$ eV) the condition is satisfied up to $\sim$ TeV $e^\pm$ energies. For scatterings on more energetic starlight ($\epsilon \approx$ 0.3 eV), the condition breaks down already above $\approx$ few GeV $e^\pm$ energies.} For large enough electron energy the IC scattering enters into the full Klein-Nishina regime, where the $\gamma e^\pm$ scattering rate becomes rapidly smaller than the Thomson approximation, and thus $R_i^{\rm KN}(E_e) <1$ (see e.g.\ fig.~2 of~\cite{Meade})
reducing $b(E)$. The synchrotron loss rate, instead, is always proportional to the square of the electron/positron energy $E^2$.

The {\em profile of the magnetic field in the Galaxy} is very uncertain and we adopt the conventional one 
\beq B(r,z) = B_0\ {\rm exp}[-(r-r_\odot)/r_B -|z|/z_B]\label{eq:Bgal}
\eeq as given in~\cite{StrongMoskalenkoReimer98}, with $B_0 = 4.78\ \mu$G, $r_B = 10$ kpc and $z_B = 2$ kpc. With these choices, the dominant energy losses are due to ICS everywhere, except in the region of the Galactic Center and for high $e^\pm$ energies, in which case synchrotron losses dominate.
All in all, the $b(E,\vec x)$ function that we obtain is sampled in fig.~\ref{fig:b} and given in numerical form on the \myurl{www.marcocirelli.net/PPPC4DMID.html}{website}~\cite{website}. In the figure, one sees the $E^2$ behaviour at low energies changing into a softer dependence as the energy increases (the transition happens earlier at the GC, where starlight is more abundant, and later at the periphery of the Galaxy, where CMB is the dominant background). At the GC, it eventually re-settles onto a $E^2$ slope at very high energies, where synchrotron losses dominate.

\medskip

{\em The diffusion coefficient} function $\mathcal{K}$ is also in principle dependent on the position, since the distribution of the diffusive  inhomogeneities of the magnetic field changes throughout the galactic halo. However, a detailed mapping of such variations is prohibitive: e.g.\ they would have different features inside/outside the galactic arms as well as inside/outside the galactic disk, so that they would depend very much on poorly known local galactic geography. Moreover, including a spatial dependence in $\mathcal{K}$ would make the semi-analytic method described below much more difficult to implement numerically. We therefore leave these possible refinements for future work~\footnote{See~\cite{perelstein2} for a recent analysis for antiprotons.} and, as customary, we adopt the parameterization $\mathcal{K}(E,\vec x)=\mathcal{K}_0 (E/\GeV)^\delta=\mathcal{K}_0\, \epsilon^\delta$. 

\smallskip

The values of the propagation parameters $\delta$, $K_0$ and $L$ (the height of the diffusion cylinder defined above) are deduced from a variety of cosmic ray data and modelizations. It is customary to adopt the sets presented in Table~\ref{tab:proparam}, which are found to minimize or maximize the final fluxes.~\footnote{We stress, however, that the determination of these parameters is a whole evolving research area, which will certainly update these values in the future as more refined modelizations and further CR data become available. See e.g.~\cite{Maurin2010,Dragon2010,Trotta2010} for recent references. The choices presented in Table~\ref{tab:proparam} should be seen as the current bracketing of sensible possibilities.} 

\begin{table}[t]
\center
\begin{tabular}{c|cc|ccc|c}
 & \multicolumn{2}{c|}{Electrons or positrons} & \multicolumn{3}{c|}{Antiprotons (and antideuterons)}  \\
Model  & $\delta$ & $\mathcal{K}_0$ [kpc$^2$/Myr] & $\delta$ & $\mathcal{K}_0$ [kpc$^2$/Myr] & $V_{\rm conv}$ [km/s] & $L$ [kpc]  \\
\hline 
MIN  & 0.55 & 0.00595 & 0.85 &  0.0016 & 13.5 & 1 \\
MED & 0.70 & 0.0112 & 0.70 &  0.0112 & 12 & 4  \\
MAX  & 0.46 & 0.0765 &  0.46 &  0.0765 & 5 & 15 
\end{tabular}
\caption{\em \small {\bfseries Propagation parameters} for charged particles in the Galaxy (from~\cite{FornengoDec2007,DonatoPRD69}). 
\label{tab:proparam}}
\end{table}

Finally, DM DM annihilations or DM decays in each point of the halo with DM density $\rho(\vec x)$ provide {\em the source term} $Q$ of eq.~(\ref{eq:diffeq}), which reads
\beq 
Q = \frac{1}{2} \left(\frac{\rho}{M_{\rm DM}}\right)^2 f^{\rm ann}_{\rm inj},\qquad f^{\rm ann}_{\rm inj} = \sum_{f} \langle \sigma v\rangle_f \frac{dN_{e^\pm}^f}{dE} \qquad {\rm (annihilation)},
\label{eq:Qann}
\eeq
\beq 
Q = \left(\frac{\rho}{M_{\rm DM}}\right) f^{\rm dec}_{\rm inj},\qquad f^{\rm dec}_{\rm inj} = \sum_{f} \Gamma_f \frac{dN_{e^\pm}^f}{dE} \qquad {\rm (decay)},
\label{eq:Qdec}
\eeq
where $f$ runs over all the channels with $e^\pm$ in the final state, with the respective thermal averaged cross sections $\sigma v$ or decay rate $\Gamma$.

\medskip

\subsubsection{Electrons or positrons: result}
\label{positronpropagationresult}

The {\em differential flux of} $e^\pm$ $d\Phi_{e^\pm}/dE = v_{e^\pm} f /4 \pi$
in each given point of our Galaxy for any injection spectrum can be written as
\beq\label{eq:positronsflux}
\frac{d\Phi_{e^\pm}}{dE}(E,\vec{x}) =
\frac{v_{e^\pm}}{4\pi \, b(E,\vec x)}\left\{
\begin{array}{ll}
\displaystyle\!\! \frac12 \left(\frac{\rho(\vec x)}{M_{\rm DM}}\right)^2 \sum_f \langle \sigma v \rangle_f \int_E^{M_{\rm DM}} dE_{\rm s} \, \frac{dN^f_{e^\pm}}{dE}(E_{\rm s}) \, {I}(E,E_{\rm s},\vec{x}) &{\rm (annihilation)} \\[5mm]
\displaystyle \phantom{\frac12 }\left(\frac{\rho(\vec x)}{M_{\rm DM}}\right) \sum_f \Gamma_f \int_E^{M_{\rm DM}/2} dE_{\rm s} \, \frac{dN^f_{e^\pm}}{dE}(E_{\rm s}) \,{I}(E,E_{\rm s},\vec{x}) &{\rm (decay)}
\end{array}
\right. 
\eeq
where $E_{\rm s}$ is the $e^\pm$ energy at production (`s' stands for `source')
and the  {\em generalized halo functions} $I(E,E_{\rm s},\vec x)$ are essentially the Green functions from a source
with fixed energy $E_{\rm s}$ to any energy $E$.
In other words, the halo functions $I$ encapsulate all the astrophysics (there is a halo function $I$ for each choice of DM distribution profile and choice of $e^\pm$ propagation parameters) and are independent of the particle physics model: convoluted with the injection spectra, they give the final spectrum searched for. 
They obey $I(E,E,\vec x) = 1$ and $I(E,E_{\rm s},\vec x) = 0$ on the boundary of the diffusion cylinder.
Neglecting diffusion (i.e.\ setting ${\cal K}= 0)$ one would have $I(E,E_{\rm s},\vec x)=1$.
These functions are provided numerically on the~\myurl{www.marcocirelli.net/PPPC4DMID.html}{website}~\cite{website} in the form of {\sc Mathematica}$^{\tiny{\textregistered}}$ interpolating functions. 
Plugged in eq.~(\ref{eq:positronsflux}), they allow to compute the $e^\pm$ flux everywhere in the Galaxy.

\begin{figure}[!hp]
\begin{center}
\includegraphics[height= 0.98 \textheight]{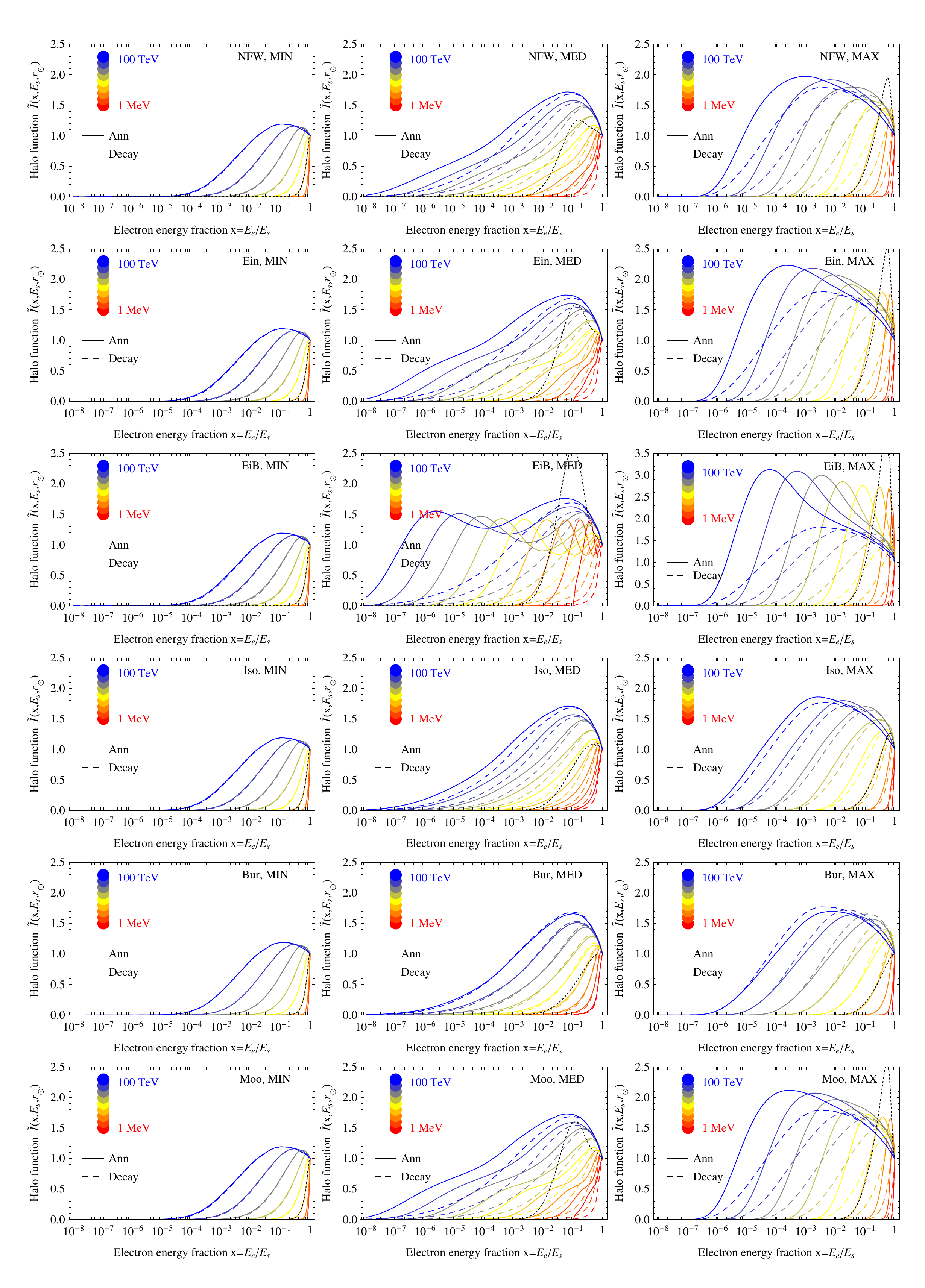}
\vspace{-1cm}
\caption{\em \small \label{fig:halofunct2} {\bfseries Generalized halo functions for electrons or positrons}, for several different values of the injection energy $\epsilon_S$ (color coded). The superimposed dotted black lines are the `reduced' halo functions discussed in~\ref{propagationapprox}.}
\end{center}
\end{figure}

The functions particularized to the location of the Earth, that is: $I(E,E_{\rm s},\vec r_\odot)$, are plotted in fig.~\ref{fig:halofunct2} and provided numerically on the~\myurl{www.marcocirelli.net/PPPC4DMID.html}{website}~\cite{website} too. 
Plugged in eq.~(\ref{eq:positronsflux}), these allow to compute the $e^\pm$ flux at the location of the Earth, $\Phi(\epsilon, r_\odot, z_\odot)$. We also provide separately the resulting fluxes (see the next subsection \ref{eppropagated}). 

\bigskip

The generalized halo functions $I$ are computed as follows (the uninterested reader can skip the rest of this section).
Due to numerical issues it is convenient to search for the solution of eq.~(\ref{eq:diffeq}) using an ansatz similar to eq.\eq{positronsflux} but somewhat different:
\beq
f(\epsilon,\vec{x}) = \frac1{b_{\rm T}(\epsilon)}\left\{
\begin{array}{ll}
\displaystyle \frac12 \left(\frac{\rho_\odot}{M_{\rm DM}}\right)^2  \int_\epsilon^{M_{\rm DM}} d\epsilon_{\rm s} \, f_{\rm inj}^{\rm ann}(\epsilon_{\rm s}) \, \tilde{I}(\epsilon,\epsilon_{\rm s},\vec{x}) \quad& {\rm (annihilation)} \\[5mm]
\displaystyle \phantom{\frac12 }\left(\frac{\rho_\odot}{M_{\rm DM}}\right)  \int_\epsilon^{M_{\rm DM}/2} d\epsilon_{\rm s} \, f_{\rm inj}^{\rm dec}(\epsilon_{\rm s}) \, \tilde{I}(\epsilon,\epsilon_{\rm s},\vec{x}) \qquad &{\rm (decay)}
\end{array}
\right. 
\label{eq:npositrons}
\eeq
where $\epsilon = E/{\rm GeV}$. 
Here we adopted the (arbitrary but convenient) normalizing factor $b_{\rm T}(\epsilon)=\epsilon^2 \, \mbox{GeV}/\tau_\odot$, with $\tau_\odot=\mbox{GeV}/b(1\,\mbox{GeV}, \vec{x}_\odot)=5.7 \times 10^{15}$ sec, which is the energy loss coefficient at Earth in the Thomson limit regime. 
Plugging now the ansatz (\ref{eq:npositrons}) in the differential equation (\ref{eq:diffeq}) one can recast (\ref{eq:diffeq}) into a partial differential equation for $\tilde{I}(\epsilon,\epsilon_{\rm s},\vec x)$ (this extends the solution method first discussed (to our knowledge) in~\cite{Zupan}). Indeed, (\ref{eq:diffeq}) becomes
\begin{eqnarray}
&-&\mathcal{K}_0 \tau_\odot \epsilon^{\delta-2} \int_\epsilon^{M_{\rm DM}(M_{\rm DM}/2)} d\epsilon_{\rm s} \, f_{\rm inj}(\epsilon_{\rm s}) \, \nabla^2 \tilde{I}(\epsilon,\epsilon_{\rm s},\vec{x}) + \left.\frac{b(\epsilon,\vec{x}) \, f_{\rm inj}(\epsilon) \tilde{I}(\epsilon,\epsilon_{\rm s},\vec{x})}{b_{\rm T}(\epsilon)}\right|_{\epsilon=\epsilon_{\rm s}} + \nonumber \\ 
&-& \int_\epsilon^{M_{\rm DM}(M_{\rm DM}/2)} d\epsilon_{\rm s} \, f(\epsilon_{\rm s}) \frac\partial{\partial\epsilon} \left(\frac{b(\epsilon,\vec{x})}{b_{\rm T}(\epsilon)} \tilde{I}(\epsilon,\epsilon_{\rm s},\vec{x})\right)=f_{\rm inj}(\epsilon)\left(\frac{\rho(\vec{x})}{\rho_\odot}\right)^\eta,
\end{eqnarray}
where $\eta=1,2$ for decay or annihilation scenarios respectively and the upper integration limit changes accordingly. One then extracts the
 partial differential equation for $\tilde{I}$:
\beq\label{6eq:diffHalo}
\nabla^2\tilde{I}(\epsilon,\epsilon_{\rm s},\vec{x})+\frac1{\mathcal{K}_0 \tau_\odot \epsilon^{\delta-2}} \frac{\partial}{\partial\epsilon}\left(\frac{b(\epsilon,\vec{x})}{b_{\rm T}(\epsilon)}  \tilde{I}(\epsilon,\epsilon_{\rm s},\vec{x})\right)=0, 
\eeq
with boundary conditions 

\beq
\label{eq:boundarycond}
\left\{
\begin{array}{l}
\displaystyle \tilde{I}(\epsilon_{\rm s},\epsilon_{\rm s},\vec{x})=\frac{b_{\rm T}(\epsilon_{\rm s})}{b(\epsilon_{\rm s},\vec{x})}\left(\frac{\rho(\vec{x})}{\rho_\odot}\right)^\eta,\\[5mm]
\displaystyle \tilde{I}(\epsilon,\epsilon_{\rm s},\vec{x}_{\rm max})=0, \quad {\rm with}\  \vec{x}_{\rm max} \equiv (R,L). 
\end{array}
\right.
\eeq
Finally the halo functions with the normalization conventions of eq.\eq{positronsflux} are obtained as
\beq
I(E,E_{\rm s},\vec x)=\tilde{I}(\epsilon,\epsilon_{\rm s},\vec{x}) \left[\frac{b_{\rm T}(\epsilon)}{b(\epsilon,\vec x)}\left(\frac{\rho(\vec{x})}{\rho_\odot}\right)^\eta\right]^{-1},
\eeq
Solving numerically eq.~(\ref{6eq:diffHalo}) with (\ref{eq:boundarycond}) allows to compute the $\tilde{I}(\epsilon,\epsilon_{\rm s},\vec x)$ and in turn the ${I}(\epsilon,\epsilon_{\rm s},\vec x)$.

\subsubsection{Electrons or positrons: approximated energy loss}
\label{propagationapprox}

The above treatment is pretty general in that it allows to compute the propagated fluxes taking into account the full energy and position dependance of $b(E,\vec x)$, as discussed above. An approximated formalism had been adopted in the past (see e.g.~\cite{FornengoDec2007} and references therein) and we report it here for completeness and to compare with our full result. 

\smallskip

\begin{figure}[!t]
\begin{minipage}{0.6\textwidth}
\tiny{
\begin{tabular}{c|c|rrrrrrrr}
\multicolumn{10}{l}{\footnotesize DM annihilation} \\
\hline
\rm{halo} & \rm{prop} & $a_0$ & $a_1$ & $a_2$ & $a_3$ & $b_1$ & $b_2$ & $c_1$ & $c_2$\\
\hline
 & \text{MIN} & 0.502 & 0.297 & -0.828 & -1.691 & 0.098 & 0.116 & -0.342 & 0.169 \\
 \text{NFW} & \text{MED} & 0.505 & 0.510 & 0.579 & 0.984 & 0.876 & 0.699 & 0.137 & 0.082 \\
 & \text{MAX} & 0.502 & 0.642 & 1.462 & 0.793 & 1.184 & 0.787 & 0.165 & 0.068 \\
 \hline
 & \text{MIN} & 0.502 & 0.299 & -0.784 & -1.678 & 0.099 & 0.119 & -0.336 & 0.167 \\
 \text{Moo} & \text{MED} & 0.507 & 0.363 & 2.207 & 1.399 & 0.909 & 0.758 & 0.165 & 0.056 \\
 & \text{MAX} & 0.505 & 0.737 & 2.096 & 0.691 & 1.223 & 0.802 & 0.138 & 0.061 \\
 \hline
  & \text{MIN} & 0.502 & 0.295 & -0.847 & -1.703 & 0.097 & 0.116 & -0.344 & 0.169 \\
 \text{Iso} & \text{MED} & 0.573 & 0.594 & 0.301 & 0.684 & 0.760 & 1.412 & 0.164 & 1.731 \\
  & \text{MAX} & 0.495 & 0.358 & 1.823 & 1.415 & 1.104 & 0.940 & 0.412 & 0.219 \\
 \hline
 & \text{MIN} & 0.502 & 0.306 & -0.785 & -1.641 & 0.100 & 0.119 & -0.341 & 0.170 \\
 \text{Ein} & \text{MED} & 0.507 & 0.345 & 2.095 & 1.469 & 0.905 & 0.741 & 0.160 & 0.063 \\
 & \text{MAX} & 0.505 & 0.793 & 1.859 & 0.644 & 1.218 & 0.796 & 0.146 & 0.066 \\
 \hline
 & \text{MIN} & 0.502 & 0.311 & -0.705 & -1.615 & 0.103 & 0.123 & -0.329 & 0.167 \\
 \text{EiB} & \text{MED} & 0.508 & 0.596 & 3.209 & 0.862 & 0.947 & 0.761 & 0.175 & 0.046 \\
 & \text{MAX} & 0.510 & 0.972 & 3.168 & 0.523 & 1.275 & 0.818 & 0.104 & 0.061 \\
 \hline
 & \text{MIN} & 0.490 & 0.727 & -0.039 & 0.708 & 0.095 & 1.444 & 0.217 & 0.933 \\
 \text{Bur} & \text{MED} & 0.500 & 0.760 & 0.255 & 0.658 & 0.696 & 0.721 & 0.302 & 0.154 \\
 & \text{MAX} & 0.503 & 0.630 & 0.238 & 0.789 & 1.021 & 1.115 & 0.350 & 0.222
\end{tabular}}
\caption{\em \small {\bfseries Reduced halo function $\mathcal{I}(\lambda_D)$ for $e^\pm$ from annihilating Dark Matter}, for the different DM profiles and sets of propagation parameters, and the corresponding fit parameters to be used in eq.~$(\ref{eq:fitpositrons})$.}
\label{fig:halofunctions}
\end{minipage}
\quad
\begin{minipage}{0.37\textwidth}
\vspace{-0.9cm}
\centering
\includegraphics[width=\textwidth]{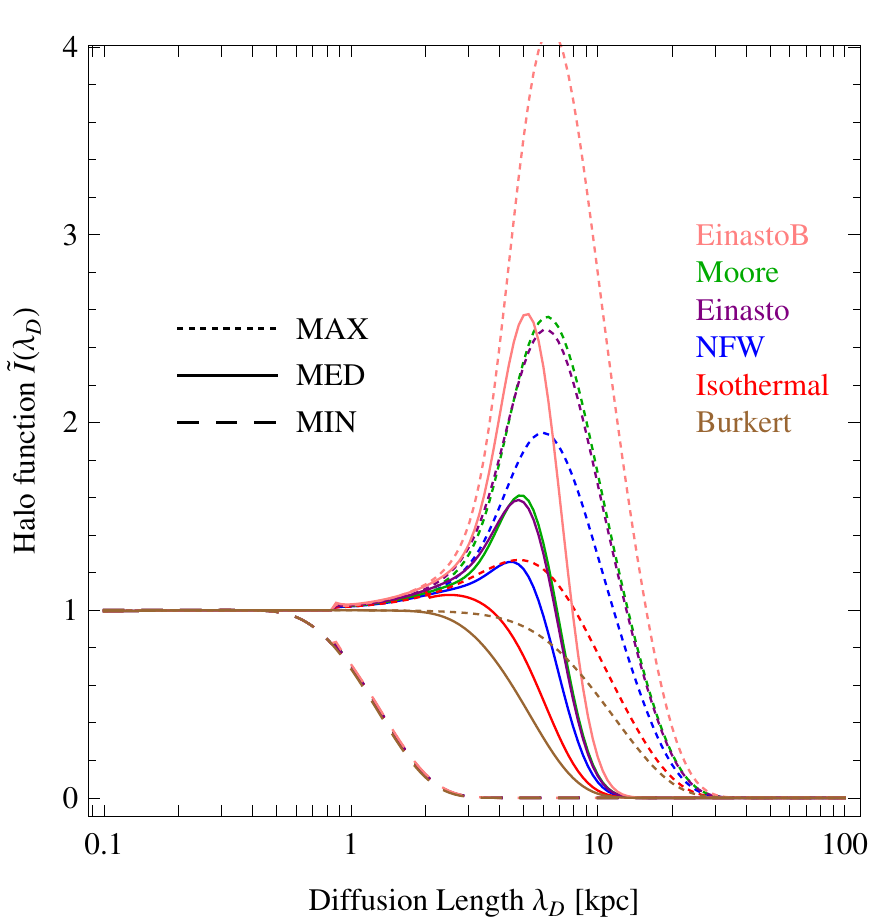}
\end{minipage}
\begin{minipage}{0.6\textwidth}
\vspace{1cm}
\tiny{
\begin{tabular}{c | c | rrrrrrrr}
\multicolumn{10}{l}{\footnotesize DM decay} \\
\hline
\rm{Halo} & \rm{Prop} & $a_0$ & $a_1$ & $a_2$ & $a_3$ & $b_1$ & $b_2$ & $c_1$ & $c_2$\\
\hline
& \text{MIN} & 0.487 & 0.558 & -0.049 & 0.917 & 0.096 & 1.444 & 0.217 & 0.903 \\
 \text{NFW} & \text{MED} & 0.501 & 0.638 & 0.313 & 0.784 & 0.715 & 0.740 & 0.299 & 0.152 \\
& \text{MAX} & 0.501 & 0.579 & 0.458 & 0.863 & 1.070 & 1.057 & 0.355 & 0.202 \\
\hline
& \text{MIN} & 0.487 & 0.555 & -0.050 & 0.922 & 0.097 & 1.444 & 0.217 & 0.908 \\
 \text{Moo} & \text{MED} & 0.501 & 0.687 & 0.276 & 0.728 & 0.721 & 0.747 & 0.291 & 0.151 \\
& \text{MAX} & 0.500 & 0.563 & 0.554 & 0.889 & 1.076 & 1.044 & 0.367 & 0.199 \\
\hline
& \text{MIN} & 0.487 & 0.558 & -0.048 & 0.916 & 0.096 & 1.444 & 0.216 & 0.898 \\
 \text{Iso} & \text{MED} & 0.501 & 0.705 & 0.310 & 0.710 & 0.700 & 0.717 & 0.311 & 0.155 \\
& \text{MAX} & 0.504 & 0.609 & 0.270 & 0.817 & 1.050 & 1.111 & 0.334 & 0.212 \\
\hline
& \text{MIN} & 0.487 & 0.552 & -0.050 & 0.927 & 0.097 & 1.444 & 0.217 & 0.907 \\
 \text{Ein} & \text{MED} & 0.501 & 0.677 & 0.277 & 0.739 & 0.725 & 0.748 & 0.287 & 0.150 \\
& \text{MAX} & 0.500 & 0.230 & 1.443 & 2.183 & 1.078 & 1.036 & 0.371 & 0.202 \\
\hline
& \text{MIN} & 0.487 & 0.414 & -0.068 & 1.238 & 0.098 & 1.444 & 0.217 & 0.915 \\
 \text{EiB} & \text{MED} & 0.502 & 0.797 & 0.189 & 0.627 & 0.737 & 0.751 & 0.265 & 0.164 \\
& \text{MAX} & 0.498 & 0.258 & 1.614 & 1.954 & 1.087 & 1.022 & 0.389 & 0.196 \\
\hline
& \text{MIN} & 0.488 & 0.691 & -0.037 & 0.739 & 0.094 & 1.444 & 0.215 & 0.885 \\
 \text{Bur} & \text{MED} & 0.501 & 0.721 & 0.283 & 0.695 & 0.679 & 0.712 & 0.318 & 0.156 \\
& \text{MAX} & 0.497 & 0.636 & 0.278 & 0.791 & 1.051 & 1.203 & 0.351 & 0.107%
\end{tabular}}
\caption{\em \small {\bfseries Reduced halo function $\mathcal{I}(\lambda_D)$ for $e^\pm$ from decaying DM}, for the different halo profiles and sets of propagation parameters, and the corresponding fit parameters to be used in eq.~$(\ref{eq:fitpositrons})$.}
\label{fig:halofunctionsdecay}
\end{minipage}
\quad
\begin{minipage}{0.37\textwidth}
\vspace{-0cm}
\centering
\includegraphics[width=\textwidth]{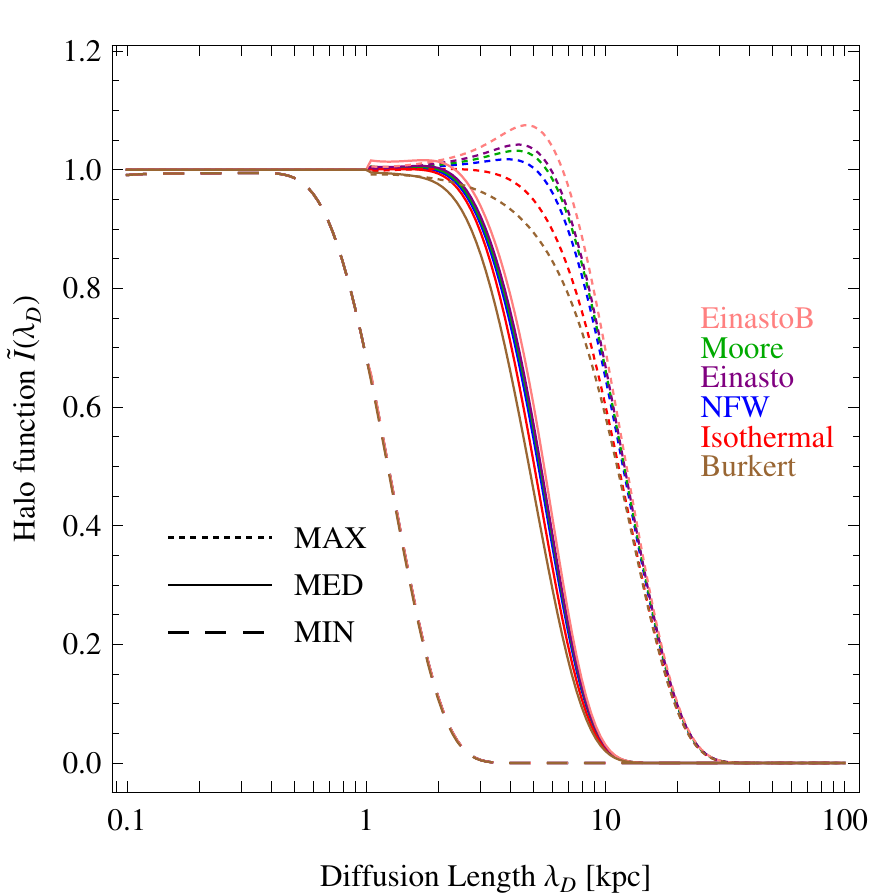}
\end{minipage}
\end{figure}

Assuming a space-independent $b = b_{\rm T}(\epsilon) =\epsilon^2 \, \mbox{GeV}/\tau_\odot$ everywhere in the Galaxy, one can define a `reduced' halo function $\mathcal{I}(\lambda_D, \vec x)$ (and a simplified differential equation for it) in terms of a single quantity $\lambda_D = \lambda_D(\epsilon,\epsilon_{\rm s}) = \sqrt{4\mathcal{K}_0 \tau_\odot\left(\epsilon^{\delta-1} - \epsilon_{\rm s}^{\delta-1}\right)/(1-\delta)}$, which represents the diffusion length of $e^\pm$ injected with energy $\epsilon_S$ and detected with energy $\epsilon$. One has (instead of the full equation (\ref{6eq:diffHalo})) 
\beq
\label{6eq:diffHalonox}
\nabla^2\mathcal{I}(\lambda_D, \vec x)-\frac2{\lambda_D} \frac{\partial}{\partial\lambda_D} \mathcal{I}(\lambda_D,\vec{x})=0, \qquad 
\left\{
\begin{array}{l}
\displaystyle \mathcal{I}(0,\vec{x})=\left(\frac{\rho(\vec{x})}{\rho_\odot}\right)^\eta,\\[5mm]
\displaystyle \mathcal{I}(\lambda_D,\vec{x}_{\rm max})=0.
\end{array}
\right.
\eeq 
Solving eq.~(\ref{6eq:diffHalonox}) provides $\mathcal{I}(\lambda_D, \vec x)$ in any given point $\vec x$.~\footnote{
Alternatively, one can find the solution for $\mathcal{I}(\lambda_D, \vec x)$ via an expansion in Bessel and Fourier series~\cite{FornengoDec2007}. Formally one finds that 
\beq
\mathcal{I}(\lambda_D, r, z) = \sum_{n,m=1}^{\infty} J_0( \zeta_n r/ R )\ \sin \left( \frac{m \pi}{ 2 L} (z+L) \right) \ {\rm exp}\left[ -\left( \left( \frac{m \pi}{2 L} \right)^2 + \left(\frac{\zeta_n}{R}\right)^2 \right) \frac{\lambda_D^2}{4} \right] R_{n,m}
\eeq
where $J_i$ is the Bessel function of the first kind (cylindrical harmonic) of order $i$, $\zeta_n$ is the $n$-th zero of the $i=0$  function and $R_{n,m}$ corresponds to the Bessel- and Fourier-transform of $(\rho/\rho_\odot)^2$:
\beq
R_{n,m} = \frac{2}{J_1(\zeta_n)^2 R^2} \int_0^Rdr\, r J_0( \zeta_n r/ R ) \frac{1}{L} \int_{-L}^{+L} dz\, \sin\Big(m \pi (z+L) / 2L\Big) \left( \frac{\rho(r,z)}{\rho_\odot} \right)^2.
\eeq
This method, however, converges much more slowly than numerically solving eq.~(\ref{6eq:diffHalonox}), and it is therefore unpractical for computing the solution in more than one point $(r,z)$. We checked anyway that the two solutions coincide.
In the numerical computations performed with this Bessel-Fourier expansion method it is convenient to smooth out the steepness of some of the profiles close to the GC, adopting the prescription discussed in~\cite{smoothing}. It simply amounts to replacing the divergent profile by a well behaved one below an arbitrarily chosen critical radius of $r_{\rm crit} = 0.5\, {\rm kpc}$ from the GC, while approximately preserving the absolute number of annihilations in that region. Such well behaved profiles are plotted in fig.~\ref{fig:DMprofiles} as dashed lines.
%More precisely, we use
%\begin{equation}
%\rho(r<r_{\rm crit}) = \rho(r_{\rm crit}) \left[1 + \frac{2 \pi^2}{3} \left( \frac{3}{3-2\gamma} -1 \right) \left( \frac{\sin(\pi r/r_{\rm crit})}{\pi r/r_{\rm crit}} \right)^2 \right]^{1/2}.
%\label{eq:smoothing}
%\end{equation}
%for the NFW and Moore profiles. 
We just cut at $r_{\rm crit}/10$ for the Einasto profiles. It can be checked that such smoothing does not affect the determination of $\mathcal{I}(\lambda_D,r_\odot,z_\odot)$.
}
With that, one can write the equivalent of eq.~(\ref{eq:positronsflux}) as
\beq
\label{eq:positronsfluxnox}
\frac{d\Phi_{e^\pm}}{dE}(\epsilon,\vec{x}) = \frac1{4\pi}\frac{v_{e^\pm}}{b_{\rm T}(\epsilon)}\left \{
\begin{array}{ll}
\displaystyle \frac12 \left(\frac{\rho_\odot}{M_{\rm DM}}\right)^2 \sum_f \langle \sigma v \rangle_f \int_\epsilon^{M_{\rm DM}} d\epsilon_{\rm s} \, \frac{dN^f_{e^\pm}}{dE}(\epsilon_{\rm s}) \, \mathcal{I}(\lambda_D(\epsilon,\epsilon_{\rm s}),\vec x) & {\rm (annihilation)} \\[5mm]
\displaystyle \left(\frac{\rho_\odot}{M_{\rm DM}}\right) \sum_k \Gamma_k \int_\epsilon^{M_{\rm DM}/2} d\epsilon_{\rm s} \, \frac{dN^f_{e^\pm}}{dE}(\epsilon_{\rm s}) \, \mathcal{I}(\lambda_D(\epsilon,\epsilon_{\rm s}),\vec x) & {\rm (decay)}
\end{array}
\right. 
\eeq 
The function $I(\lambda_D, r_\odot, z_\odot)$ at the location of the Earth is well reproduced in terms of the fit 
\beq\mathcal{I}(\lambda_D) = a_0 + a_1 \tanh\left(\frac{b_1-\ell}{c_1}\right) \left[ a_2 \exp \left( -\frac{(\ell - b_2)^2}{c_2}\right) + a_3 \right]
\label{eq:fitpositrons}
\eeq
with $\ell = \log_{10}(\lambda_D/{\rm kpc})$ and the coefficients given in the tables in Figure~\ref{fig:halofunctions} and~\ref{fig:halofunctionsdecay} and also reported on the \myurl{www.marcocirelli.net/PPPC4DMID.html}{website}~\cite{website}.\footnote{The fit functions reproduce the results of our numerical calculation to better than 5\%, with the exception of the EinastoB case, for which the accuracy drops to a still acceptable 10\%, over the whole range $\lambda_D = 0.1 \to 100\ {\rm kpc}$. The fit functions should not be used outside of this range.}

These approximated halo functions are also superimposed (black dotted lines) in the plots of fig.~\ref{fig:halofunct2}, having chosen $\epsilon_S =$ 1 GeV and the annihilation case. It is evident that the reduced halo functions miss much of the richer structure of the full ones, especially for MED and MAX parameters and for peaked profiles. On the other hand, the zeroing of the functions in the two approaches occurs at similar values of $x$ in each plot (compare the black dotted lines and the orange ones corresponding to $\epsilon_S =$ 1 GeV in the color coding). In computing the convolution as an integral over $\epsilon_S$, the different shapes and the similar zeroing somewhat counteract one each other, so that the final spectra as computed by the full formalism will be different, but not drastically, with respect to the approximated result. We illustrate later in specific examples (see fig.~\ref{fig:positronspropagated}) the quantitative impact. In points of the Galaxy other than the location of the Earth, e.g.\ close to the Galactic Center, the difference between the two computations would however be more important. This affects the spectra of ICS gamma rays produced from these regions.

%%%%%%%%

\subsubsection{Antiprotons}
\label{antiprotonpropagation}

The propagation of antiprotons through the galaxy is described by a diffusion equation analogous to the one for positrons.
Again, the number density of antiprotons per unit energy $f(t,\vec x,K) = dN_{\bar p}/dK$ vanishes on the surface of the cylinder at $z=\pm L$ and $r=R$. $K=E-m_p$ is the $\bar p$ kinetic energy, conveniently used instead of the total energy $E$ (a distinction which is of course not particularly relevant when one looks at fluxes originating from TeV-scale DM, i.e. at energies much larger than the proton mass $m_p$, but important for the low energy tails and in the case of small DM masses). Since $m_p \gg m_e$ one can neglect the energy loss term that was instead important for positrons. But new terms appear in the diffusion equation for $f$, which reads
\beq 
\label{eq:diffeqp}
\frac{\partial f}{\partial t} - \mathcal{K}(K)\cdot \nabla^2f + \frac{\partial}{\partial z}\left( {\rm sign}(z)\, f\, V_{\rm conv} \right) = Q-2h\, \delta(z)\, (\Gamma_{\rm ann} + \Gamma_{\rm non-ann}) f  , 
\eeq
where:
\begin{itemize}
\item[-] The pure {\em diffusion term} can again be written as $\mathcal{K}(K) = \mathcal{K}_0 \beta \, (p/\GeV)^\delta$, where $p = (K^2 +2 m_p K)^{1/2}$ and 
$\beta = v_{\bar p}/c = \left(1-m_p^2/(K+m_p)^2\right)^{1/2}$ are the antiproton momentum and velocity. $\delta$ and $\mathcal{K}_0$ are given in Table~\ref{tab:proparam}.

\item[-] The $V_{\rm conv}$ term corresponds to a {\em convective wind}, assumed to be constant and directed outward from the galactic plane, that tends to push away $\bar p$ with energy $T \circa{<}10\, m_p$. Its value is given in Table~\ref{tab:proparam}.

\item[-] The {\em source term} $Q$ due to DM DM annihilations or DM decay has a form fully analogous to eq.\eq{Qann} or\eq{Qdec}, with $E$ now formally replaced by $K$.

\item[-]
The first part of the last term in eq.\eq{diffeqp} describes the annihilations of $\bar p$ on interstellar protons in the galactic plane
(with a thickness of $h=0.1\,{\rm kpc} \ll L$) with rate
$\Gamma_{\rm ann} = (n_{\rm H} + 4^{2/3} n_{\rm He}) \sigma^{\rm ann}_{p\bar{p}} v_{\bar{p}}$,
where $n_{\rm H}\approx 1/{\rm cm}^3$ is the hydrogen density, $n_{\rm He}\approx 0.07\, n_{\rm H}$ is the Helium density (the factor $4^{2/3}$ accounting for the different geometrical cross section in an effective way)
and $ \sigma^{\rm ann}_{p\bar{p}}$ is given by \cite{crosssection,HisanoAntiparticles}
\beq
\sigma_{p \bar p}^{\rm ann} = \left\{ 
\begin{array}{ll}
661\, (1+0.0115\, K^{-0.774} - 0.984\, K^{0.0151})\ {\rm mbarn}, & {\rm for}\ K < 15.5\, {\rm GeV} \\ 
36\, K^{-0.5}\ {\rm mbarn}, & {\rm for}\ K \geq 15.5\, {\rm GeV}
\end{array}
 \right. .
\label{eq:sigmaann}
\eeq
The second part, similarly, describes the interactions on interstellar protons in the galactic plane in which the $\bar p$'s do not annihilate but lose a significant fraction of their energy. Technically, one should keep them in the flux, with a degraded energy: they are referred to as ``tertiary antiprotons''. We adopt instead the simplifying approximation of treating them as if they were removed from the flux. The cross section that we need for the whole last term of eq.\eq{diffeqp} is then the sum of $\sigma^{\rm ann}_{p\bar{p}} + \sigma^{\rm non-ann}_{p\bar{p}} = \sigma^{\rm inel}_{p\bar{p}}$. It is given in~\cite{crosssection} as
\beq
\sigma^{\rm inel}_{p\bar{p}}(K) = 24.7\, (1 + 0.584\, K^{-0.115} + 0.856\, K^{-0.566})\ {\rm mbarn}
\label{eq:sigmaann}
\eeq
(at large energies this expression has to be replaced by a better approximation~\cite{DonatoApJ563}). We find, anyway, that the precise expressions adopted for these cross sections do not significantly impact the final results.  
\item[-]
We neglect, as just said, the effect of ``tertiary antiprotons''. It can be re-included in terms of an absorption term proportional to a different $\sigma^{\rm non-ann}$, and of a re-injection term $Q^{\rm tert}$ proportional to the integrated cross section over $f(K)$. The full solution of the resulting integro-differential equation can be found in~\cite{DonatoApJ563}. The effect of tertiaries is mainly relevant at low energies $K \lesssim$ few GeV.
\item[-] 
Finally notice that we do not include diffusive reacceleration. It does not play a major role for $\bar p$ (as it was not for $e^\pm$) at the high energies in which we are mostly interested, say larger than tens of GeV. It can instead affect the spectrum at $\sim$GeV energies, and it can be reintroduced in an effective way by adding an effective energy-loss coefficient and/or modifying the energy dependance of the spatial diffusion coefficient. We refer to~\cite{DonatoPRD69,DonatoApJ563,MaurinApJ555} for details.
%\item[-] Finally notice that diffusive reacceleration does not play a major role for $\bar p$, as it was not for $e^\pm$~\cite{SalatiCargese}. 
 \end{itemize}

\begin{figure}[!t]
\begin{minipage}{0.6\textwidth}
\tiny{
\begin{tabular}{c|c|rrrrrr}
\multicolumn{8}{l}{\footnotesize DM annihilation} \\
\hline
\rm{halo} & \rm{prop} & $a_0$ & $a_1$ & $a_2$ & $a_3$ & $a_4$ & $a_5$ \\[0.3mm]
\hline
 & \text{MIN} & 0.9251 & 0.6464 & -0.3105 & -0.0917 & 0.0406 & -0.0039 \\
 \text{NFW} & \text{MED} & 1.8398 & 0.5345 & -0.2930 & -0.0439 & 0.0252 & -0.0025 \\
 & \text{MAX} & 2.6877 & -0.1339 & -0.1105 & 0.0170 & -0.0010 & 0.0000 \\
\hline
 & \text{MIN} & 0.9455 & 0.6282 & -0.3175 & -0.0807 & 0.0373 & -0.0036 \\
 \text{Moo} & \text{MED} & 1.8714 & 0.5922 & -0.2815 & -0.0629 & 0.0304 & -0.0029 \\
 & \text{MAX} & 2.7960 & -0.1231 & -0.1164 & 0.0188 & -0.0013 & 0.0000 \\
\hline
 & \text{MIN} & 0.9191 & 0.6493 & -0.3104 & -0.0922 & 0.0407 & -0.0039 \\
 \text{Iso} & \text{MED} & 1.7969 & 0.4441 & -0.3128 & -0.0159 & 0.0179 & -0.0019 \\
 & \text{MAX} & 2.5071 & -0.1557 & -0.0990 & 0.0135 & -0.0005 & -0.0000 \\
\hline
 & \text{MIN} & 0.9104 & 0.6564 & -0.3067 & -0.0962 & 0.0418 & -0.0040 \\
 \text{Ein} & \text{MED} & 1.8804 & 0.5813 & -0.2960 & -0.0502 & 0.0271 & -0.0027 \\
 & \text{MAX} & 2.7914 & -0.1294 & -0.1115 & 0.0165 & -0.0008 & -0.0000 \\
\hline
 & \text{MIN} & 0.9104 & 0.6564 & -0.3067 & -0.0962 & 0.0418 & -0.0040 \\
 \text{EiB} & \text{MED} & 1.9505 & 0.6984 & -0.3038 & -0.0689 & 0.0331 & -0.0032 \\
 & \text{MAX} & 3.0003 & -0.1225 & -0.1102 & 0.0138 & -0.0000 & -0.0001 \\
\hline
 & \text{MIN} & 0.9175 & 0.6429 & -0.3133 & -0.0902 & 0.0404 & -0.0039 \\
 \text{Bur} & \text{MED} & 1.7426 & 0.3736 & -0.3127 & -0.0027 & 0.0140 & -0.0016 \\
 & \text{MAX} & 2.3763 & -0.1637 & -0.0948 & 0.0123 & -0.0003 & -0.0000\end{tabular}}
\caption{\em \small {\bfseries Propagation function for antiprotons from annihilating DM}, for the different halo profiles and sets of propagation parameters, and the corresponding fit parameters to be used in eq.~$(\ref{eq:fitantiprotons})$.}
\label{fig:Rfunctions}
\end{minipage}
\quad
\begin{minipage}{0.37\textwidth}
\vspace{-0.9cm}
\centering
\includegraphics[width=\textwidth]{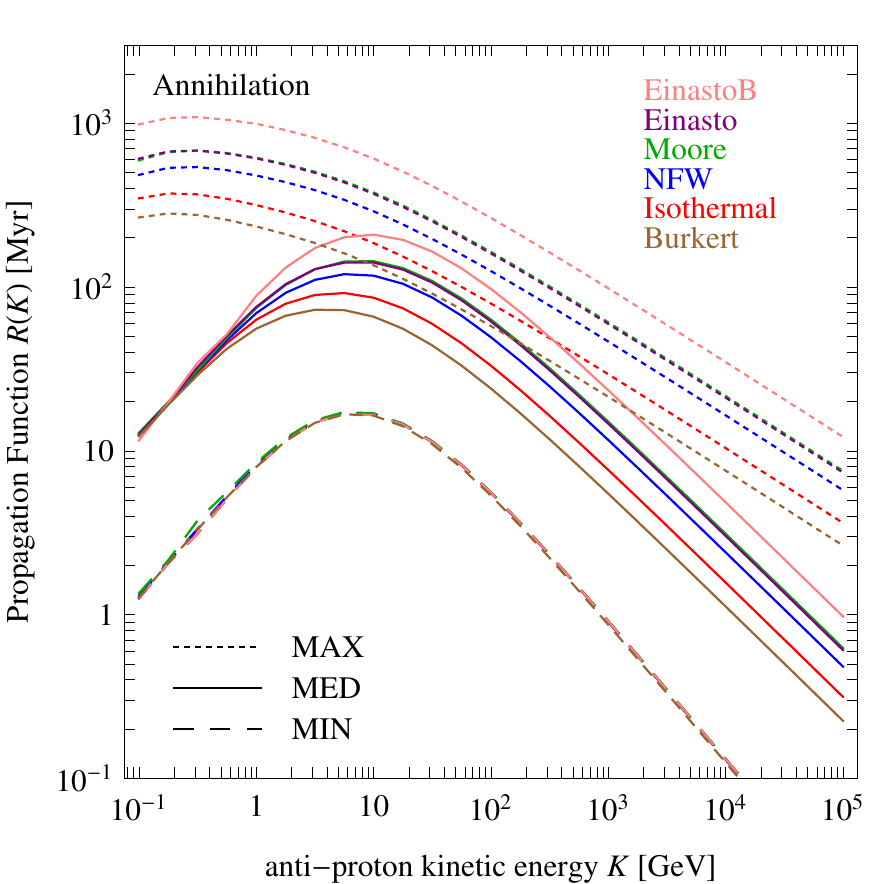}
\end{minipage}
\begin{minipage}{0.6\textwidth}
\vspace{1cm}
\tiny{
\begin{tabular}{c | c | rrrrrr}
\multicolumn{8}{l}{\footnotesize DM decay} \\
\hline
\rm{halo} & \rm{prop} & $a_0$ & $a_1$ & $a_2$ & $a_3$ & $a_4$ & $a_5$ \\[0.3mm]
\hline
 & \text{MIN} & 0.9175 & 0.6429 & -0.3121 & -0.0910 & 0.0406 & -0.0039 \\
 \text{NFW} & \text{MED} & 1.7507 & 0.3892 & -0.3101 & -0.0070 & 0.0151 & -0.0016 \\
 & \text{MAX} & 2.4353 & -0.1549 & -0.1004 & 0.0144 & -0.0007 & -0.0000 \\
  \hline
 & \text{MIN} & 0.9179 & 0.6431 & -0.3132 & -0.0902 & 0.0404 & -0.0038 \\
 \text{Moo} & \text{MED} & 1.7534 & 0.3933 & -0.3095 & -0.0077 & 0.0153 & -0.0017 \\
 & \text{MAX} & 2.4430 & -0.1541 & -0.1001 & 0.0144 & -0.0009 & 0.0000 \\
  \hline
 & \text{MIN} & 0.9171 & 0.6434 & -0.3134 & -0.0903 & 0.0404 & -0.0039 \\
 \text{Iso} & \text{MED} & 1.7450 & 0.3770 & -0.3128 & -0.0033 & 0.0142 & -0.0016 \\
 & \text{MAX} & 2.4095 & -0.1580 & -0.0988 & 0.0139 & -0.0006 & -0.0000 \\
  \hline
 & \text{MIN} & 0.9177 & 0.6436 & -0.3130 & -0.0904 & 0.0404 & -0.0039 \\
 \text{Ein} & \text{MED} & 1.7554 & 0.3959 & -0.3096 & -0.0082 & 0.0154 & -0.0017 \\
 & \text{MAX} & 2.4466 & -0.1543 & -0.1007 & 0.0144 & -0.0007 & -0.0000 \\
  \hline
 & \text{MIN} & 0.9177 & 0.6436 & -0.3130 & -0.0904 & 0.0404 & -0.0039 \\
 \text{EiB} & \text{MED} & 1.7600 & 0.4044 & -0.3083 & -0.0104 & 0.0160 & -0.0017 \\
 & \text{MAX} & 2.4612 & -0.1530 & -0.1012 & 0.0145 & -0.0007 & -0.0000 \\
  \hline
 & \text{MIN} & 0.9165 & 0.6419 & -0.3141 & -0.0898 & 0.0403 & -0.0039 \\
 \text{Bur} & \text{MED} & 1.7352 & 0.3616 & -0.3150 & 0.0006 & 0.0132 & -0.0015 \\
 & \text{MAX} & 2.3956 & -0.1570 & -0.0998 & 0.0144 & -0.0007 & 0.0000
\end{tabular}}
\caption{\em \small {\bfseries Propagation function for antiprotons from decaying DM}, for the different halo profiles and sets of propagation parameters, and the corresponding fit parameters to be used in eq.~$(\ref{eq:fitantiprotons})$.}
\label{fig:Rfunctionsdecay}
\end{minipage}
\quad
\begin{minipage}{0.37\textwidth}
\vspace{-0cm}
\centering
\includegraphics[width=\textwidth]{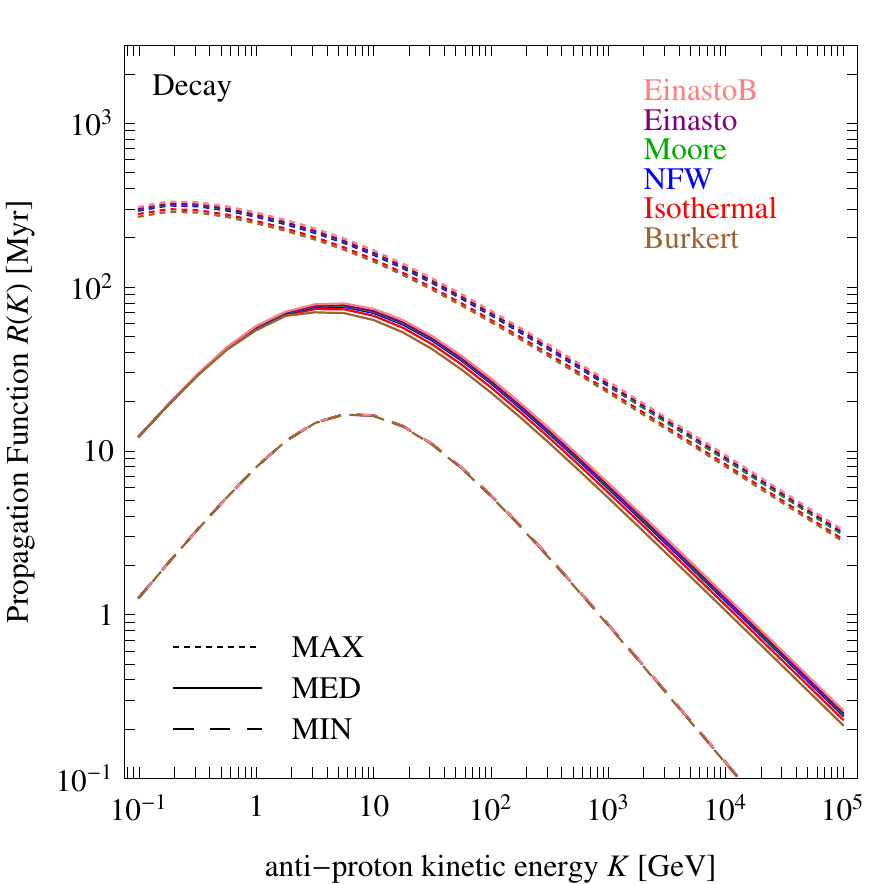}
\end{minipage}
\end{figure}

Assuming again steady state conditions the first term in the diffusion equation vanishes, and the equation can be solved analytically~\cite{MaurinApJ555, methodPbar, TailletRRDA}. In the ``no-tertiaries" approximation that we adopt, the solution for the antiproton differential flux at the position of the Earth $ d\Phi_{\bar p}/dK\,(K,\vec r_\odot) = v_{\bar p}/(4\pi) f $ acquires a simple factorized form (see e.g.~\cite{DonatoPRD69})
\beq
\frac{d\Phi_{\bar p}}{dK}(K,\vec r_\odot) = \frac{v_{\bar p}}{4\pi} \left\{
\begin{array}{l}
 \displaystyle  \left(\frac{\rho_\odot}{M_{\rm DM}}\right)^2 R(K)   \sum_f \frac{1}{2} \langle \sigma v\rangle_f \frac{dN^f_{\bar p}}{dK} \quad {\rm (annihilation)} \\[3mm]
 \displaystyle  \left(\frac{\rho_\odot}{M_{\rm DM}}\right) R(K)   \sum_f \Gamma_f \frac{dN^f_{\bar p}}{dK}  \quad {\rm (decay)}
  \end{array}
  \right.
  .
\label{eq:fluxpbar}
\eeq
The $f$ index runs over all the annihilation channels with antiprotons in the final state, with the respective cross sections or decay rates; this part contains the particle physics input. The function $R(K)$ encodes all the astrophysics of production and propagation.~\footnote{Formally, it is given by
\beq
R(K) = \sum_{n=1}^\infty J_0\left(\zeta_n \frac{r_\odot}{R}\right) 
{\rm exp}\left[ -\frac{V_{\rm conv} L}{2 \mathcal{K}(K)} \right]
\frac{y_n(L)}{A_n \sinh(S_n L/2)}
\label{eq:R}
\eeq
with 
\beq
y_n(Z) = \frac{4}{J_1^2(\zeta_n) R^2} \int_0^R dr\, r \, J_0( \zeta_n r/ R ) 
\int_0^Z dz \, {\rm exp}\left[ \frac{V_{\rm conv} (Z-z)}{2 \mathcal{K}(K)} \right] {\rm sinh}\left(S_n (Z-z)/2\right)
\left( \frac{\rho(r,z)}{\rho_\odot} \right)^2 
\label{eq:y}
\eeq
The coefficients $A_n = 2 h \Gamma_{\rm ann} + V_{\rm conv} + \mathcal{K}(K)\, S_n\coth(S_n L/2)$ with $S_n=\left(V_{\rm conv}^2/\mathcal{K}(K)^2 + 4\zeta_n^2/R^2\right)^{1/2}$ encode the effects of diffusion.}
There is such a `propagation function' for annihilations and for decays for any choice of DM galactic profile and for any choice of set of propagation parameters among those in Table~\ref{tab:proparam}. 
We provide $R(K)$ for all these cases in terms of a fit function  
\beq
{\rm log}_{10}\left[R(K)/{\rm Myr}\right] = a_0 + a_1\, \kappa + a_2\,  \kappa^2 + a_3\,  \kappa^3 + a_4\,  \kappa^4 + a_5\,  \kappa^5,
\label{eq:fitantiprotons}
\eeq
with $\kappa = \log_{10}K/\GeV$ and the coefficients reported in the tables in fig.~\ref{fig:Rfunctions} and~\ref{fig:Rfunctionsdecay} (and also reported on the \myurl{www.marcocirelli.net/PPPC4DMID.html}{website}~\cite{website}).\footnote{The fit functions reproduce the results of our numerical calculation to better than 6\% (with the exception of the EinastoB MED case (for annihilations), for which the accuracy drops to a still acceptable 11\%) over the whole range $K = 100\ {\rm MeV} \to 100\ {\rm TeV}$. The fit functions should not be used outside of this range.}

\medskip

Finally, for completeness we also mention the {\em average solar modulation effect}, although it is mainly relevant for non-relativistic $\bar p$'s: the solar wind decreases the kinetic energy $K$ and momentum $p$ of charged cosmic rays such that the energy spectrum $d\Phi_{\bar{p}\oplus}/dK_\oplus$
of antiprotons that reach the Earth with energy $K_\oplus$ and momentum $p_\oplus$ (sometimes referred to as Top of the Atmosphere `ToA' fluxes) is approximatively related to their energy spectrum in the interstellar medium, $d\Phi_{\bar p}/dK$,
as~\cite{GA}
\beq \frac{d\Phi_{{\bar p}\oplus}}{dK_\oplus} = \frac{p_\oplus^2}{p^2} \frac{d\Phi_{\bar p}}{dK},\qquad
K= K_\oplus + |Ze| \phi_F, \qquad
p^2 = 2m_p K+K^2.
\eeq
The so-called Fisk potential $\phi_F$ parameterizes in this effective formalism the kinetic energy loss. A value of $\phi_F= 0.5\, {\rm GV}$ is characteristic of a minimum of the solar cyclic activity, corresponding to the period in which most of the observations have been done in the second half of the 90's and at the end of the years 2000's.

\subsubsection{Antideuterons}
\label{antideuteronpropagation}

\begin{figure}[!t]
\begin{minipage}{0.6\textwidth}
\tiny{
\begin{tabular}{c|c|rrrrrr}
\multicolumn{8}{l}{\footnotesize DM annihilation} \\
\hline
\rm{halo} & \rm{prop} & $a_0$ & $a_1$ & $a_2$ & $a_3$ & $a_4$ & $a_5$ \\[0.3mm]
\hline
 & \text{MIN} & 1.0175 & 0.4312 & -0.3546 & -0.0390 & 0.0298 & -0.0032 \\
 \text{NFW} & \text{MED} & 1.8728 & 0.3547 & -0.2974 & -0.0130 & 0.0175 & -0.0020 \\
 & \text{MAX} & 2.5460 & -0.1613 & -0.0925 & 0.0171 & -0.0027 & 0.0003 \\
 \hline
 & \text{MIN} & 1.0175 & 0.4312 & -0.3546 & -0.0390 & 0.0298 & -0.0032 \\
 \text{Moo} & \text{MED} & 1.9175 & 0.4110 & -0.2968 & -0.0260 & 0.0216 & -0.0023 \\
 & \text{MAX} & 2.6520 & -0.1496 & -0.0979 & 0.0186 & -0.0030 & 0.0003 \\
\hline
 & \text{MIN} & 1.0175 & 0.4312 & -0.3546 & -0.0390 & 0.0298 & -0.0032 \\
 \text{Iso} & \text{MED} & 1.8111 & 0.2651 & -0.3008 & 0.0058 & 0.0120 & -0.0015 \\
 & \text{MAX} & 2.3694 & -0.1842 & -0.0824 & 0.0146 & -0.0024 & 0.0002 \\
 \hline
 & \text{MIN} & 1.0175 & 0.4312 & -0.3546 & -0.0390 & 0.0298 & -0.0032 \\
 \text{Ein} & \text{MED} & 1.9215 & 0.3930 & -0.2964 & -0.0206 & 0.0197 & -0.0021 \\
 & \text{MAX} & 2.6484 & -0.1558 & -0.0931 & 0.0160 & -0.0023 & 0.0002 \\
 \hline
 & \text{MIN} & 1.0175 & 0.4312 & -0.3546 & -0.0390 & 0.0298 & -0.0032 \\
 \text{EiB} & \text{MED} & 2.0149 & 0.5045 & -0.3185 & -0.0317 & 0.0245 & -0.0026 \\
 & \text{MAX} & 2.8543 & -0.1457 & -0.0909 & 0.0111 & -0.0007 & 0.0000 \\
 \hline
 & \text{MIN} & 1.0175 & 0.4312 & -0.3546 & -0.0390 & 0.0298 & -0.0032 \\
 \text{Bur} & \text{MED} & 1.7412 & 0.2003 & -0.2932 & 0.0141 & 0.0090 & -0.0012 \\
 & \text{MAX} & 2.2415 & -0.1935 & -0.0787 & 0.0137 & -0.0022 & 0.0002
 \end{tabular}}
\caption{\em \small {\bfseries Propagation function for antideuterons from annihilating DM}, for the different halo profiles and sets of propagation parameters, and the corresponding fit parameters to be used in eq.~$(\ref{eq:fitantideuterons})$.}
\label{fig:RfunctionsD}
\end{minipage}
\quad
\begin{minipage}{0.37\textwidth}
\vspace{-0.9cm}
\centering
\includegraphics[width=\textwidth]{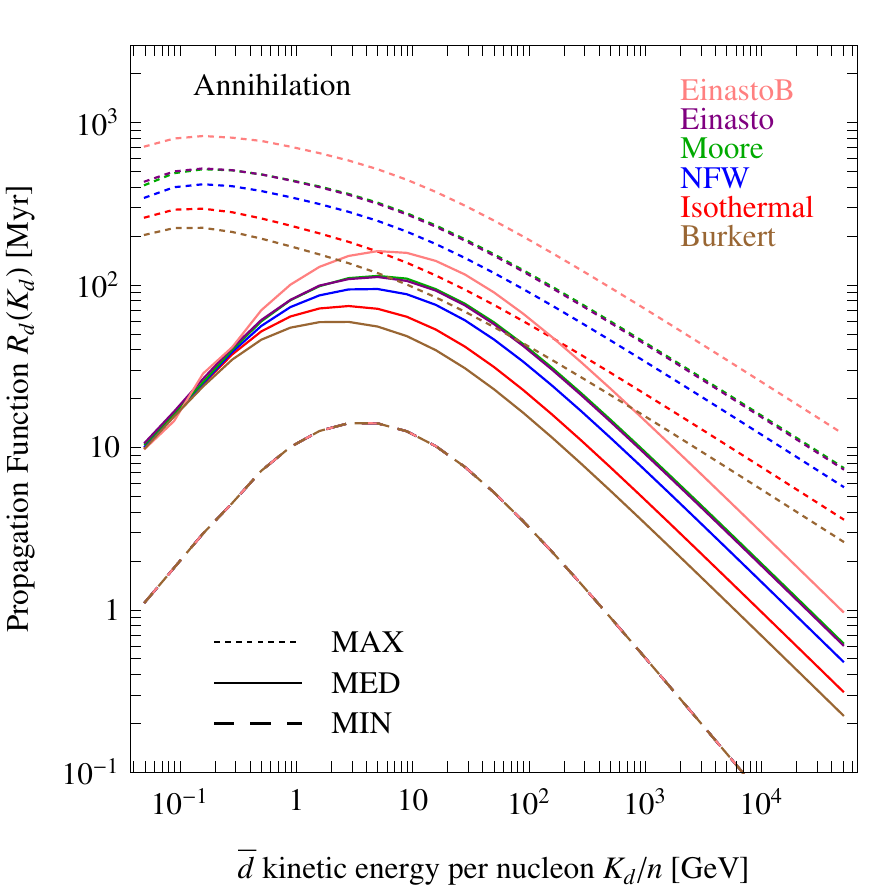}
\end{minipage}
\begin{minipage}{0.6\textwidth}
\vspace{1cm}
\tiny{
\begin{tabular}{c | c | rrrrrr}
\multicolumn{8}{l}{\footnotesize DM decay} \\
\hline
\rm{halo} & \rm{prop} & $a_0$ & $a_1$ & $a_2$ & $a_3$ & $a_4$ & $a_5$ \\[0.3mm]
\hline
 & \text{MIN} & 1.0097 & 0.4282 & -0.3575 & -0.0378 & 0.0297 & -0.0033 \\
 \text{NFW} & \text{MED} & 1.7536 & 0.2154 & -0.2937 & 0.0116 & 0.0098 & -0.0013 \\
 & \text{MAX} & 2.3010 & -0.1861 & -0.0833 & 0.0155 & -0.0026 & 0.0003 \\
 \hline
 & \text{MIN} & 1.0097 & 0.4282 & -0.3575 & -0.0378 & 0.0297 & -0.0033 \\
 \text{Moo} & \text{MED} & 1.7564 & 0.2193 & -0.2931 & 0.0110 & 0.0099 & -0.0013 \\
 & \text{MAX} & 2.3085 & -0.1848 & -0.0830 & 0.0153 & -0.0028 & 0.0003 \\
 \hline
 & \text{MIN} & 1.0097 & 0.4282 & -0.3575 & -0.0378 & 0.0297 & -0.0033 \\
 \text{Iso} & \text{MED} & 1.7445 & 0.2035 & -0.2936 & 0.0138 & 0.0091 & -0.0012 \\
 & \text{MAX} & 2.2758 & -0.1893 & -0.0818 & 0.0151 & -0.0026 & 0.0002 \\
 \hline
 & \text{MIN} & 1.0097 & 0.4282 & -0.3575 & -0.0378 & 0.0297 & -0.0033 \\
 \text{Ein} & \text{MED} & 1.7591 & 0.2216 & -0.2935 & 0.0107 & 0.0100 & -0.0013 \\
 & \text{MAX} & 2.3119 & -0.1852 & -0.0835 & 0.0155 & -0.0026 & 0.0002 \\
 \hline
 & \text{MIN} & 1.0097 & 0.4282 & -0.3575 & -0.0378 & 0.0297 & -0.0033 \\
 \text{EiB} & \text{MED} & 1.7657 & 0.2297 & -0.2936 & 0.0093 & 0.0104 & -0.0013 \\
 & \text{MAX} & 2.3260 & -0.1837 & -0.0840 & 0.0156 & -0.0026 & 0.0002 \\
 \hline
 & \text{MIN} & 1.0097 & 0.4282 & -0.3575 & -0.0378 & 0.0297 & -0.0033 \\
 \text{Bur} & \text{MED} & 1.7291 & 0.1891 & -0.2922 & 0.0156 & 0.0085 & -0.0011 \\
 & \text{MAX} & 2.2631 & -0.1893 & -0.0826 & 0.0157 & -0.0027 & 0.0003
\end{tabular}}
\caption{\em \small {\bfseries Propagation function for antideuterons from decaying DM}, for the different halo profiles and sets of propagation parameters, and the corresponding fit parameters to be used in eq.~$(\ref{eq:fitantideuterons})$.}
\label{fig:RfunctionsdecayD}
\end{minipage}
\quad
\begin{minipage}{0.37\textwidth}
\vspace{-0cm}
\centering
\includegraphics[width=\textwidth]{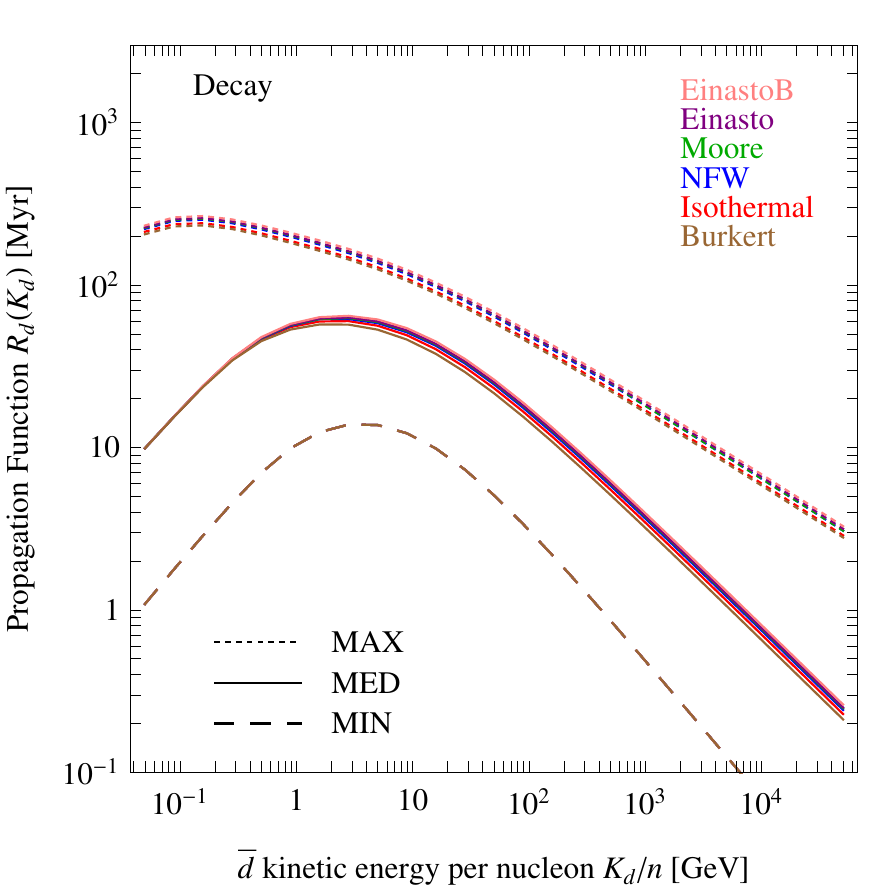}
\end{minipage}
\end{figure}

The propagation of antideuterons through the Galaxy follows closely that of antiprotons discussed above, with a few trivial changes. 
The diffusion equation is still the one in eq.~(\ref{eq:diffeqp}). In it:
\begin{itemize}
\item[-] Diffusion, being governed by the electromagnetic properties of the particles, is the same for antideuterons as for antiprotons, but of course the deuteron mass $m_d$ should replace the proton mass in the expression for the kinetic energy and the momentum. 
\item[-] It is actually customary for low $Z$ nuclei to use as a variable $K_d/n$: the kinetic energy per nucleon ($n=2$ in this case). We will present all results as functions of this quantity.
\item[-] The treatment of spallations of $\bar d$ on the interstellar gas (`annihilating' and `non-annihilating' reactions) is less straightforward than for $\bar p$, essentially for the scarcity of experimental nuclear data on $\bar d$. We still write $\Gamma_{\rm (non-)ann} = (n_{\rm H} + 4^{2/3} n_{\rm He}) \sigma^{\rm (non-)ann}_{p\bar{d}} v_{\bar{d}}$ and so we now need $\sigma^{\rm inel}_{p\bar{d}}$. This can be obtained from related experimental measurements with the charge conjugated reaction $\bar{p}d$ or with the reaction $p\bar{p}$: we refer for more details to~\cite{DonatoDbar, DMantideuterium} and references therein. All in all, we find that a good approximation is to effectively adopt $\sigma^{\rm inel}_{p\bar{d}} \simeq 2\, \sigma^{\rm inel}_{p\bar{p}}$.
\end{itemize}
With the ingredients above one can compute, exactly as for antiprotons, an antideuteron propagation function $R_d(K_d/n)$ for annihilations and for decays for any choice of DM galactic profile and for any choice of set of propagation parameters among those in Table~\ref{tab:proparam}. 
We provide $R_d(K_d/n)$ for all these cases in terms of a fit function  
\beq
{\rm log}_{10}\left[R_d(K_d/n)/{\rm Myr}\right] = a_0 + a_1\, \kappa_d + a_2\,  \kappa_d^2 + a_3\,  \kappa_d^3 + a_4\,  \kappa_d^4 + a_5\,  \kappa_d^5,
\label{eq:fitantideuterons}
\eeq
with $\kappa_d = \log_{10}(K_d/n/\GeV)$ and the coefficients reported in the tables in fig.~\ref{fig:RfunctionsD} and~\ref{fig:RfunctionsdecayD} (and also reported on the \myurl{www.marcocirelli.net/PPPC4DMID.html}{website}~\cite{website}).\footnote{The fit functions reproduce the results of our numerical calculation to better than 5\% (with the exception of the EinastoB case, for which the accuracy drops to a still acceptable 10\%) over the whole range $K_d/n = 50\ {\rm MeV} \to 50\ {\rm TeV}$. The fit functions should not be used outside of this range.} Not suprisingly, since the changes are so minimal with respect to antiprotons and affecting subdominant processes only, the propagation functions resemble those for antiprotons closely. 

With these ingredients, it is straightforward to compute the antideuteron differential flux at the position of the Earth as
\beq
\frac{d\Phi_{\bar d}}{dE}(K,\vec r_\odot) = \frac{v_{\bar d}}{4\pi} \left\{
\begin{array}{l}
 \displaystyle  \left(\frac{\rho_\odot}{M_{\rm DM}}\right)^2 R_d(K_d/n)   \sum_f \frac{1}{2} \langle \sigma v\rangle_f \frac{dN^f_{\bar d}}{dK_d} \quad {\rm (annihilation)} \\[3mm]
 \displaystyle  \left(\frac{\rho_\odot}{M_{\rm DM}}\right) R_d(K_d/n)   \sum_f \Gamma_f \frac{dN^f_{\bar d}}{dK_d}  \quad {\rm (decay)}
  \end{array}
  \right.
  .
\label{eq:fluxdbar}
\eeq
(notice that the primary $\bar d$ fluxes at production are given in Sec.~\ref{primary} as a function of $K_d$ and not $K_d/n$). Solar modulation can be applied as discussed for antiprotons, with the due changes.

%%%%%%%%%%%%%%%%%%%%%%%%%%%%%%%%%%%%%%%%%%%%%%%%%

\subsection{Fluxes after propagation at the location of the Earth}
\label{chargedpropagated}

In this section we present the resulting fluxes of charged cosmic rays from DM as they would be observed at Earth, computed applying the formalism of the previous section.

\subsubsection{Electrons or positrons}
\label{eppropagated}

\begin{figure}[!t]
\begin{center}
\includegraphics[width=0.48 \textwidth]{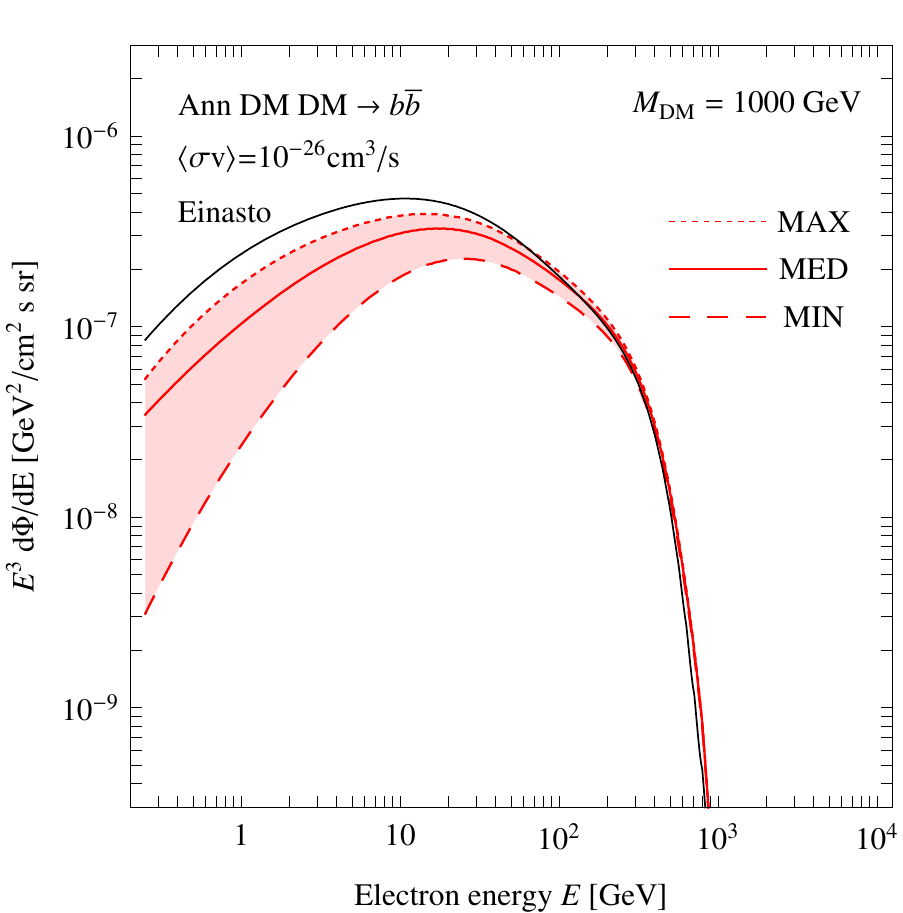} \quad \includegraphics[width=0.48 \textwidth]{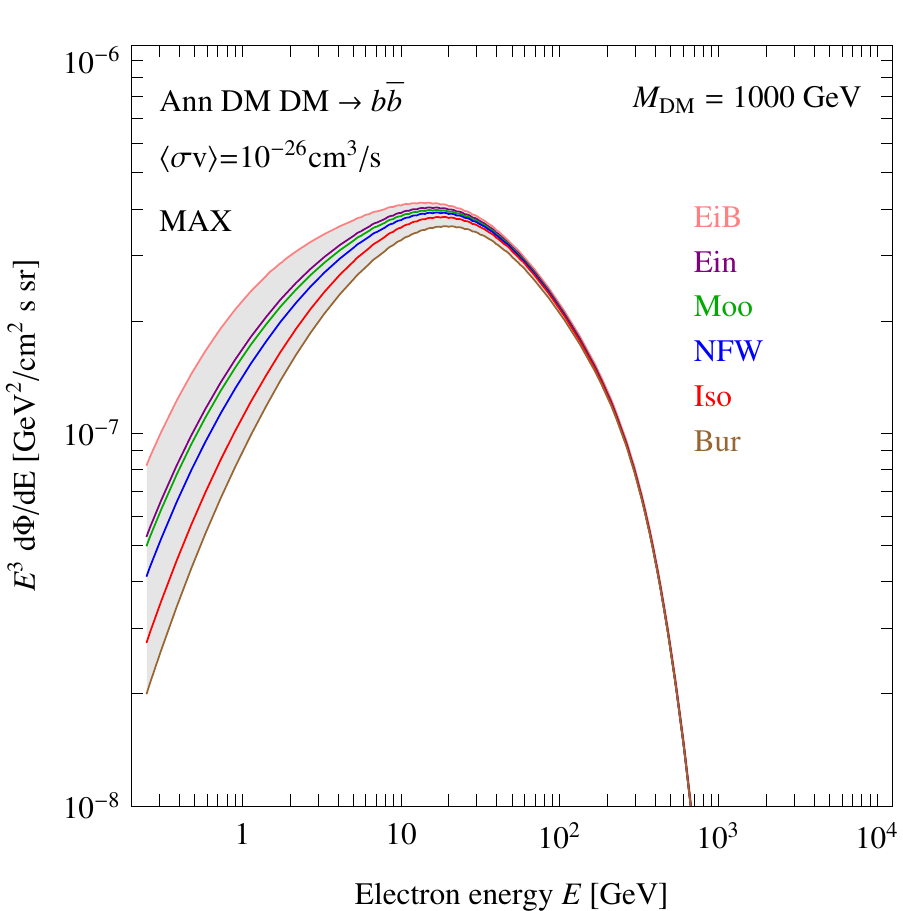}\\[2mm]
\includegraphics[width=0.48 \textwidth]{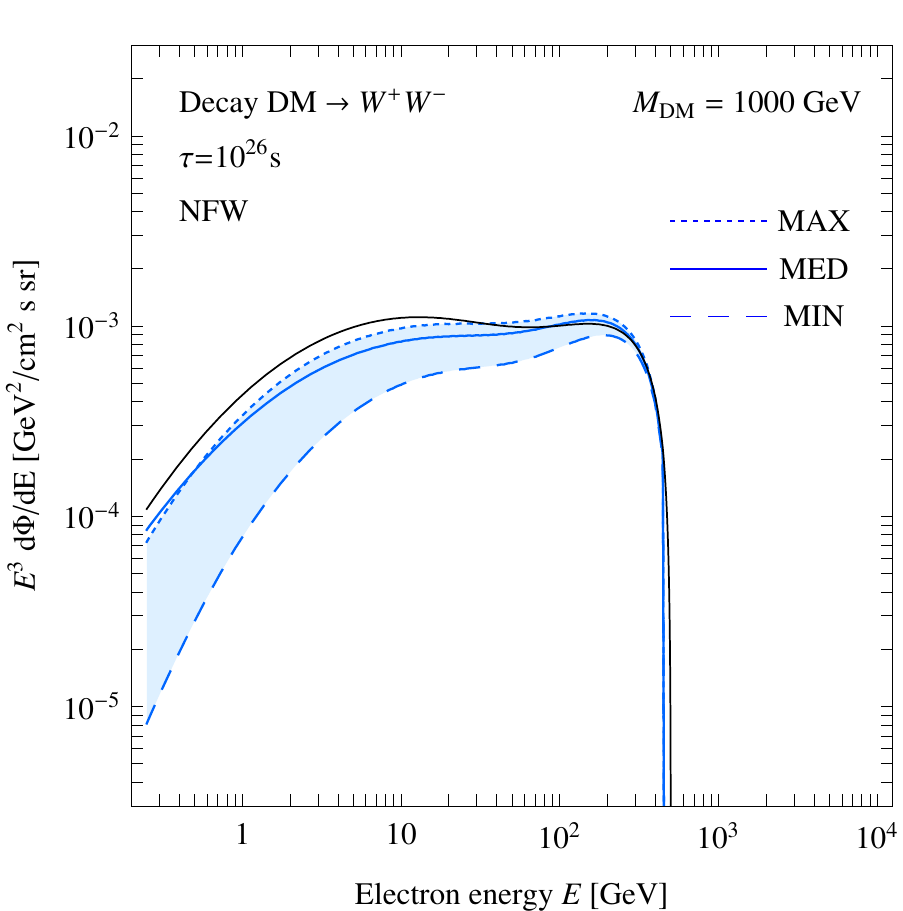} \quad \includegraphics[width=0.48 \textwidth]{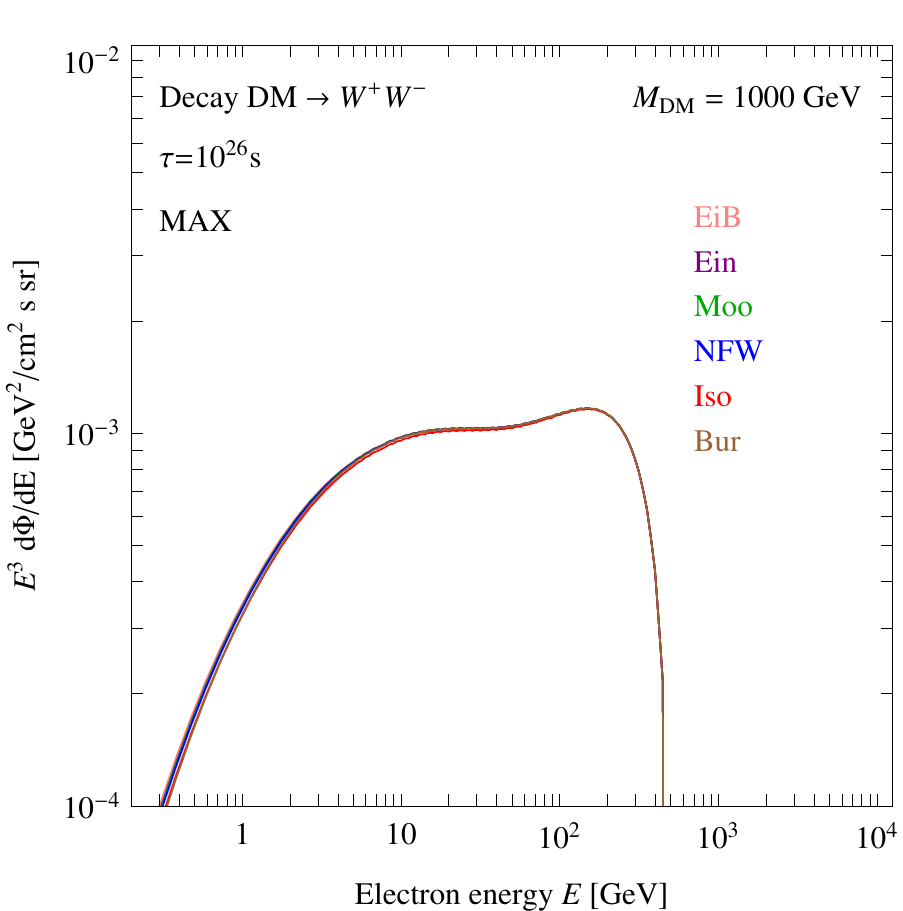}
\caption{\em \small \label{fig:positronspropagated} {\bfseries Fluxes of electrons or positrons at the Earth, after propagation}, for the case of annihilations (top row) and decay (bottom row). In the left panels the propagation parameters are variated, while the halo profile is kept fixed. The opposite is done for the right panels. The choices of annihilation or decay channels and parameters are indicated.}
\end{center}
\end{figure}

Applying the recipe of eq.~(\ref{eq:positronsflux}) it is straightforward to compute the fluxes of electrons and positrons at Earth. We provide them in numerical form on the~\myurl{www.marcocirelli.net/PPPC4DMID.html}{website}~\cite{website}, both in the form of {\sc Mathematica}$^{\tiny{\textregistered}}$ interpolating functions and numerical tables.

\medskip

Fig.~\ref{fig:positronspropagated} presents some examples of such fluxes, for the cases of annihilation and decay.
The choice of propagation parameters (MIN, MED or MAX) affects the final results up to one order of magnitude, especially at energies below tens of GeV and determine somewhat the spectral shape.  For annihilating models, the choice of DM halo profile has a limited impact and again it gives rise to a spread in the prediction mainly at small $E$. Instead it is imperceptible for the decaying case. 
The black solid lines in the left panels of fig.~\ref{fig:positronspropagated} represent the positron flux at Earth computed with the approximated method discussed in sec.~\ref{propagationapprox} (and considering MED propagation parameters). As one can see, at the location of the Earth there are some differences in the flux up to a factor 2 due to the different shape of the halo function, but not much more than that. Therefore we can infer that the new full computation of the $e^\pm$ propagated flux has a limited impact, making the old one quite safe.
Important differences, instead, are appreciable in other regions of the sky, especially close to the galactic center where one has larger energy losses. As a consequence we expect that the diffuse $\gamma$ rays, produced by these propagated electrons/positrons everywhere (see sec.~\ref{ICSgamma}), will be more sensitive to the difference between the two methods.

\begin{figure}[!t]
\begin{center}
\includegraphics[width=0.48 \textwidth]{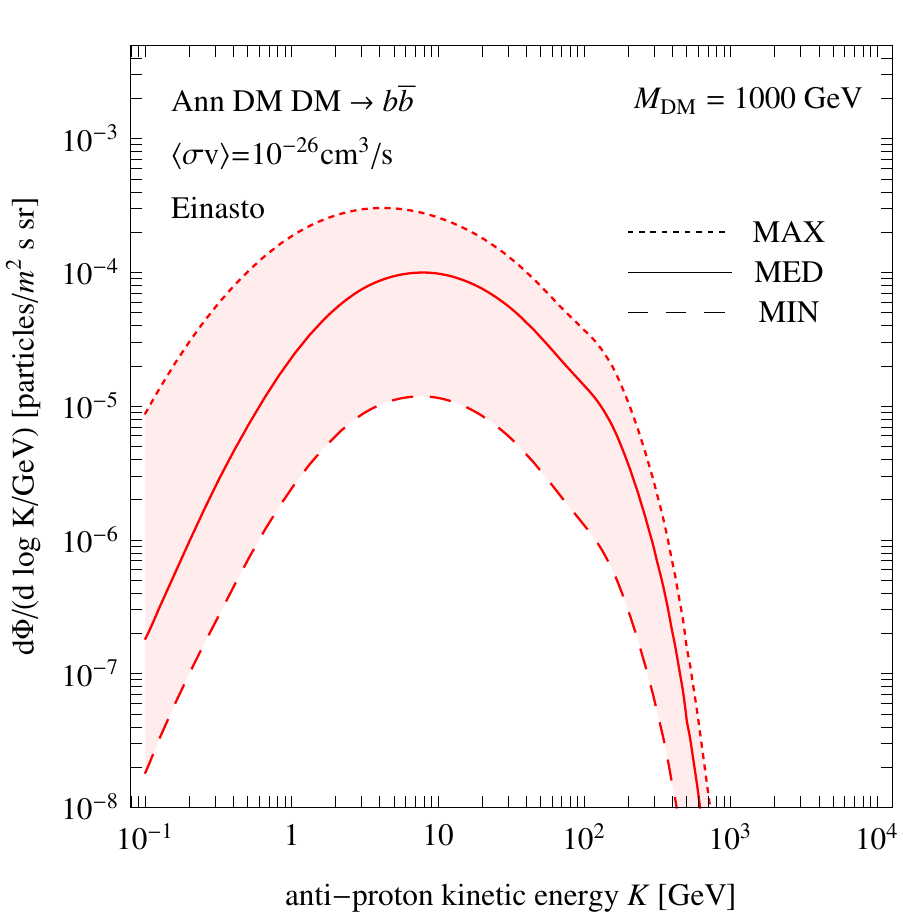} \quad \includegraphics[width=0.48 \textwidth]{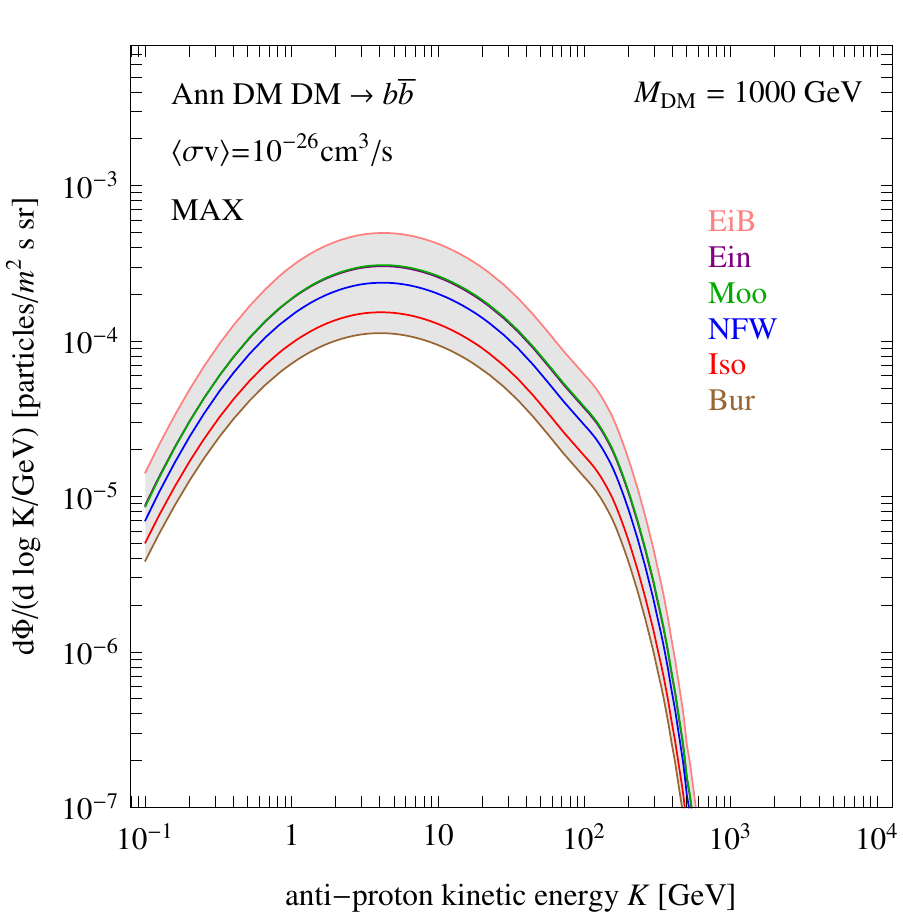}\\[2mm]
\includegraphics[width=0.48 \textwidth]{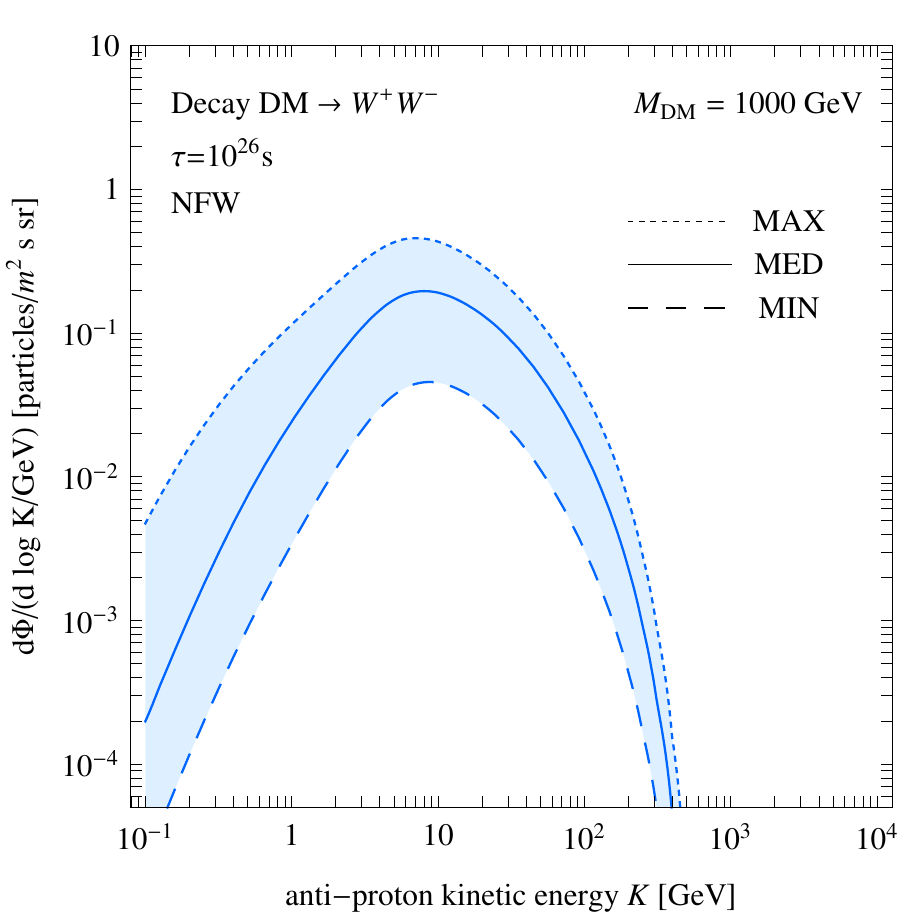} \quad \includegraphics[width=0.48 \textwidth]{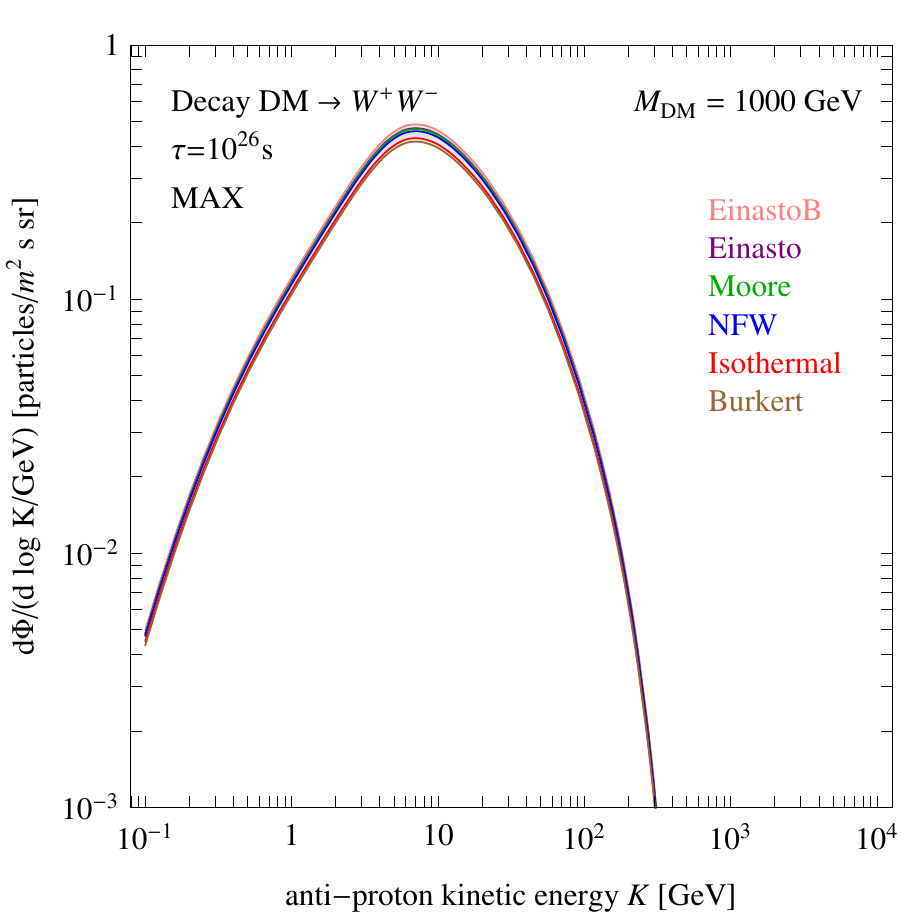}
\caption{\em \small \label{fig:pbarpropagated} {\bfseries Fluxes of antiprotons at the Earth, after propagation}, for the case of annihilations (top row) and decay (bottom row). In the left panels the propagation parameters are variated, while the halo profile is kept fixed. The opposite is done for the right panels. The choices of annihilation or decay channels and parameters are indicated.}
\end{center}
\end{figure}

\subsubsection{Antiprotons}
\label{pbarpropagated}

Applying the recipe of eq.~(\ref{eq:fluxpbar}) it is straightforward to compute the fluxes of antiprotons at Earth, for a given choice of halo profile and propagation parameters. We provide them in numerical form on the~\myurl{www.marcocirelli.net/PPPC4DMID.html}{website}~\cite{website}, both in the form of {\sc Mathematica}$^{\tiny{\textregistered}}$ interpolating functions and numerical tables.

\medskip

Fig.\fig{pbarpropagated} presents some examples of such fluxes, for the cases of annihilation and decay. We do not correct for any solar modulation. It is apparent that the choice of propagation parameters (MIN, MED or MAX) affects in a relevant way the final result, up to a couple of orders of magnitude, even if the spectral shapes are not sensibly modified. The choice of the
DM halo profile, instead, has a limited impact and it is barely visible for the decay case. This is already evident of course in the little variations of the halo function in Fig.~\ref{fig:Rfunctionsdecay} and can be traced back to the fact that the decay signal, being proportional to the first power of the DM density, is mainly sensitive to the local DM halo, where the profiles do not differ sensibly.

\subsubsection{Antideuterons}
\label{dbarpropagated}

Applying the recipe of eq.~(\ref{eq:fluxdbar}) it is straightforward to compute the fluxes of antideuterons at Earth, for a given choice of halo profile and propagation parameters. As usual, we provide them in numerical form on the~\myurl{www.marcocirelli.net/PPPC4DMID.html}{website}~\cite{website}, both in the form of {\sc Mathematica}$^{\tiny{\textregistered}}$ interpolating functions and numerical tables, but we do not present example plots in this case. The qualitative features are very similar to the case of antiprotons, with the necessary changes in scales.

%%%%%%%%%%%%%%%%%%%%%%%%%%%%%%%%%%%%%%%%%%%%%%%%%

\section{Prompt gamma rays}
\label{promptgamma}

Dark Matter produces high energy gamma rays both by direct (`prompt') emission during annihilation or decay and by 
 Inverse Compton Scattering of $e^\pm$ produced by DM on the ambient light (`secondary').
In this section we focus on prompt gamma rays, while in the next one we deal with secondary emissions.

\medskip

The {\em differential flux of photons} from a given angular direction $d\Omega$ produced by the annihilation of self-conjugated DM particles (e.g.\ Majorana fermions) is
\beq 
\label{gammaflux}
\frac{d \Phi_\gamma}{d\Omega\,dE} = \frac{1}{2}\frac{r_\odot}{4\pi} \left(\frac{\rho_\odot}{M_{\rm DM}}\right)^2 J
\sum _f\langle \sigma v\rangle_f \frac{dN_\gamma^f}{dE}  ,\qquad
J = \int_{\rm l.o.s.} \frac{ds}{r_\odot} \left(\frac{\rho(r(s,\theta))}{\rho_\odot}\right)^2 \qquad {\rm (annihilation)}
\eeq
where $dN_\gamma^f/dE$ is the energy spectrum of photons produced per one annihilation~\footnote{Not per initial state particle; not per final state primary particle.} in the channel with final state $f$. If DM is not constituted by self-conjugated particles (e.g.\ in the case of Dirac fermions), then $\sigma v$ must be averaged over DM particles and antiparticles: in practice, the equation above has to be divided by an additional factor of 2 if only particle-antiparticle annihilations are present.

In the case of DM decay, an analogous equation holds
\beq 
\label{gammafluxdec}
\frac{d \Phi_\gamma}{d\Omega\,dE} = \frac{r_\odot}{4\pi} \frac{\rho_\odot}{M_{\rm DM}} J
\sum _f \Gamma_f  \frac{dN_\gamma^f}{dE}  ,\qquad
J = \int_{\rm l.o.s.} \frac{ds}{r_\odot} \left(\frac{\rho(r(s,\theta))}{\rho_\odot}\right)  \qquad {\rm (decay)}
\eeq
Here the coordinate $r$, centered on the Galactic Center, reads $r(s,\theta)=(r_\odot^2+s^2-2\,r_\odot\,s\cos\theta)^{1/2}$, and $\theta$ is the aperture angle between the direction of the line of sight and the axis connecting the Earth to the Galactic Center.

\subsection{J factors}
\label{J}

The $J$ {\em factor} in eq.~(\ref{gammaflux}) and eq.~(\ref{gammafluxdec}) integrates the intervening matter along the line of sight (along which the variable $s$ runs) individuated by the angular direction, and it is conventionally weighted by $r_\odot$ (here assumed to be 8.33 kpc) and the appropriate power of $\rho_\odot$ (here assumed to be 0.3 GeV/cm$^3$) so to be adimensional.\footnote{Alternatively, sometimes an analogous factor is defined as ${\mathcal J} = \int_{\rm l.o.s.} \rho^2(r) = r_\odot \rho_\odot^2\, J$ in units of GeV$^2$/cm$^5$ (annihilation) or ${\mathcal J} = \int_{\rm l.o.s.} \rho(r) = r_\odot \rho_\odot \, J$ in units of GeV/cm$^2$ (decay).} $J (\theta)$ is of course invariant under rotations around the axis which connects the Earth to the GC, due to the assumed spherical symmetry of the DM distribution $\rho(r)$. 

\begin{figure}[!t]
\begin{center}
\includegraphics[width=0.48 \textwidth]{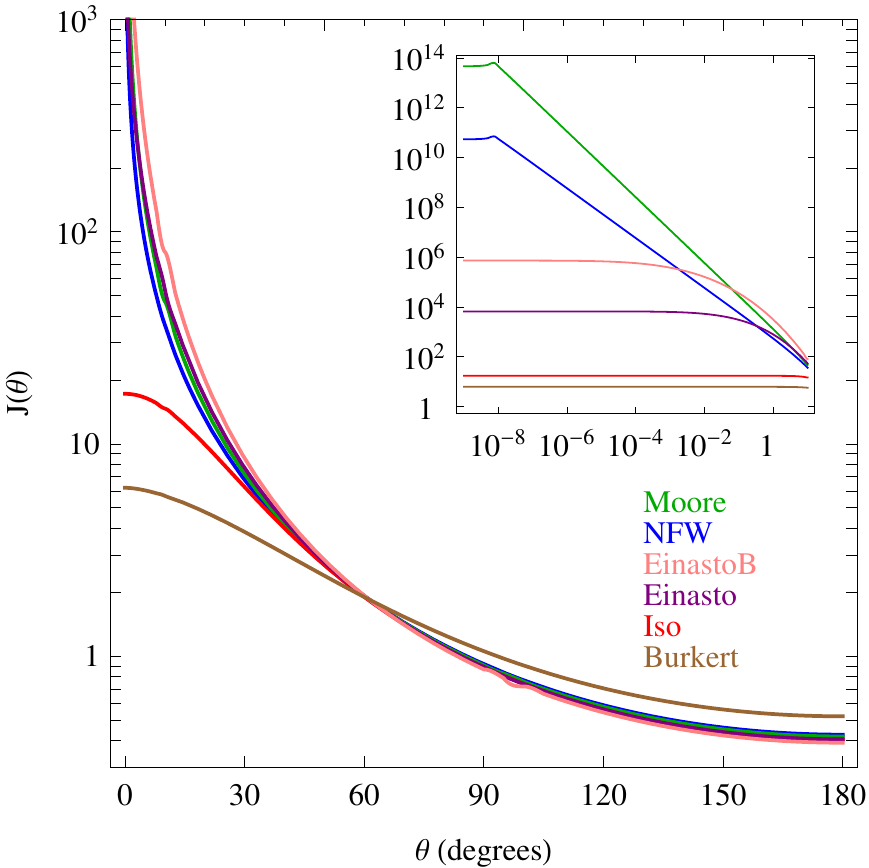} \quad \includegraphics[width=0.48 \textwidth]{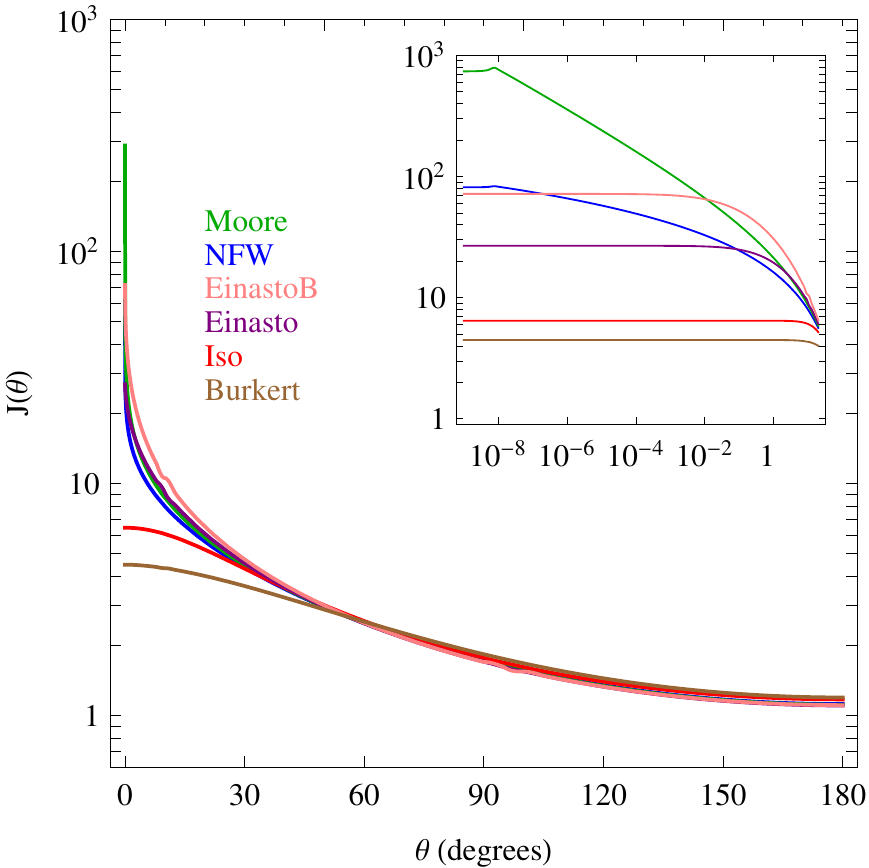}
\end{center}
\caption{\em \small \label{fig:Jtheta} $\boldsymbol{J(\theta)}$ {\bfseries for annihilating} (left) {\bfseries and decaying} (right) {\bfseries Dark Matter}, for the different DM profiles. The color code individuates the profiles (Burkert, Isothermal, Einasto, EinastoB, NFW, Moore  from bottom to top in the inset). }
\end{figure}

The $J$ factors are plotted in fig.~\ref{fig:Jtheta} as a function of $\theta$. We provide them in terms of  {\sc Mathematica}$^{\tiny{\textregistered}}$ interpolating functions on the~\myurl{www.marcocirelli.net/PPPC4DMID.html}{website}~\cite{website}.

\medskip

The recipes (\ref{gammaflux}) and (\ref{gammafluxdec}) are ready for consumption if one needs the flux of gamma rays from a given direction. More often, of course, one needs the {\em integrated flux} over a region $\Delta \Omega$, corresponding e.g.\ to the window of observation or the resolution of the telescope. The $J$ factor is then replaced by the {\em average $J$ factor} for such region, simply defined as $\bar J (\Delta \Omega) = \left( \int_{\Delta \Omega} J\, d\Omega \right) /\Delta \Omega$.
The following simple formul\ae\ hold for regions that are disks of aperture $\theta_{\rm max}$ centered around the GC, annuli $\theta_{\rm min} < \theta < \theta_{\rm max}$ centered around the GC or generic regions defined in terms of galactic latitude $b$ and longitude $\ell$~\footnote{Galactic polar coordinates ($d,\ell,b$) are defined as
$$x = d \cos \ell \cos b,\qquad y = d\sin \ell \cos b,\qquad  z = d \sin b$$
where the Earth is located at $\vec x = 0$ (such that $d$ is the distance from us);
the Galactic Center at $x=r_\odot$, $y=z=0$; and the Galactic plane corresponds to $z\approx 0$.
Consequently $\cos \theta= x/d = \cos b \cdot \cos \ell$.}  (provided they are symmetric around the GC): 
\beq
\begin{array}{llll}
& \displaystyle \Delta \Omega= 2 \pi \int_0^{\theta_{\rm max}} d\theta\, \sin \theta, \quad & \displaystyle \bar J = \frac{2 \pi}{\Delta \Omega} \int d\theta\, \sin \theta\, J(\theta), \quad  &{\rm (disk)} \\[4mm]
& \displaystyle \Delta \Omega= 2 \pi \int_{\theta_{\rm min}}^{\theta_{\rm max}} d\theta\, \sin \theta, \quad & \displaystyle \bar J = \frac{2 \pi}{\Delta \Omega} \int d\theta\, \sin \theta\, J(\theta), \quad & {\rm (annulus)} \\[4mm]
& \displaystyle \Delta \Omega= 4 \int_{b_{\rm min}}^{b_{\rm max}} \int_{\ell_{\rm min}}^{\ell_{\rm max}} db\, d\ell\, \cos b, \quad & \displaystyle \bar J = \frac{4}{\Delta \Omega} \iint db\, d\ell\, \cos b\, J(\theta(b,\ell)), \quad & (b\times \ell\ {\rm region}) 
\end{array}
\label{eq:formulaeJfactors}
\eeq
\begin{sidewaystable}
\centering
\footnotesize{
\begin{tabular}{l|c|c|c|cccccc|ccccccc}
Region & \multicolumn{1}{c|}{latitude $b$}  & \multicolumn{1}{c|}{longitude $l$} & $\Delta\Omega$ &  \multicolumn{6}{c|}{$\bar J_{\rm ann}$} &\multicolumn{6}{c}{$\bar J_{\rm dec}$}  \\
 & or \, aperture $\theta$& & [steradians] & NFW  & Ein & EinB & Iso & Bur & Moore & NFW  & Ein & EinB & Iso & Bur & Moore \\
\hline
\vspace{-3mm} & & & & & & & & & & & & \\

`GC 0.1$^\circ$' & $\theta <0.1^\circ$ &  $-$  & 0.96\, $10^{-5}$ & 11579 & 3579 & 55665 & 17.2 & 6.21 & 81751 & 26.3 & 25.4 & 55.3 & 6.45 & 4.47 & 44.9  \\
`GC 0.14$^\circ$' & $\theta <0.14^\circ$ &  $-$  & 0.19\, $10^{-4}$ & 8255 & 3206 & 43306 & 17.2 & 6.21 & 52395 & 25.1 & 25.0 & 52.9 & 6.45 & 4.47 & 41.5  \\
`GC 1$^\circ$' & $\theta <1^\circ$ &  $-$  & 0.96\, $10^{-3}$ & 1118 & 1196 & 6945 & 17.2 & 6.21 & 3855 & 18.0 & 21.0 & 35.5 & 6.45 & 4.47 & 24.9  \\
`GC 2$^\circ$' & $\theta <2^\circ$ &  $-$  & 0.004 & 542 & 711 & 3103 & 17.2 & 6.19 & 1521 & 15.5 & 18.6 & 28.7 & 6.45 & 4.47 & 20.2  \\
`Gal Ridge' & 0$^\circ < |b| <$0.3$^\circ$  & 0$^\circ < |\ell| <$0.8$^\circ$  & 0.29\, $10^{-3}$ & 1904 & 1605 & 11828 & 17.2 & 6.21 & 7927 & 19.6 & 22.2 & 39.9 & 6.45 & 4.47 & 28.4  \\
`3 $\times$ 3' & 0$^\circ < |b| <$3$^\circ$  & 0$^\circ < |\ell| <$3$^\circ$  & 0.011 & 306 & 443 & 1577 & 17.1 & 6.16 & 741 & 13.6 & 16.4 & 23.6 & 6.43 & 4.46 & 16.9  \\
`5 $\times$ 5' & 0$^\circ < |b| <$5$^\circ$  & 0$^\circ < |\ell| <$5$^\circ$  & 0.030 & 174 & 264 & 783 & 16.8 & 6.10 & 367 & 11.8 & 14.1 & 19.0 & 6.39 & 4.44 & 14.0  \\
`5 $\times$ 30' & 0$^\circ < |b| <$5$^\circ$  & 0$^\circ < |\ell| <$30$^\circ$  & 0.183 & 47.7 & 70.5 & 170 & 12.1 & 5.16 & 84.8 & 7.27 & 8.24 & 9.74 & 5.55 & 4.12 & 7.99  \\
`10 $\times$ 10' & 0$^\circ < |b| <$10$^\circ$  & 0$^\circ < |\ell| <$10$^\circ$  & 0.121 & 77.7 & 118 & 280 & 15.5 & 5.85 & 138 & 9.30 & 10.9 & 13.4 & 6.19 & 4.37 & 10.5  \\
`10 $\times$ 30' & 0$^\circ < |b| <$10$^\circ$  & 0$^\circ < |\ell| <$30$^\circ$  & 0.364 & 35.5 & 51.8 & 109 & 11.7 & 5.09 & 57.2 & 6.86 & 7.71 & 8.89 & 5.48 & 4.10 & 7.44  \\
`10 $\times$ 60' & 0$^\circ < |b| <$10$^\circ$  & 0$^\circ < |\ell| <$60$^\circ$  & 0.727 & 19.5 & 27.8 & 56.7 & 7.59 & 3.91 & 30.4 & 5.06 & 5.51 & 6.13 & 4.39 & 3.57 & 5.36  \\
`GP w/o GC' & 0$^\circ < |b| <$5$^\circ$  & 30$^\circ < |\ell| <$180$^\circ$  & 0.913 & 1.32 & 1.35 & 1.38 & 1.28 & 1.24 & 1.33 & 1.85 & 1.85 & 1.86 & 1.90 & 1.88 & 1.85  \\
`sides of GC' & 0$^\circ < |b| <$10$^\circ$  & 10$^\circ < |\ell| <$30$^\circ$  & 0.242 & 14.4 & 18.5 & 24.0 & 9.79 & 4.70 & 16.7 & 5.64 & 6.12 & 6.62 & 5.12 & 3.96 & 5.90  \\
`outer Galaxy' & 0$^\circ < |b| <$10$^\circ$  & 90$^\circ < |\ell| <$180$^\circ$  & 1.091 & 0.560 & 0.535 & 0.518 & 0.535 & 0.671 & 0.551 & 1.30 & 1.29 & 1.29 & 1.35 & 1.39 & 1.30  \\
`10$-$20' & 10$^\circ < |b| <$20$^\circ$  & 0$^\circ < |\ell| <$180$^\circ$  & 2.116 & 3.23 & 3.85 & 4.62 & 2.57 & 1.77 & 3.58 & 2.43 & 2.50 & 2.58 & 2.40 & 2.21 & 2.47  \\
`20$-$60' & 20$^\circ < |b| <$60$^\circ$  & 0$^\circ < |\ell| <$180$^\circ$  & 6.585 & 1.64 & 1.71 & 1.78 & 1.57 & 1.44 & 1.67 & 2.05 & 2.06 & 2.08 & 2.10 & 2.05 & 2.06  \\
`Gal Poles' & 60$^\circ < |b| <$90$^\circ$  & 0$^\circ < |\ell| <$180$^\circ$  & 1.684 & 0.992 & 0.965 & 0.947 & 0.964 & 1.11 & 0.982 & 1.77 & 1.75 & 1.75 & 1.82 & 1.88 & 1.77  \\

\end{tabular}
\caption{\em \small Some popular observational regions, their angular area and the corresponding values of the {\bfseries average $\boldsymbol{\bar J}$ factor} for different DM halo profiles, in the case of annihilation and decay. `GP' stands for Galactic Plane and `GC' for Galactic Center. With a slight abuse of notation we indicate the absolute value of the longitude $|\ell|$ to signify that the considered regions are always symmetrical with respect to the $\ell=0$ axis (for instance $0^\circ < \ell < 3^\circ$ actually means $\ell > 357^\circ$, $\ell < 3^\circ$). \label{tab:Jfactors}}}
\end{sidewaystable}
where the integration limits in the formul\ae\ for $\bar J$ are left implicit for simplicity but obviously correspond to those in $\Delta\Omega$. For the `$b \times \ell$ region' the limits of the integration region are intended to be in one quadrant (e.g.\ the $b >0^\circ$, $0<\ell<90^\circ$ one for definiteness), hence the factor of 4 to report it to the four quadrants.

The values of the $\bar J$ factors and $\Delta \Omega$ for some popular observational regions are reported in table~\ref{tab:Jfactors}, for the cases of annihilating and decaying DM and for the different halo profiles. Any other region can be computed by using the formul\ae{} in eq.~(\ref{eq:formulaeJfactors}) and the $J(\theta)$ functions provided above.

With these ingredients, one explicitly has for the differential $\gamma$ ray flux from a region $\Delta\Omega$
\beq
\label{gammafluxI}
\frac{d \Phi_\gamma}{dE}(E_\gamma) =  \frac{r_\odot}{4\pi} \left\{ 
\begin{array}{ll}
\displaystyle \frac{1}{2} \left(\frac{\rho_\odot}{M_{\rm DM}}\right)^2 \bar J \, \Delta \Omega \sum _f\langle \sigma v\rangle_f \frac{dN_\gamma^f}{dE_\gamma} & {\rm (annihilation)} \\[3mm]
\displaystyle \frac{\rho_\odot}{M_{\rm DM}} \bar J \, \Delta\Omega \sum _f \Gamma_f  \frac{dN_\gamma^f}{dE_\gamma} & {\rm (decay)}
\end{array}
\right.
\eeq

%%%%%%%%%%%%%%%%%%%%%%%%%%%%%%%%%%%%%%%%%%%%%%%%%%

\section{Photons from electrons and positrons in the Galaxy}
\label{ICSgamma}

Galactic $e^\pm$ generated by DM in the diffusion volume
lose essentially all their energy into photons by means of two processes: Inverse Compton and synchrotron radiation.

The resulting fluxes of ICS $\gamma$ rays and of microwave synchrotron radiation are thus possible signatures of DM. The ICS flux is particularly promising. One of its best features is that it originates from `everywhere' in the diffusion volume of the galactic halo, including regions where the astrophysical background is reduced (e.g.\ at high latitudes). 
Moreover, essentially everywhere synchrotron energy losses are sub-dominant with respect to Inverse Compton energy losses (as discussed in Sec.~\ref{positronpropagation}), so that, thanks to energy conservation, the resulting ICS $\gamma$ flux suffers only moderate astrophysical uncertainties. 

The microwave synchrotron emission, on the other hand, is generated in significant amount from where the intensity of the magnetic field and of Dark Matter is highest, close to the Galactic Center, and therefore is plagued by more uncertainty and more background.

\medskip

We first present the detailed recipe to compute ICS $\gamma$ rays. We leave synchrotron radiation for the next subsection.

\subsection{Inverse Compton gamma rays}
\label{ICS}

The differential flux of ICS photons within an angular region $\Delta \Omega$ can be written in terms of the emissivity $j(E_\gamma,r)$ of a cell located at a distance $r\equiv|\vec x|$ from the Galactic Center as

\beq
\frac{d\Phi_{{\rm IC}\gamma}}{dE_\gamma \, d\Omega} = \frac 1{E_\gamma} \int_{\rm l.o.s.} ds\, \frac{j(E_\gamma,r(s,\theta))}{4\pi}
\eeq
In general, for any radiative process, the emissivity is obtained as a convolution of the spatial density of the emitting medium with the power that it radiates (see e.g.~\cite{rybicki}). In this case therefore 
\beq
\label{Rademissivity}
j(E_\gamma,r)=2\int_{m_e}^{M_{\rm DM}(/2)}dE_e\ \mathcal{P}_{\rm IC}(E_\gamma,E_e,r)\ \frac{dn_{e^\pm}}{dE_e}(r,E_e),
\eeq
where $\mathcal P=\sum_i \mathcal P_{\rm IC}^i$ is the differential power emitted into photons due to ICS radiative processes (the sum runs over the different components of the photon bath: CMB, dust-rescattered light and starlight) and $dn_{e^\pm}/dE_e$ is the electron (or positron) number density after diffusion and energy losses, as computed in subsection \ref{positronpropagationresult} (notice that there it was denoted as $f$ for simplicity, see page~\pageref{eq:diffeq}; $dn_{e^\pm}/dE_e$ just corresponds to eq.~(\ref{eq:positronsflux}) removing the $v_{e^\pm}/4\pi$ factor). The minimal and maximal energies of the electrons are determined by the electron mass $m_e$ and the mass of the DM particle $M_{\rm DM}$. The `$/2$' notation applies to the decay case. The overall factor of 2 takes into account the fact that, beside the electrons, an equal population of positrons is produced by DM annihilations/decays and radiates.\footnote{Recall from footnote \ref{footnotee+e-} that with the notation $e^\pm$ we always refer to the independent fluxes of electrons $e^-$ {\it or} positrons $e^+$ and not to the sum.}  

\medskip

%\xxx{AS} \xxx{MA HAI CAMBIATO TUTTE LE FORMULE NON CI CAPISCO PIU'NULLA, ORA SONO PIU'COMPLICATE E FORSE $I_{IC}$ NON E'PIU'ADIMENSIONALE.............}

The radiated power $\mathcal P_{\rm IC}$, in the full Klein-Kishina case, is given by (we refer the reader to~\cite{CP,Meade} and references therein for more details on the derivation)
\begin{equation}
\begin{split}
&\mathcal{P}_{\rm IC}^i(E_\gamma,E_e,\vec x)  = \\
&\frac{3 \sigma_{\rm T}}{4\gamma^2}  \hspace{-0.2cm} \int_{1/4\gamma^2}^1\hspace{-0.65cm} dq  \left(E_\gamma-\frac{E_\gamma}{4q\gamma^2 (1-\epsilon)} \right) \frac{n_i\big(E_\gamma^0(q),\vec x \big)}{q} \left[ 2q\ln q+q+1-2q^2+\frac{1}{2}\frac{\epsilon^2}{1-\epsilon}(1-q) \right].
\end{split}
\label{eq:power}
\end{equation}
where $\gamma = E_e/m_e$ is the Lorentz factor of the scattering electron and the integrand is expressed in terms of
\beq
q=\frac{\epsilon}{\Gamma_E(1-\epsilon)}, \qquad {\rm with}\  \Gamma_E=\frac{4E_\gamma^0E_e}{m_e^2},\quad \epsilon=\frac{E_\gamma}{E_e},  \quad {\rm in} \ \frac{1}{4\gamma^2}\simeq 0 \le q \le 1. 
\eeq
Here $E_\gamma^0$ is the initial energy of the photon in the background bath. 
Correspondingly, $E_{\gamma}$ lies in the range $E_\gamma^0/E_e \le E_\gamma \le E_e\, \Gamma_E/(1+\Gamma_E)$.
The non-relativistic (Thompson) limit corresponds to $\Gamma_E\ll1$, so that $\epsilon \ll 1$, the last term in the integrand of $\mathcal P$ is negligible, and $q \to y = E_\gamma/(4\gamma^2 E_\gamma^0)$ with $0\le y \le 1$.
Thus in the Thomson limit
\beq
\mathcal{P}_{\rm IC}^i(E_\gamma,E_e,\vec x) = \frac{3 \sigma_{\rm T}}{4\gamma^2} E_\gamma  \int_{0}^1 \hspace{-0.2cm} dy   \frac{n_i\big(E_\gamma^0(y),\vec x \big)}{y} \left[ 2y\ln y+y+1-2y^2 \right]\qquad {\rm [Thomson\ limit]}.
\label{eq:powerThomson}
\eeq

Plugging now $\mathcal{P}_{\rm IC}$ and $n_{e^\pm}$ in eq.~(\ref{Rademissivity}), we can write the IC differential flux in the following convenient form: 
\beq
\frac{d\Phi_{{\rm IC}\gamma}}{dE_\gamma\, d\Omega} = \frac1{E_\gamma^2}\frac{r_\odot}{4\pi} 
\left\{\begin{array}{l}
\displaystyle \frac12 \left(\frac{\rho_\odot}{M_{\rm DM}}\right)^2 \int_{m_e}^{M_{\rm DM}} \hspace{-0.45cm} dE_{\rm s}  \sum_f \langle \sigma v \rangle_f \frac{dN^f_{e^\pm}}{dE}(E_{\rm s})  \, I_{\rm IC}(E_\gamma,E_{\rm s},b,\ell) \quad {\rm (annihilation)} \\[5mm]
\displaystyle \frac{\rho_\odot}{M_{\rm DM}} \int_{m_e}^{M_{\rm DM}/2}\hspace{-0.45cm} dE_{\rm s}  \sum_f \Gamma_f  \frac{dN^f_{e^\pm}}{dE}(E_{\rm s})  \, I_{\rm IC}(E_\gamma,E_{\rm s},b,\ell) \qquad \,\,\,\,\,{\rm (decay)}
\end{array}\right.
\label{eq:summaryIC}
\eeq
%\beq
%\frac{d\Phi_{{\rm IC}\gamma}}{dE_\gamma\, d\Omega} = \frac1{E_\gamma^2}\frac{r_\odot}{4\pi} 
%\left\{\begin{array}{l}
%\displaystyle \frac12 \left(\frac{\rho_\odot}{M_{\rm DM}}\right)^2 \int_{m_e}^{M_{\rm DM}} \hspace{-0.45cm} dE_{\rm s}  \sum_f \langle \sigma v \rangle_f \frac{dN^f_{e^\pm}}{dE}(E_{\rm s})  \left[ E_\gamma \, I_{\rm IC}(E_\gamma,E_{\rm s},b,\ell)\right] \quad {\rm (annihilation)} \\[5mm]
%\displaystyle \frac{\rho_\odot}{M_{\rm DM}} \int_{m_e}^{M_{\rm DM}/2}\hspace{-0.45cm} dE_{\rm s}  \sum_k \Gamma_k  \frac{dN^f_{e^\pm}}{dE}(E_{\rm s})  \left[ E_\gamma \, I_{\rm IC}(E_\gamma,E_{\rm s},b,\ell)\right] \qquad \,\,\,\,\,{\rm (decay)}
%\end{array}\right.
%\label{eq:summaryIC}
%\eeq
where $E_{\rm s}$ is the $e^\pm$ injection energy and $I_{\rm IC}(E_\gamma,E_{\rm s},b,\ell)$ (with the dimension of an energy) is a halo function for the IC radiative process. 
This formalism allows therefore to express the flux of ICS $\gamma$ as the convolution of the electron injection spectrum $dN_{e^\pm}/dE$ and this new kind of halo functions, in close analogy with the formalism for charged particles.
Indeed, we can explicitly express $I_{\rm IC}$ in terms of the ICS ingredients discussed above and the generalized halo functions for $e^\pm$ that we introduced in Sec.~\ref{positronpropagationresult}. We get 
\beq\label{HaloICS}
I_{\rm IC}(E_\gamma,E_{\rm s},b,\ell)= 2\, E_\gamma \int_{\rm l.o.s.} \frac{ds}{r_\odot} \left( \frac{\rho(r(s,\theta))}{\rho_\odot} \right)^\eta \hspace{-0.2cm} \int_{m_e}^{E_{\rm s}}\hspace{-0.3cm}dE\,  \frac{{\sum_i \mathcal P_{\rm IC}^i(E_\gamma,E,r(s,\theta))}}{b(E,r(s,\theta))}\,  I(E,E_{\rm s},r(s,\theta)),
\eeq
where again $\eta=1,2$  for the decay or annihilation scenarios respectively. 
The intensity of the interstellar radiation $\sum_i n_i$ cancels out in the ratio  $\sum \mathcal{P}_i/b$, up to the sub-leading synchrotron contribution and provided that we are not interested in the contributions from the individual light baths. 

\medskip

\begin{figure}[!p]
\begin{center}
\includegraphics[width=1 \textwidth]{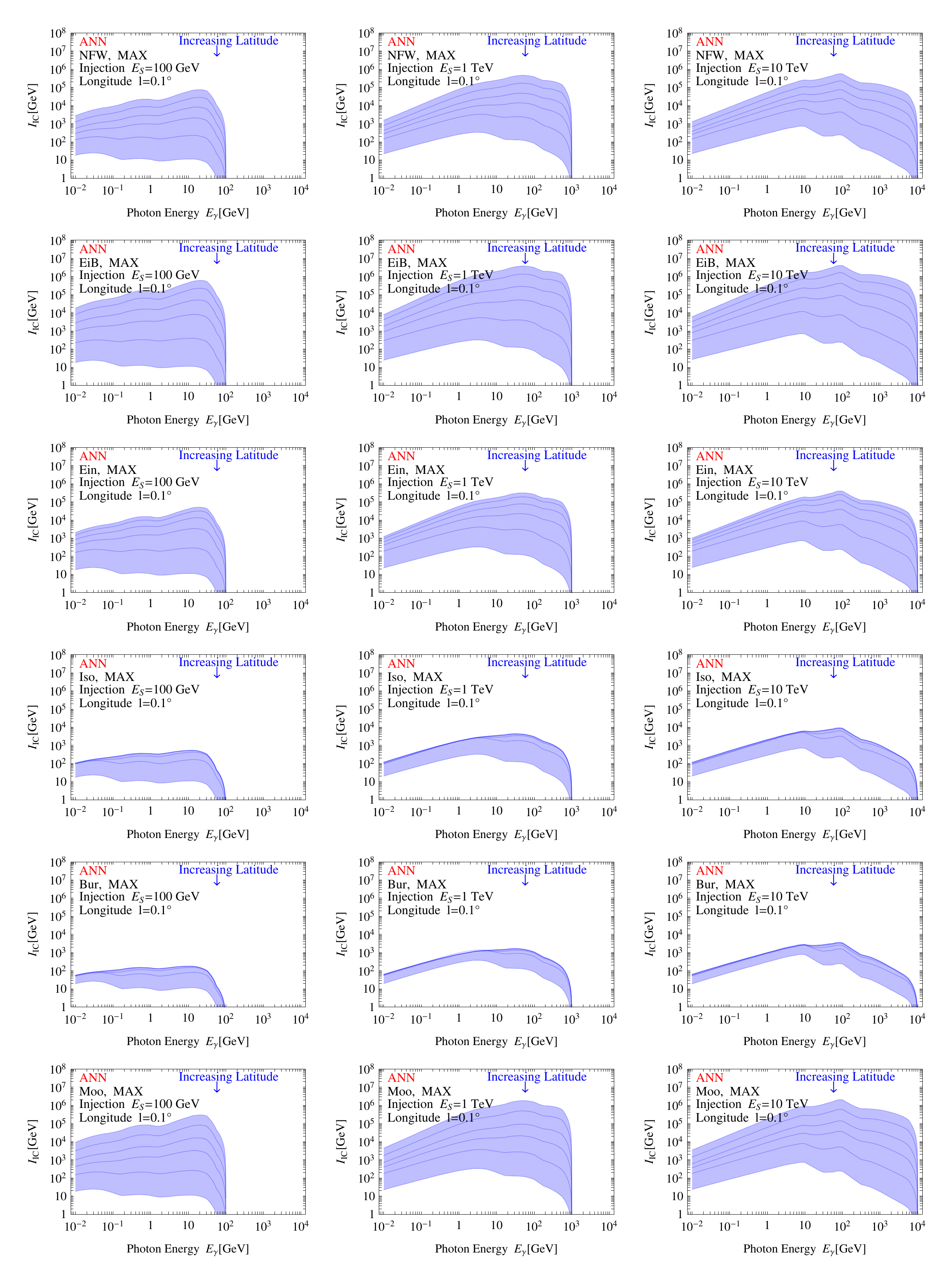}
\caption{\em \small \label{fig:IICPlot} {\bfseries Halo functions for Inverse Compton Scattering}, for the Dark Matter annihilation case and varying DM halo profiles and electron injection energies.}
\end{center}
\end{figure}

\begin{figure}[!p]
\begin{center}
\includegraphics[width=1 \textwidth]{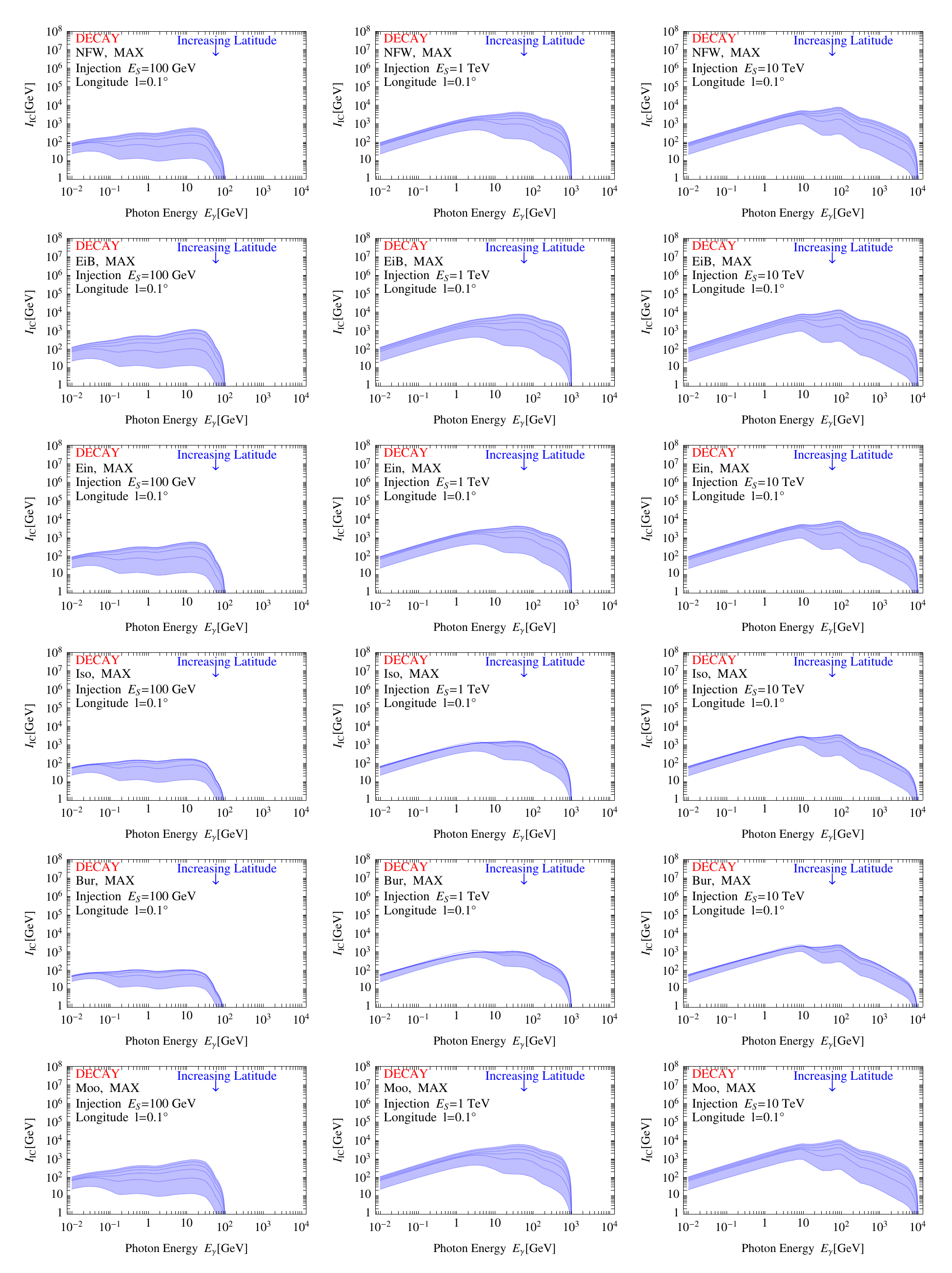}
\caption{\em \small \label{fig:IICPlotD} {\bfseries Halo functions for Inverse Compton Scattering}, for the Dark Matter decay case.}
\end{center}
\end{figure}

The class of functions $I_{\rm IC}(E_\gamma,E_{\rm s},b,\ell)$ are plotted in fig.~\ref{fig:IICPlot}  for the annihilation case (and in fig.~\ref{fig:IICPlotD}  for the decay case) by varying the $e^\pm$ injection energy and the latitude $b$ for a fixed longitude $\ell=0.1^\circ$. As one can see, increasing the latitude (from $0.1^\circ$ to $90^\circ$) the spectrum becomes less step due to two combined effects: the decreasing abundance of star-light and dust, that provide the more energetic part of the target photon bath, and the always present Klein-Nishina relativistic effect, that blunts the high energy part of the spectrum especially for large injection energy and in regions close to the GC where more star-light and dust are present. Furthermore, at small latitudes, the difference in the normalization among different profiles becomes appreciable. 

All these classes of functions for decay and annihilation scenarios  are provided  in {\sc Mathematica}$^{\tiny{\textregistered}}$ interpolated functions form at the \myurl{www.marcocirelli.net/PPPC4DMID.html}{website}~\cite{website}.
There is one halo function for each DM density profile and $e^\pm$ propagation model (MIN, MED, MAX); each function depends on 4 variables:
the energy $E_e^0$ of $e^\pm$ as produced by DM before diffusion; the final photon energy $E_{\gamma'}$; the two galactic coordinates $b,\ell$ that specify the observed direction.

\begin{figure}[!t]
\includegraphics[width=0.48 \textwidth]{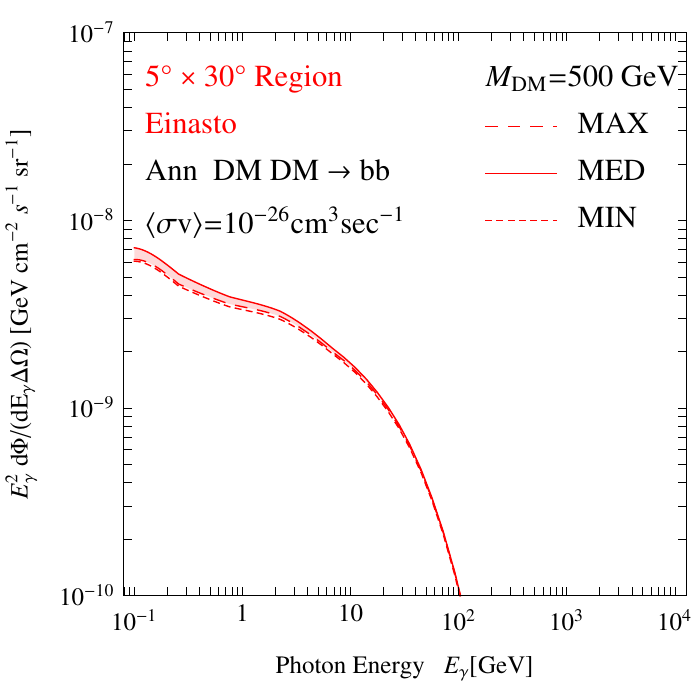} \quad \includegraphics[width=0.48 \textwidth]{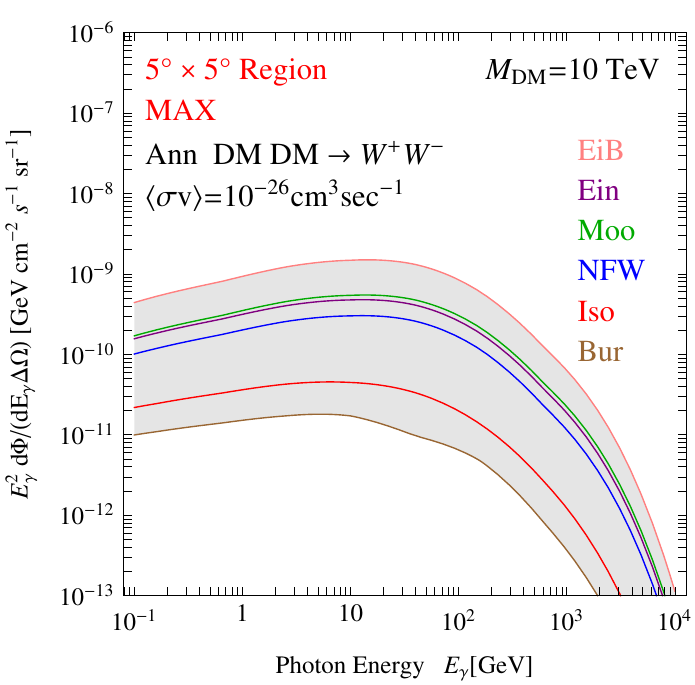}
\includegraphics[width=0.48 \textwidth]{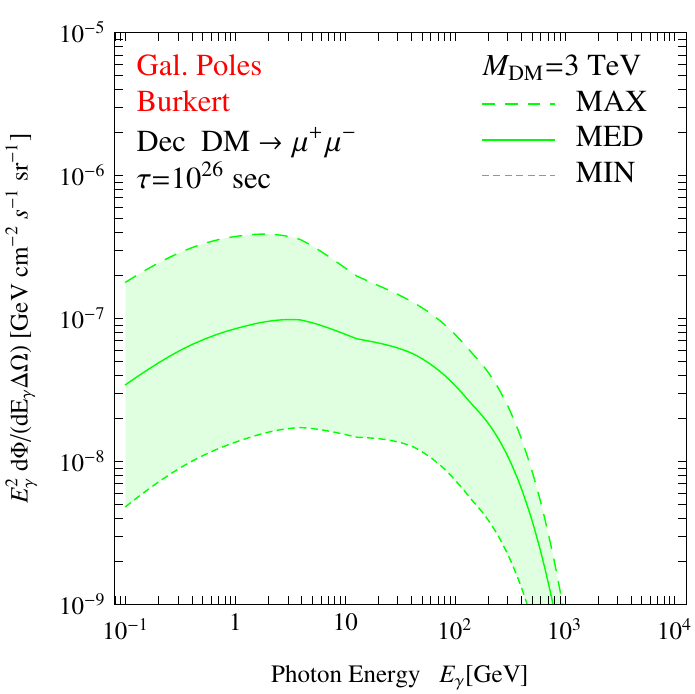} \quad \includegraphics[width=0.48 \textwidth]{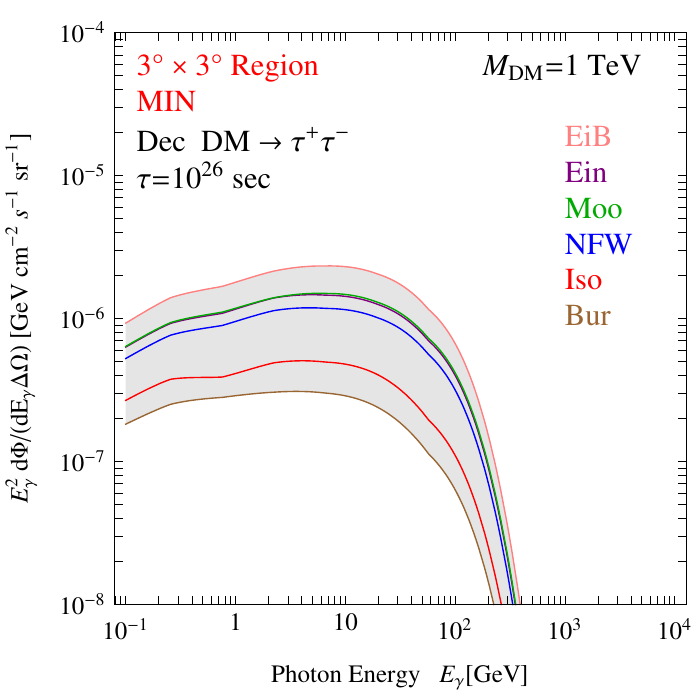}
\caption{\em \small \label{fig:ICgammas} {\bfseries Fluxes of galactic Inverse Compton gamma rays}, for the case of annihilations (upper panels) and decay (lower panels).}
\end{figure}

\medskip

The final step to obtain the differential ICS $\gamma$ flux $d\Phi_{\rm IC \gamma}/dE_\gamma d\Omega$ consists in performing the convolution integral over $E_{\rm s}$ with any desired prompt $e^\pm$ energy spectrum from DM.  We do not provide pre-compiled ICS $\gamma$ fluxes (due to file-size limitations), but we provide the code that performs such final integral on the~\myurl{www.marcocirelli.net/PPPC4DMID.html}{website}~\cite{website}.

\medskip

Finally one can compute the differential flux from a region $\Delta \Omega$ by integrating over $b$ and $\ell$ as discussed in Sec.~\ref{promptgamma} 
\beq
\frac{d\Phi_{\rm IC \gamma}}{dE_\gamma} = \iint db\, d\ell\, \cos b\, \frac{d\Phi_{\rm IC \gamma}}{d\Omega\, dE_\gamma}.
\label{eq:summaryICI}
\eeq

Due to the intertwined dependence on $b$ and $\ell$ and on $E_\gamma$ and $E_s$ of the halo functions, here the geometrical integral cannot be pulled out as for prompt $\gamma$ rays, so a $\bar J$ factor cannot be defined in a simple way.

\bigskip

Figure~\ref{fig:ICgammas} presents some examples of IC $\gamma$ ray fluxes computed as discussed above. In the top left panel (for the case of annihilations into $b \bar b$) we plot the integrated flux in a region which is large but quite close to the GC, and we vary the propagation parameters of the electrons and positrons: the impact on the IC $\gamma$ fluxes is small but visible at small energies, and it is due to the difference in the resulting populations of $e^\pm$. In the top right panel we vary the DM profiles in the `5 $\times$ 5' region: since the profiles differ sensibly in such a relatively small region, the impact on the fluxes is of the order of more than an order of magnitude. The lower panels refer to decay cases and focus on leptonic channels. When looking for the signal in the region of the Galactic Poles, the variation of the propagation parameters (left plot) has a significative impact mainly because the height of the diffusion cylinder (and therefore the contribution to the integrated emission) increases as one moves from MIN to MAX. When looking for the signal from a small `3 $\times$ 3 region around the GC (right panel), the variation of the choice of profile has a non-negligible impact (even in the decay case) because the observation is concentrated towards where profiles differ most.

\subsection{Synchroton radiation from $e^\pm$}
\label{sync}

In this section we briefly discuss the synchrotron emission from DM produced energetic $e^\pm$ immersed in strong magnetic fields (see e.g.~\cite{BCST} and references therein). We do not provide, however, numerical results for this signature. 

\bigskip

An electron or positron with momentum $p$ in  a turbulent magnetic field $B$ 
generates the following spectrum of  synchrotron  power $W_{\rm syn}$ into photons with frequency $\nu$:
\beq\label{eq:syn}
\frac{dW_{\rm syn}}{d\nu}=\frac{\sqrt{3}}{6\pi}\frac{e^3B}{m_e} F(\frac{\nu}{\nu_{\rm syn}}),\qquad
F(x)=x \int_x^\infty K_{5/3}(\xi)d\xi  \approx \frac{8\pi}{9\sqrt{3}} \delta(x-1/3)\eeq
so that, to a good approximation, it generates photons with frequency $\nu_{\rm syn}/3$, where
\beq\label{eq:nusyn}
\nu_{\rm syn} = 
\frac{3eB p^2}{4\pi m_e^3}=
4.2\,{\rm MHz} \frac{B}{\rm G} \left(\frac{p}{m_e}\right)^2\eeq
such that the total energy radiated is proportional to $B^2$.
Galactic magnetic fields are highly uncertain; we adopted the simplest `conventional' choice of eq.\eq{Bgal}.
While we included in eq.\eq{b(E)} the magnetic field contribution to $e^\pm$ energy losses, it is sub-dominant everywhere with respect to the Inverse Compton contribution.
Consequently, as long as one considers any other magnetic field profile which is similarly sub-dominant, it is not necessary to 
recompute in each case the halo functions describing the diffusion of $e^\pm$.
(The diffusion coefficient, which also depends on the turbulence spectrum magnetic field, is in practice 
extracted by fitting models of cosmic rays as discussed above).

Consequently a galactic distribution $dn_{e^\pm}/dE_e$ generates
\beq  \frac{dW_{\rm syn}}{d\nu\, d\Omega} =\frac{2}{4\pi}\int_{\rm l.o.s.}ds \int dE_e  \frac{dn_{e^\pm}}{dE_e}\frac{dW_{\rm syn}}{d\nu}
 \label{synflux} \eeq
 Inserting $dn_{e^\pm}/dE_e$ from\eq{positronsflux} and $dW_{\rm syn}/d\nu$ 
 from eq.s\eq{syn} and  using the $\delta$-function approximation such that
 $E_e\approx p=\sqrt{4\pi m_e^3\nu/eB}$ $=0.43\GeV (\nu/{\rm GHz})^{1/2}(B/{\rm mG})^{-1/2}$
we get
 \beq 
 \label{synspectrumfinal}
 \begin{split}
& \nu \frac{dW_{\rm syn}}{d\nu\, d\Omega}= \\
& \frac{1}{4 \pi }  
\int_{\rm l.o.s.} ds \, p\, \mathcal{F}_{\rm syn} \, \left\{
\begin{array}{cl}
\displaystyle \frac{1}{2}  \left(\frac{\rho (\vec x)}{M_{\rm DM}}\right)^2 \sum_f \langle \sigma v\rangle_f  \int_p^{M_{\rm DM}} dE_{\rm s} \frac{dN_{e^\pm}}{dE_{\rm s}} I( E_e, E_{\rm s},\vec{x}) & \text{(annihilation)}\\[4mm]
\displaystyle \frac{ \rho(\vec x)}{M_{\rm DM}} \sum_f \Gamma_f \int_p^{M_{\rm DM}/2} dE_{\rm s} \frac{dN_{e^\pm}}{dE_{\rm s}} I( E_e, E_{\rm s},\vec{x}) & \text{(decay)}
\end{array}
\right.
\end{split}
\eeq
where $e$ is the electron charge and $\mathcal{F}_{\rm syn}=u_B/\tilde{u}$ is the fraction of the $e^\pm$ energy lost into synchrotron,
with an uncertain spatial profile ($\tilde{u}$ is defined in eq.\eq{b(E)}).

%A lower $B$ leads to a higher synchrotron flux at the low frequency we consider.

\section{Extragalactic gamma rays}
\label{extragalactic}

The $\gamma$-rays emitted by DM annihilations or decays in all the extragalactic structures and (in principle) all along the history of the Universe reach us in the form of an isotropic contribution to the total $\gamma$-ray intensity. In this section we discuss how to compute it, in terms of many of the ingredients introduced above. With respect to the galactic case, however, we have to include the effect of  the `cosmological dimming' due to the expansion of the Universe and the fact that, unlike in the galactic environment, on cosmologically large distances one can not neglect the absorption of gamma-rays. We mainly follow the formalism presented in~\cite{Huetsi:2009ex,Hutsi:2010ai} (see also~\cite{CPS}).

%\subsection{Extragalactic radiation transfer of gamma rays}

\medskip

In full generality, the differential flux of extragalactic gamma rays perceived at a certain redshift $z$ is given by 
\begin{equation}
\label{eq:EGflux}
\frac{d\Phi_{{\rm EG}\gamma}}{dE_\gamma}(E_\gamma,z)  
= c \frac{1}{E_\gamma}  \int\limits_z^{\infty}dz'\frac{1}{H(z')(1+z')}\left(\frac{1+z}{1+z'}\right)^{3} j_{{\rm EG}\gamma}(E'_\gamma,z')\ e^{-\tau(E_\gamma,z,z')}.
\end{equation}
The factor $[H(z')(1+z')]^{-1}$, where $H(z) \equiv H_0 \sqrt{\Omega_m(1 + z)^3 + (1-\Omega_m)}$ is the Hubble function, converts the redshift interval to a proper distance interval (in other words, the integral can be thought to be the integration over time of all the photons emitted in the past, then converted into redshift).The factor $[(1+z)/(1+z')]^3$ accounts for the cosmological dimming of intensities due to the dilution of the source.
$E'$ is the energy of a photon at redshift $z'$, assuming it has energy $E$ at redshift $z$: $E' \equiv E (1+z')/(1+z)$.
The function $\tau(E_\gamma,z,z')$ is the optical depth describing the absorption between the redshifts $z$ and $z'$. To calculate the Hubble function we assume a flat $\Lambda$CDM ``concordance'' cosmology with $\Omega_m=0.27$,  $\Omega_\Lambda=0.73$ and $h=0.7$. 

Because we observe the gamma rays at $z=0$, the formula for the flux simply reduces to 
\begin{equation}
\label{eq:EGfluxtoday}
\frac{d\Phi_{{\rm EG}\gamma}}{dE_\gamma}(E_\gamma)  =c \frac{1}{E_\gamma} \int\limits_0^{\infty}dz'\frac{1}{H(z')(1+z')^{4}} \, j_{{\rm EG}\gamma}(E'_\gamma,z') \ e^{-\tau(E_\gamma,z')}.
\end{equation}

%\subsection{Extragalactic emission coefficients}
\medskip

As in the Galaxy, the local gamma-ray emissivity for the annihilating/decaying DM models is the sum of the prompt and IC radiation parts
\begin{equation}
j_{{\rm EG}\gamma}(E'_\gamma,z') = j_{{\rm EG}\gamma}^{{\rm prompt}}(E'_\gamma,z') + j_{{\rm EG}\gamma}^{{\rm IC}}(E'_\gamma,z').
\end{equation}

The emissivity $j_{{\rm EG}\gamma}^{{\rm prompt}}$ is given by
\begin{equation}
\label{jEGprompt}
j_{{\rm EG}\gamma}^{{\rm prompt}}(E'_\gamma,z') = E'_\gamma \begin{cases}
\displaystyle \frac{1}{2} B(z') \left(\frac{\bar{\rho}(z')}{M_{\rm DM}}\right)^2 \sum_f \langle \sigma v \rangle_f  \frac{dN^f_\gamma}{dE_\gamma}(E'_\gamma)& \text{(annihilation)}\\[4mm]
\displaystyle  \frac{\bar{\rho}(z')}{M_{\rm DM}}  \sum_f  \Gamma_f \frac{dN^f_\gamma}{dE_\gamma}(E'_\gamma)& \text{(decay)}
\end{cases}
\end{equation}
where the average cosmological DM density is $\bar{\rho}(z) = \bar{\rho}_0(1+z)^3$ and $\bar{\rho}_0 \simeq 1.15\, 10^{-6} \,{\rm GeV} \, {\rm cm}^{-3}$ is its value today ($z=0$); $dN_{\gamma}/dE_\gamma$ is the spectrum of the prompt photons as computed in section~\ref{primary}. In the case of annihilations, DM clustering enhances the rate: the factor $B(z)$ effectively takes that into account  (as described in detail below). 

\medskip

The computation of the emission coefficient for Inverse Compton radiation $j_{{\rm EG}\gamma}^{{\rm IC}}$ is more involved: IC radiation depends on the population of the $e^\pm$ produced by annihilation/decay of DM at redshift $z'$ and the background photon bath $n(E',z')$ at the same redshift. Actually, more precisely, the population of the $e^\pm$ at a certain redshift $z'$ is the result of the diffusion-and-loss processes (in space and `time') from preceding redshifts. However, the mean free path of $e^\pm$ in the intergalactic medium is very short compared to cosmological length scales (dominantly due to the presence of the CMB). Thus we can approximate the IC spectrum as generated ``on-spot'' and neglect the effect of such diffusion. (Technically, remember that this corresponds to setting the equivalent of the `generalized halo function' defined in Sec.~\ref{positronpropagationresult} to 1.)
In turn, concerning the background photons, it is a good approximation to assume that they only consist in the CMB: one can safely neglect the IR, light and UV photons from starlight and secondary dust radiation that were instead important in the Galactic medium, where ICS on CMB, IR and light photons give comparable contributions.

\medskip

%\subsubsection{Extragalactic Inverse Compton radiation}
Under these approximations the IC emissivity can be expressed as
\begin{equation}
\label{jEGICS}
\begin{split}
&j_{{\rm EG}\gamma}^{{\rm IC}}(E'_\gamma,z') = \\
& 2 \int\limits_{m_e}^{M_{\rm DM}(/2)}{\rm d}E_e\frac{\mathcal{P_{\rm IC}^{\rm CMB}}(E'_\gamma,E_e,z')}{b_{\rm IC}^{\rm CMB}(E_e,z')}\int\limits_{E_e}^{M_{\rm DM}(/2)}{\rm d}\widetilde{E}_e\frac{{\rm d}N_e}{{\rm d}\widetilde{E}_e} \left\{
\begin{array}{cl}
\displaystyle \frac{1}{2} B(z') \left(\frac{\bar{\rho}(z')}{M_{\rm DM}}\right)^2 \sum_f \langle \sigma v\rangle_f & \text{(annihilation)}\\[4mm]
\displaystyle \frac{\bar{\rho}(z')}{M_{\rm DM}} \sum_f \Gamma_f & \text{(decay)}
\end{array}
\right.
\end{split}
\end{equation}
where the overall factor of 2 takes into account that equal populations of electrons and positrons radiate (the `$/2$' notation applies to decay).
The functions $\mathcal{P}_{\rm IC}^{\rm CMB}(E'_\gamma,E_e,z')$ and $b_{\rm IC}^{\rm CMB}(E_e,z')$ are the radiated power and the energy loss coefficient function for $e^\pm$, exactly analogous to those defined in Sec.~\ref{ICS} (eq.~(\ref{eq:power})) and Sec.~\ref{positronpropagation} (eq.~(\ref{bIC})), the only difference being that they are now functions of the redshift and not of the position in the Galaxy.
They can be computed in the full Klein-Nishina case or in Thomson limit. 
As the intergalactic medium is dominated by low energy CMB photons, the Klein-Nishina formalism is needed only for the extreme mass region of DM, above $M_{\rm DM}>20$ TeV (see the comparison between the black and colored dotted lines in the last panel in Fig.~\ref{fig:extragammas}) (the situation is therefore different from the case of the Galaxy, where IR and stellar light gives a relevant contribution to the target photon bath, and thus the Klein-Nishina formalism should be used already at energies above $M_{\rm DM} \sim 100$ GeV).

%\subsubsection{Thomson approximation for the calculation of IC radiation}

%\medskip

%The Thomson approximation behaves in a good manner up to the mass of DM particle $20-30$ TeV (see Fig.~\ref{fig:EXgammas}). It differs from the Galactic case where there are plenty of eV-range photons around in the Galaxy. As mentioned the intergalactic medium is dominated by soft CMB photons~\footnote{Due to the strong cosmological dimming the low redshift gamma-ray intensity is mostly originating from a rather local neighborhood only, and thus one does not have to worry about CMB photons becoming energetic once $z$ increases enough.}.

The advantage of using the Thomson approximation, when valid, %presented by equations (\ref{eq:ExGThomson}) and (\ref{eq:ExGThomson2}) 
is that one needs to calculate the IC emissivity only for one fixed redshift, due to the fact that $n(E,z) \equiv n_{\rm CMB}(E,z) = n_{\rm Pl}[E,T_0(1+z)]$, where $n_{\rm Pl}$ is the black body spectrum and $T_0$ is the temperature of CMB at $z=0$. 
For example, one can calculate the IC spectrum at $z=0$ and then for other redshifts use simple scaling properties of $n_{\rm Pl}$. 
Heuristically, this can just be understood as follows: the IC cross section in the Thomson regime is energy-independent; since the energy spectrum of the injected $e^\pm$ is the same at any $z$, the IC photon at the production epoch $z$ will have an energy proportional to the one of the upscattered background photon. The latter is $(1+z)$ times the current one, which is exactly the factor compensated by the subsequent redshifting. So, the IC spectrum is universal and equal to the one calculated at $z=0$.
In formulae, this means that the ICS emissivity is just given by $j^{\rm IC}_{\rm EG\gamma}(E'_\gamma,z')=B(z')(1+z')^6\, j^{{\rm IC}}_{\rm EG\gamma}\left(E_\gamma,0\right)$ for the case of annihilating DM and by $j^{\rm IC}_{\rm EG\gamma}(E'_\gamma,z')=(1+z')^3\, j^{{\rm IC}}_{\rm EG\gamma}\left(E_\gamma,0\right)$ for the case of decaying DM.

%In this regime, the total emission coefficient for the case of annihilating DM is given by
%\begin{equation}\label{eq_3.5}
%j^{\rm annih}_{\rm EG\gamma}(E'_\gamma,z')=B(z')(1+z')^6\left[ {\color{red} (1+z')}\ \bar{\jmath}^{{\rm prompt}}({\color{red} E_\gamma \ \cancel{E'_\gamma}},0) + {\color{red} \cancel{\frac{1}{1+z'} \ }} \bar{\jmath}^{{\rm IC}}\left({\color{red} E_\gamma},0\right)  \right]\,,
%\end{equation}
%where $\bar{\jmath}^{{\rm prompt}}(\cdot,0)$ and $\bar{\jmath}^{{\rm IC}}(\cdot,0)$  are emissivities at $z=0$ assuming an average DM density $\bar{\rho}$. Due to the structure formation the typical value for $\rho^2$ is not simply $\bar{\rho}^2$ but is boosted by a large factor $B(z)$, i.e. $\langle \rho^2 \rangle=B(z)\, \bar{\rho}^2$, as described below.
%For decaying DM, as the energy input is simply proportional to the DM density, we should replace $(1+z')^6$ with $(1+z')^3$ in Eq. (\ref{eq_3.5}) and also take $B(z')\equiv 1$, i.e.          
%\begin{equation}
%j^{\rm decay}_{\rm EG\gamma}(E'_\gamma,z')=(1+z')^3\left[ {\color{red} (1+z')} \, \bar{\jmath}^{{\rm prompt}}({\color{red} E_\gamma \ \cancel{E'_\gamma}},0) + {\color{red} \cancel{\frac{1}{1+z'}\ }}\bar{\jmath}^{{\rm IC}}\left({\color{red} E_\gamma} ,0\right)\right]\,.
%\end{equation}
%
When the Thomson approximation is invalid and one has to apply the full Klein-Nishina formalism for the calculation of IC radiation, the emission coefficient of IC radiation $\bar{\jmath}^{{\rm IC}}$ has to be calculated for all the relevant region of redshift.

\subsection{Effect of DM clustering}
\label{subsec:EGclusering}

The emission coefficients for annihilating DM described above contain a {\em cosmological boost factor} $B(z)$ which multiplies the contribution from the homogeneous distribution of DM $\bar{\rho}_{DM}(z)$. Technically, the true square density that enters the formul\ae{} is thus $\langle\rho_{\rm DM}^2(z)\rangle \equiv \bar{\rho}_{DM}^2(z)\langle(1+\delta(z))^2\rangle \equiv B(z)\bar{\rho}_{DM}^2(z)$. Here $\delta \equiv \rho/\bar{\rho}-1$ is the density contrast and the cosmological boost factor $B(z) \equiv \langle(1+\delta(z))^2\rangle = 1+\langle\delta^2(z)\rangle$. 
%To include the effect of DM clustering for the annihilation $\rho_{DM}^2(z)$ should be replaced by $\langle\rho_{\rm DM}^2(z)\rangle \equiv \bar{\rho}_{DM}^2(z)\langle(1+\delta(z))^2\rangle \equiv B(z)\bar{\rho}_{DM}^2(z)$. Here $\delta \equiv \rho/\bar{\rho}-1$ is the density contrast and we have also defined a halo boost factor $B(z) \equiv \langle(1+\delta(z))^2\rangle = 1+\langle\delta^2(z)\rangle$. 
To calculate the boost factor we adopt a halo model which approximates the matter distribution in the Universe as a superposition of DM halos (e.g.~\cite{2002PhR...372....1C}). Within this model $B(z)$ can be given as
\begin{equation}
B(z,M_{\min}) = 1 + \frac{\Delta_c}{3\bar{\rho}_{m,0}} \int_{M_{\min}}^{\infty} dM\, M \frac{dn}{dM}(M,z) \, f\left[c(M,z)\right]
\end{equation} 
%\begin{equation}
%B(z) =                  1+  \frac{\Delta_c}{3\bar{\rho}_{m,0}}\int\limits_{M_{\rm min}}^{\infty}d M\frac{dn}{dM}(M,z)f\left[c(M,z)\right],
%\end{equation} 
where $\bar{\rho}_{m,0}$ is the matter density at $z=0$, $\Delta_c\simeq 200$ is the overdensity at which the halos are defined and $M_{\rm min}$ is the minimum halo mass. $\frac{dn}{dM}(M,z)$ is the halo mass function, that can be cast in the universal form~\cite{PressSchechter} 
\begin{equation}
\frac{dn}{dM}(M,z) = \frac{\bar{\rho}_{m,0} }{M^2}\nu f(\nu) \frac{d \log \nu}{d \log M}
\end{equation}
where the parameter $\nu=\left[\delta_{\rm c}(z)/\sigma(M)\right]^2$ is defined as the ratio between the critical overdensity $\delta_{\rm c}(z)$ and the quantity $\sigma(M)$ which is the variance of the linear density field in spheres containing a mean mass $M$. For the multiplicity function $f(\nu)$ we use the form in \cite{1999MNRAS.308..119S}

\beq\label{Mulfunct}
\nu f(\nu)=A\left(1+\frac1{\nu'^{p}}\right)\left(\frac{\nu'}{2\pi}\right)^\frac12 \, e^{-\nu'/2}
\eeq
where $\nu'=a\,\nu$ with $a=0.707$, $p=0.3$ and $A$ is determined by requiring that the integral $\int d\nu f(\nu)\equiv1$. 
Eq.~(\ref{Mulfunct}) reduces to  the original Press-Schechter formula~\cite{PressSchechter} taking $a=1$, $p=0$ and $A=1/2$. 

\begin{figure}[t]
\begin{center}
\includegraphics[width=0.49 \textwidth]{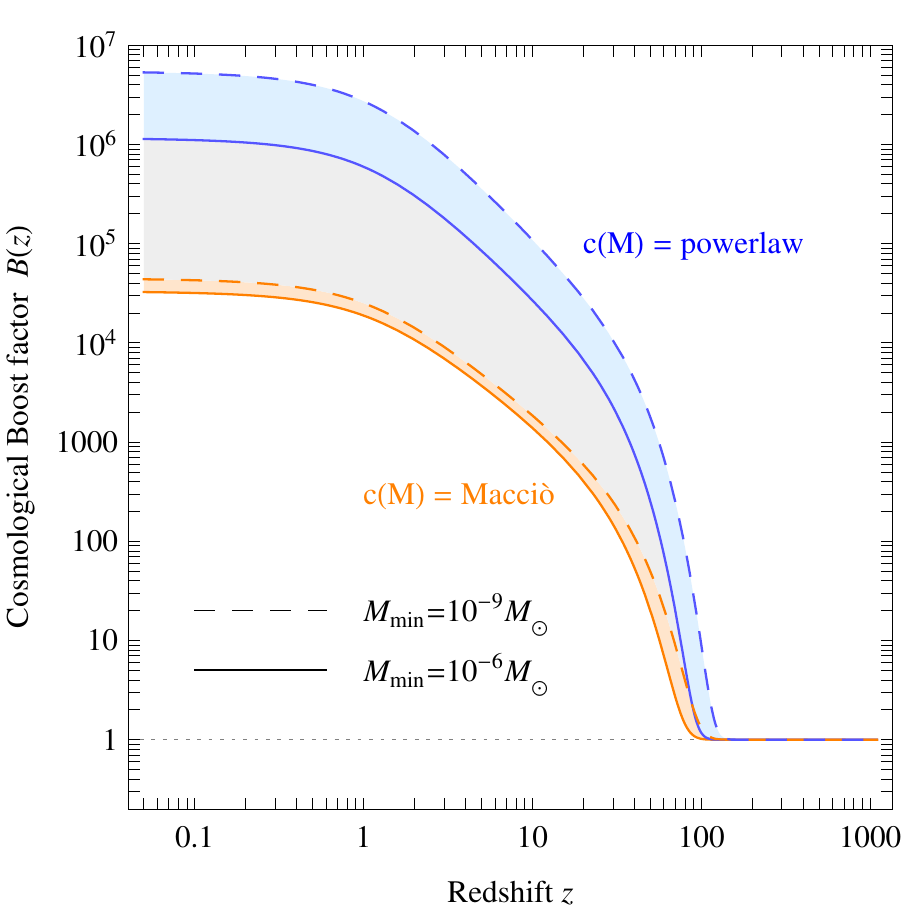}
\includegraphics[width=0.49 \textwidth]{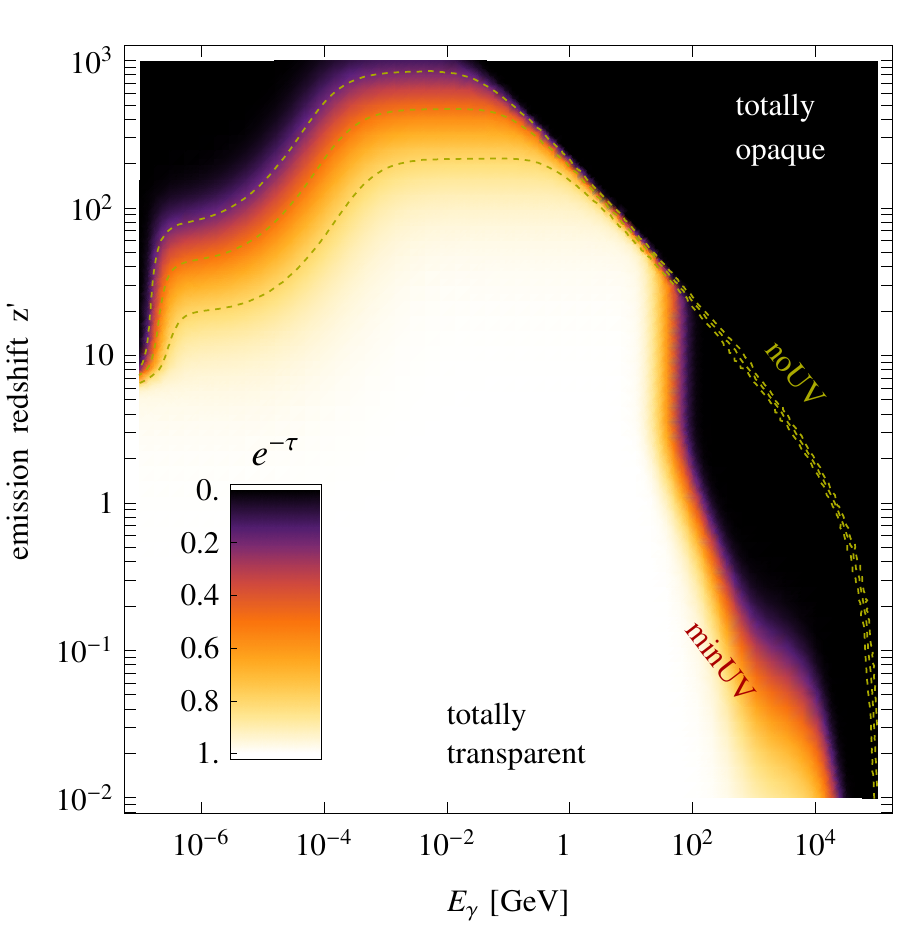}
\caption{\em \small \label{fig:EGamma} Left: {\bfseries Extragalactic (cosmological) boost factors} $B(z)$ for various concentration models and for various lower cut-offs of DM halo mass. 
Right: Contour plot of the {\bfseries opacity of the Universe}, for the `min UV' background case (color plot). The short-dashed curves correspond to `no UV' background case.}
\end{center}
\end{figure}

 $c(M,z)$ represents the halo concentration parameter function and the function $f(c)$ for the halos with the NFW density profile \cite{1997ApJ...490..493N} is given as
\begin{equation}
\label{eq:fc}
f(c) = \frac{c^3}{3}\left[1-\frac{1}{(1+c)^3}\right]\left[\log(1+c)-\frac{c}{1+c}\right]^{-2}.
\end{equation}
For the concentration parameter function $c(M,z)$ we use two different models: 
\begin{itemize}
\item the {\em `Macci{\`o} et al.'} model~\cite{2008MNRAS.391.1940M}, in which $c(M,z)=k_{200} \left(\mathcal{H}(z_f(M))/\mathcal{H}(z)\right)^{2/3}$, where $k_{200} \simeq 3.9$, $\mathcal{H}(z)=H(z)/H_0$ and $z_f(M)$ is the effective redshift for the formation of a halo with mass $M$;
\item a {\em `power law'} model (inferred from the results in~\cite{Neto:2007vq,2008MNRAS.391.1940M}) with $c(M,z)=6.5\, \mathcal{H}(z)^{-2/3}$ $(M/M_*)^{-0.1}$, $M_*=3.37\, 10^{12} h^{-1}M_\odot$, which gives a good fit within the mass range resolved by the simulations. 
\end{itemize}
Moreover, we consider two typical values for the minumum halo mass (motivated e.g.\ by~\cite{Martinez:2009jh,Bringmann:2009vf}) : 
\begin{itemize}
\item $M_{\rm min} =10^{-6}\, M_{\odot}$ and 
\item $M_{\rm min} =10^{-9}\, M_{\odot}$. 
\end{itemize}
The resulting four models for the cosmological boost factor are given in form of a {\sc Mathema-tica}$^{\tiny{\textregistered}}$ interpolating function on the~\myurl{www.marcocirelli.net/PPPC4DMID.html}{website}~\cite{website} and are plotted in Fig.~\ref{fig:EGamma}a. We see that the boost factor shoots up at $z\sim100$, which corresponds to the redshift where the DM halos with the assumed lowest masses start to form. As expected, the curves for the $10^{-9}$ $M_{\odot}$ models start to deviate slightly earlier. 

It is clear from the figure that currently the biggest uncertainty influencing the magnitude of the boost factor is related to the concentration parameter function $c(M,z)$ and in particular to the way one chooses to extend its behavior below the mass scales directly probed by the simulations. At low redshifts the variation between the models reaches two orders of magnitude. 
On top of that there is an extra uncertainty related to the profiles of the DM halos: we have assumed the NFW profile as a benchmark, but expressions analogous to eq.~(\ref{eq:fc}) (but more cumbersome) can be easily calculated for the Einasto and Burkert profiles. Comparing the $f(c)$ functions for different profiles along a large range of $c_{\rm vir}$, we find that this adds an uncertainty of roughly one order of magnitude.
%One actually finds that  the concentration functions for these latter profiles can be parameterized as $f_{\rm Ein} = C_{\rm Ein}^3 f_{\rm NFW}$ with $C_{\rm Ein} \simeq 2.0$ and $f_{\rm Bur} = C_{\rm Bur}^3 f_{\rm NFW}$ with $C_{\rm Bur} \simeq 0.7$, along a large range of $c_{\rm vir}$. As one can infer from the numerical values of the $k_{\rm i}$ coefficients, the adoption of an Einasto(/Burkert) profile would lead to $B(z)$ that are roughly a factor of 8 stronger (/3 weaker) with respect to NFW. \xxx{PAOLO: compute the values for the other profiles.}

\subsection{Absorption of gamma rays in the intergalactic medium}
\label{subsec:absorption}

The factor $\exp[-\tau(E_\gamma,z,z')]$ in eq.~(\ref{eq:EGflux}) expresses the attenuation of the flux of $\gamma$-rays having an energy at collection $E_\gamma =E'_\gamma \ (1+z)/(1+z')$ as they propagate from the emission redshift $z'$ (where they are emitted with energy $E'$) to the collection redshift $z$. 
In turn, the notation $\tau(E_\gamma,z')$ identifies the optical depth for gamma rays collected today with energy $E_\gamma =E'_\gamma/(1+z')$. 
In this section we briefly discuss the relevant physics processes and provide the ingredients to compute $\tau(E_\gamma,z,z')$ explicitly. 
We provide $\tau(E_\gamma,z')$ in the form of a {\sc Mathematica}$^{\tiny{\textregistered}}$ interpolating function on the~\myurl{www.marcocirelli.net/PPPC4DMID.html}{website}~\cite{website}.

\bigskip

The processes relevant to the absorption of energetic photons in cosmological length scales ($>10$ Mpc) and in the energy range roughly spanning from MeVs to TeVs are
\begin{itemize}
\label{absorptionlist}
% \item[-] Compton scattering off intervening gas
 \item[-] pair production on baryonic matter
 \item[-] photon-photon scattering on ambient Photon Background Radiation (PBR)
 \item[-] pair production on ambient PBR
\end{itemize}
Fig.~\ref{fig:EGamma}b illustrates the value of $e^{-\tau}$ as a function of the energy at detection $E_\gamma$ and the emission redshift $z'$.
In the lowest energy section of the plot, before the first small plateau located at KeV-ish energies, the absorption is dominated by photoionizations. Beyond that, up to the beginning of the large plateau, Compton losses are dominating. In the large flat plateau region the absorption is dominated by pair production on matter. In the final falling part of the curves the absorption is determined by photon-photon pair production. The photon-photon scattering also gives some contribution in the region where the flat plateau turns over to falling curves.

The PBR is mainly composed by the CMB, the intergalactic stellar light and secondary IR radiation. The intergalactic stellar light notably consists of the UV background produced in the low redshift Universe once the first (massive and hot) stars start to light up. 
Since this latter part is the most uncertain, we introduce three distinct modelizations:

\begin{itemize}
\item {\em `no UV'} assumes that no UV background is present.

\item {\em `minimal UV'} takes into account that recent studies of blazars, e.g.~\cite{Aliu:2008ay}, suggest signicantly lower values for the UV photon densities than estimated in many of the previous investigations. We use the UV background model as given in~\cite{Dominguez:2010bv}, which is fully consistent with the earlier study~\cite{Franceschini}.

\item {\em `maximal UV'} assumes the UV background as given by `minimal UV' is multiplied with factor $1.5$. It can considered as an upper limit of possible systematic errors of the model given by~\cite{Dominguez:2010bv}.

%\item {\em `realistic UV'} takes into account that recent studies of blazars, e.g.~\cite{Aliu:2008ay}, suggest significantly lower values for the UV photon densities than estimated in many of the previous investigations. It thus reduces the UV background level by an order of magnitude for $z>4$ and connects this model smoothly to the $z=0$ model of Stecker et al. \cite{Stecker:2005qs}.
\end{itemize}

We will see that the choice of UV background has a certain impact on the flux of very high energy extragalactic gamma rays.

%\subsection{Absorption coefficients and optical depth for gamma photons}

\bigskip

We now proceed to discuss a compact formalism allowing to calculate the optical depth $\tau(E'_\gamma,z,z^{'})$, focussing in particular on the last three processes in the list at page \pageref{absorptionlist}, since, as discussed, they are most relevant for GeV gamma ray astronomy telescopes such as FERMI. For further details see \cite{Zdziarski:1989aa} and \cite{1988ApJ...335..786Z}. 
%The processes relevant for the absorption of gamma-rays in the energy range probed by the Fermi-LAT are: (i) pair production on matter, (ii) photon-photon scattering, and (iii) pair production on PBR. At those energies the dominating one is the pair production on PBR. Here we provide a compact formalism needed to calculate the optical depth $\tau(E,z,z^{'})$ in Eq.~(\ref{eq:EGtransfer}). 
In what follows we express the photon energy in units of electron rest mass, i.e. $\epsilon\equiv E_\gamma/(m_ec^2)$.
\begin{itemize}
\item[$\circ$] The absorption coefficient at redshift $z$ for pair production on neutral matter with mass fractions of hydrogen $X=0.75$ and helium $Y=0.25$ can be approximated as
\begin{eqnarray}
\alpha_{{\rm mat-pair}}(\epsilon,z)&\simeq&\alpha_0 \, (1+z)^3\ln\left(\frac{513\, \epsilon}{\epsilon+825}\right),\quad\epsilon>6\,,\\
\alpha_0\simeq 5.3 \, n_e^0 \, \alpha_f \,  r_0^2&\simeq& 2.05\times 10^{-9}\left(\frac{\Omega_b}{0.045}\right)\left(\frac{h}{0.7}\right)^2\,{\rm Mpc}^{-1}\,,
\end{eqnarray}
where the average electron number density at $z=0$ is
\begin{equation}
n_e^0 \simeq 2.17\times 10^{-7} \left(\frac{\Omega_b}{0.045}\right)\left(\frac{h}{0.7}\right)^2\,{\rm cm}^{-3}\,.
\end{equation}
Here $\Omega_b = 0.0448 \pm 0.0011$ is the density parameter for baryons~\cite{cosmoDM}, $h$ is the reduced Hubble parameter, $\alpha_f= 7.29735\times10^{-3}$ is the fine structure constant, and $r_0= 2.8179\times10^{-13}$ cm the classical electron radius.

Below redshift $z\simeq6$ the Universe is reionized \cite{Fan:2006dp}. The absorption coefficient for pair production on fully ionized matter can be approximately given by
\begin{eqnarray}
\alpha_{{\rm ionmat-pair}}(\epsilon,z)&\simeq&\alpha_0 \, (1+z)^3\left[\ln(2\epsilon)-\frac{109}{42}\right],\quad\epsilon \gg 1\,,\\
\alpha_0\simeq \frac{20}{3}n_e^0 \, \alpha_f \, r_0^2&\simeq& 2.58\times 10^{-9}\left(\frac{\Omega_b}{0.045}\right)\left(\frac{h}{0.7}\right)^2\,{\rm Mpc}^{-1}\,.
\end{eqnarray}
\item[$\circ$] The absorption coefficient for photon-photon scattering at redshift $z$ can be given by
\begin{eqnarray}
\alpha_{{\rm \gamma\gamma-scat}}(\epsilon,z)&=&\alpha_0 \, (1+z)^6\epsilon^3\,,\\
\alpha_0= \frac{4448 \, \pi^4}{455625}\frac{\alpha_f^4 \, \Theta_0^6}{\lambda_c}&\simeq& 3.23\times 10^{-31}\left(\frac{T_0}{2.725\,{\rm K}}\right)^6\,{\rm Mpc}^{-1}\,,
\end{eqnarray}
where $\Theta_0\equiv \frac{k_B \, T_0}{m_e \, c^2}$ is the CMB temperature at $z=0$ in electron rest mass units, and $\lambda_c= 2.4263\times 10^{-10}$ cm is the electron Compton wavelength. Here we consider the CMB only as target photons: the contribution of the UV photons from stars is completely negligible.
\item[$\circ$] For photon-photon pair production processes, instead, the contribution of the intergalactic UV background is not negligible, so we have to include those together with the CMB. The target photon number density is thus $n(\epsilon,z)=n_{{\rm CMB}}(\epsilon,z)+n_{{\rm UV}}(\epsilon,z)$. As there are no simple analytical forms available for $n_{{\rm UV}}(\epsilon,z)$, we have to use a fully numerical treatment for the photon-photon pair production process. The absorption coefficient for photon-photon pair production at redshift $z$ is then computed as
\begin{eqnarray}
\alpha_{{\rm \gamma\gamma-pair}}(\epsilon,z)&=&\frac{\lambda_c^2\, \alpha_f^2}{4\pi}\int\limits_{1/\epsilon}^{\infty}{\rm d}\tilde{\epsilon}\,n(\tilde{\epsilon},z)\frac{\phi(v)}{(\epsilon\tilde{\epsilon})^2}\,,\\
{\rm with} \quad \phi(v) &=&  \frac{1+2v+2v^2}{1+v}\ln w - \frac{2\sqrt{v}(1+2v)}{\sqrt{1+v}} -\ln^2 w + 2\ln^2 (1+w) + \nonumber \\
&+&4\,{\rm Li}_2\left(\frac{1}{1+w}\right) - \frac{\pi^2}{3}\,,\\
{\rm where} \quad v&\equiv& (\epsilon \, \tilde{\epsilon}-1)\ge0\,,\quad w \equiv \frac{\sqrt{1+v} + \sqrt{v}}{\sqrt{1+v} - \sqrt{v}}\,.
\end{eqnarray}
Here ${\rm Li}_2$ is the dilogarithm function.
\end{itemize}
With the above ingredients, the total absorption coefficient  $\alpha(\epsilon,z)$ is given by
\begin{equation}
\alpha(\epsilon,z)=\left\{
\begin{array}{rl}
\alpha_{{\rm mat-pair}}(\epsilon,z)\ &{\rm if}\ 6\lesssim z \lesssim 1000\\
\alpha_{{\rm ionmat-pair}}(\epsilon,z)\ &{\rm if}\ z \lesssim 6
\end{array}
\right\} \, + \, \alpha_{{\rm \gamma\gamma-scat}}(\epsilon,z) \, + \, \alpha_{{\rm \gamma\gamma-pair}}(\epsilon,z).
\end{equation}
So finally the optical depth $\tau(E_\gamma,z,z')$ between redshifts $z$ and $z'$ can be computed as
\begin{equation}
\tau(E_\gamma,z,z')=c\int\limits_{z}^{z'}{\rm d}\tilde{z}\frac{\alpha(\tilde{\epsilon},\tilde{z})}{H(\tilde{z})(1+\tilde{z})}\,,
\end{equation}
where $\displaystyle \tilde{\epsilon}=\frac{1+\tilde{z}}{1+z}\frac{E_\gamma}{m_ec^2}$. 
%To calculate the Hubble function $H(z)$ we assume a flat $\Lambda$CDM model with $\Omega_m=0.27$ and $h=0.7$.

\begin{figure}[!t]
\includegraphics[width=0.48 \textwidth]{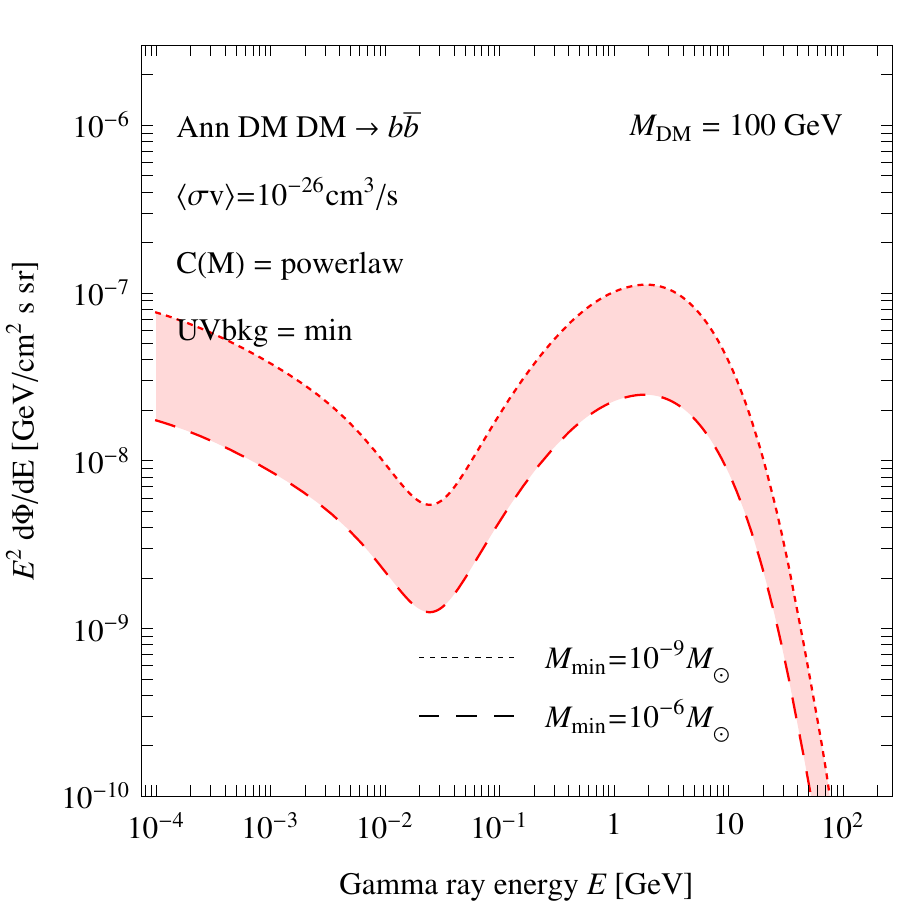} \quad \includegraphics[width=0.48 \textwidth]{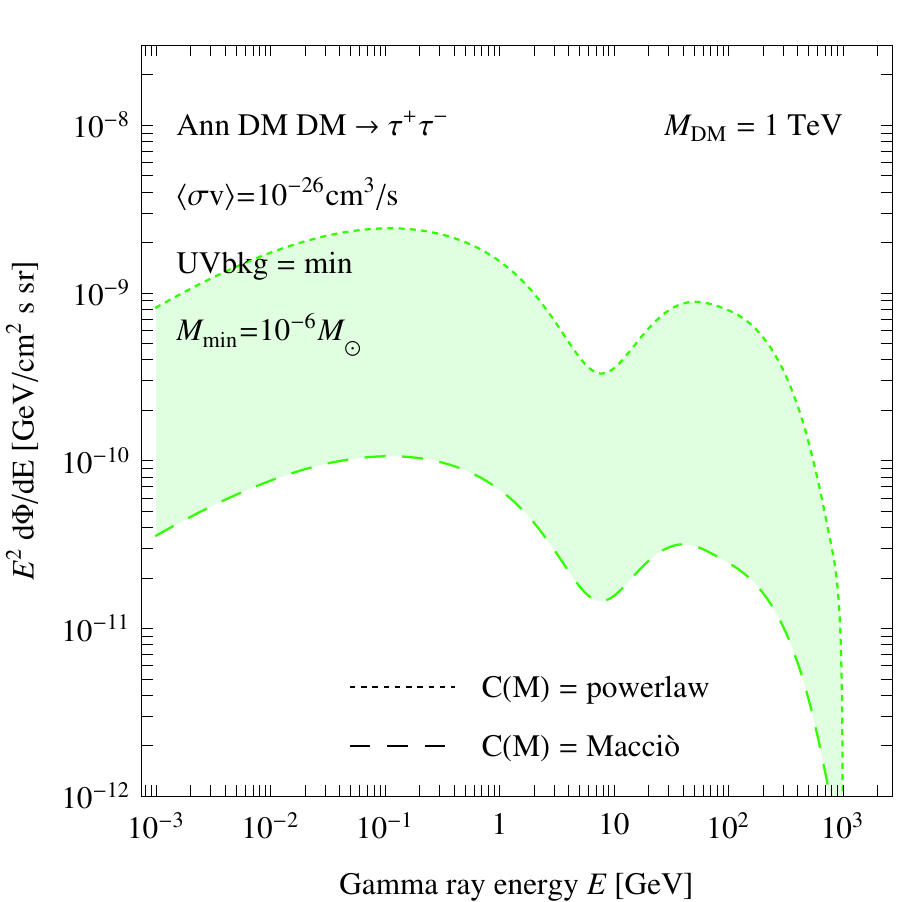}
\includegraphics[width=0.48 \textwidth]{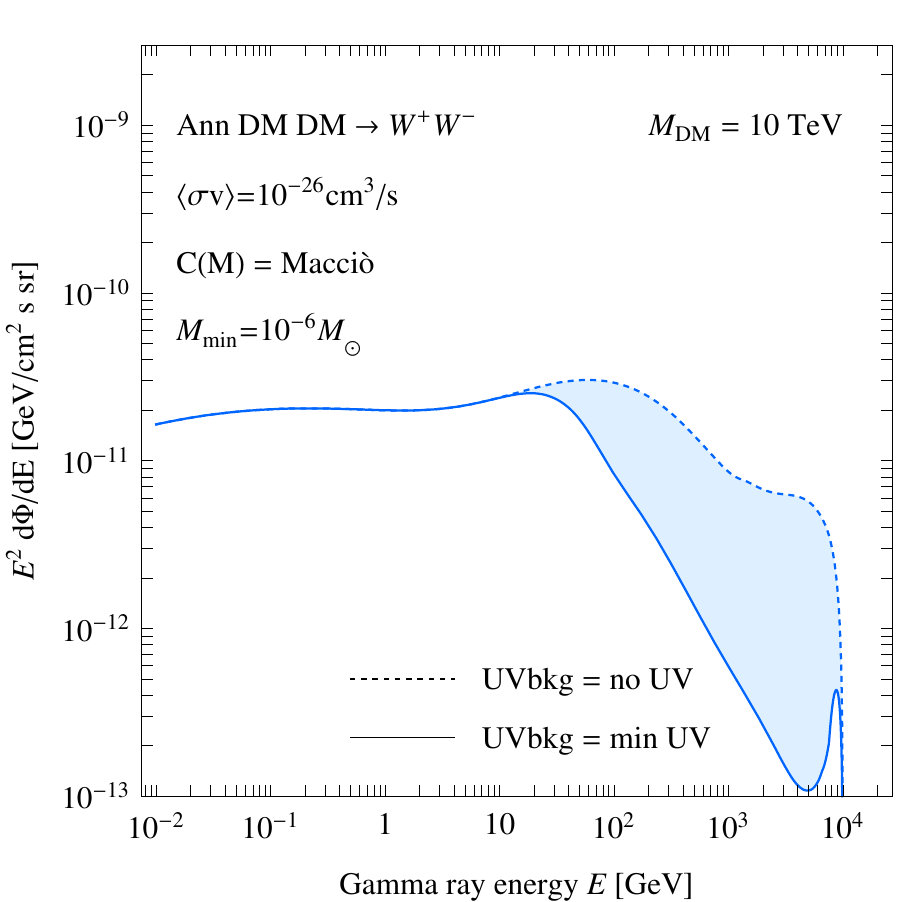} \quad \includegraphics[width=0.48 \textwidth]{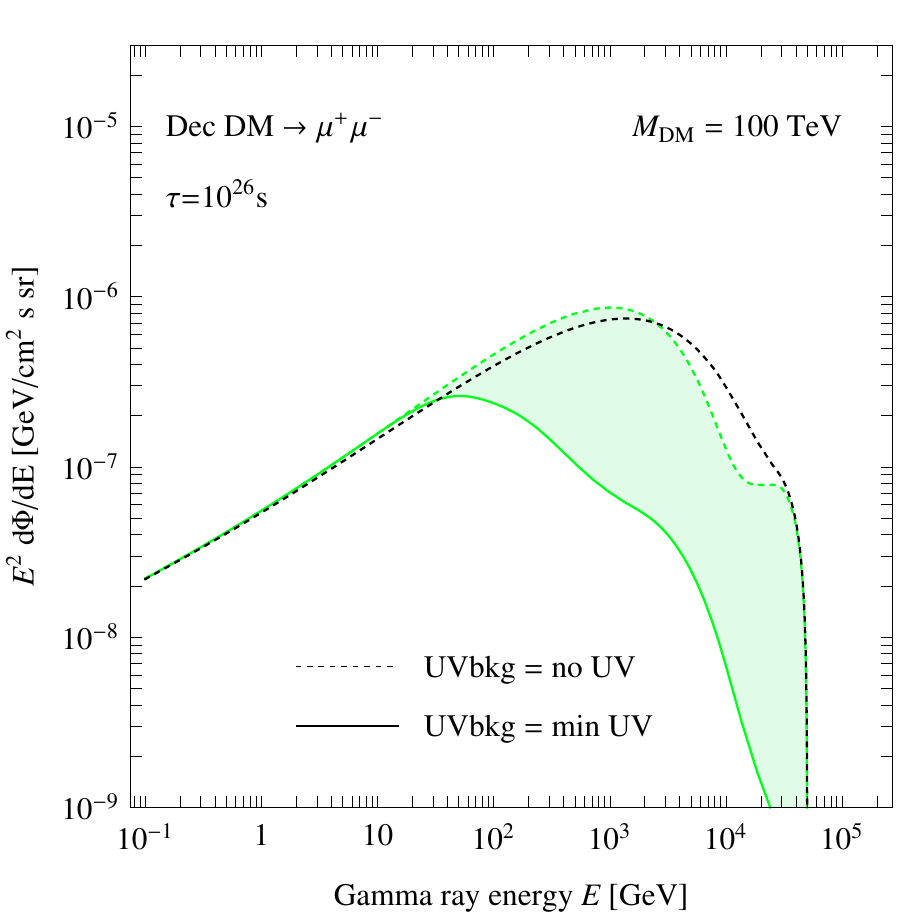}
\caption{\em \small \label{fig:extragammas} {\bfseries Fluxes of extragalactic gamma rays}, for the case of annihilations (first 3 panels) and decay (last panel). In each panel one of the astrophysical model assumptions is variated. The choices of annihilation or decay channels and particle physics parameters are indicated. On the last panel, the black dotted line indicates the flux for the {\em `no UV'} case computed in Thomson approximation.}
\end{figure}

\subsection{Extragalactic gamma rays: results}

Applying the recipe of eq.~(\ref{eq:EGflux}), with all the ingredients discussed in this section, we compute the fluxes of extragalactic gamma rays detected at Earth, for any given choice of the concentration parameter function, minimal halo mass and UV background. We provide them in numerical form on the~\myurl{www.marcocirelli.net/PPPC4DMID.html}{website}~\cite{website} in the form of {\sc Mathematica}$^{\tiny{\textregistered}}$ interpolating functions. 

\medskip

Fig.~\ref{fig:extragammas} presents some examples of such fluxes, for the case of annihilation and decay and for different primary channels. In the top left and top right panels the minimal halo mass $M_{\rm min}$ and the choice of $c(M,z)$ are variated respectively. As anticipated in the discussion of the cosmological boost factor (see~\ref{subsec:EGclusering}), this affects the normalization of the spectra by a factor of a few (varying $M_{\rm min}$) or almost by two orders of magnitude (varying the $c(M,z)$ model). 
In the bottom panels the UV background (relevant in the gamma rays absorption, see~\ref{subsec:absorption}) is varied. The effect is visible at energies above a few TeV. In the last panel we also plot for illustraton the flux (for the `noUV' case) that one would obtain if adopting the Thomson approximation discussed in Sec.~\ref{extragalactic}. For smaller DM masses the effect would be even less important.

\section{Summary}
\label{summary}

\subsection{Ingredients}

On the~\myurl{www.marcocirelli.net/PPPC4DMID.html}{website}~\cite{website}, we provide the following:

\begin{enumerate}[label=\fbox{\arabic*}]

\item  \label{dlNdlxEW} {\tt dlNdlxIEW[primary->final][DMmass,log$_{10}$x]}: spectrum $d\ln N/d\ln x$ in $x = E/M_{\rm DM}$ of {\tt final} particles
generated by DM annihilations into a pair of {\tt primary} particles.\\
Also in terms of numerical tables.

\item \label{dlNdlxPythia} {\tt dlNdlxI[primary->final][DMmass,log$_{10}$x]}: same as \ref{dlNdlxEW}, but without EW corrections.\\
Also in terms of numerical tables.

\item \label{b[E,r,z]} {\tt b[E,r,z]}: energy loss coefficient function $b(E,\vec x)$ for $e^\pm$ of energy {\tt E} at the position {\tt (r,z)} in the Galaxy.

\item \label{ElectronHaloFunctGalaxyAnnI} {\tt ElectronHaloFunctGalaxyAnnI[halo,propag][log$_{10}$x,log$_{10}$E$_s$,r,z]}: generalized  halo functions $I(E,E_{\rm s},r,z)$ for $e^\pm$ for annihilations, in any given point ({\tt r}, {\tt z}) of the Galaxy, expressed as a function of $x = E/E_{\rm s}$.\\
Analogous for decay.

\item \label{ElectronHaloFunctEarthAnnI} {\tt ElectronHaloFunctEarthAnnI[halo,propag][log$_{10}$x,log$_{10}$E$_s$]}: generalized halo functions $I(E,E_{\rm s},\vec r_\odot)$ for $e^\pm$ for annihilations, at the location of the Earth, expressed as a function of $x = E/E_{\rm s}$.\\
Analogous for decay.

\item \label{fitcoefficientspos} Tables of {\tt fit coefficients} for the reduced halo functions for $e^\pm$ for annihilations $\mathcal{I}(\lambda,\vec r_\odot)$ at the location of the Earth.\\
Analogous for decay.

\item \label{fitcoefficientspbar} Tables of {\tt fit coefficients} for the propagation functions for $\bar p$ for annihilations $R(K)$ at the location of the Earth.\\
Analogous for decay.

\item \label{fitcoefficientsdbar} Tables of {\tt fit coefficients} for the propagation functions for $\bar d$ for annihilations $R(K_d/n)$ at the location of the Earth.\\
Analogous for decay.

\item \label{ElectronFluxAnn} {\tt ElectronFluxAnn[primary,halo,propag][DMmass,$\sigma$v,log$_{10}$E$_e$]}: differential flux $\displaystyle\frac{d\Phi_{e^\pm}}{dE}$ at Earth.\\
Analogous for decay.

\item \label{ProtonFluxAnn} {\tt ProtonFluxAnn[primary,halo,propag][mass,$\sigma$v,log$_{10}$K]}: differential flux $\displaystyle\frac{d\Phi_{\bar p}}{dK}$ at Earth.\\ 
Analogous for decay.

\item \label{DeuteronFluxAnn} {\tt DeuteronFluxAnn[primary,halo,propag][mass,$\sigma$v,log$_{10}$K$_d$]}: differential flux $\displaystyle\frac{d\Phi_{\bar d}}{dK_d}$ at Earth.\\ 
Analogous for decay.

\item \label{Jave} {\tt Jave[halo][Log$_{10}\theta$]}: factor $J(\theta)$ for prompt gamma rays.\\ 
Analogous for decay.

\item \label{IIC} {\tt IICAnnI[halo,propag][Log$_{10}E_s$,Log$_{10}$E$_\gamma$,Log$_{10}\ell$,Log$_{10}b$]}: halo functions $I_{\rm IC}(E_\gamma,E_{\rm s},b,\ell)$ for Inverse Compton, for annihilations.\\ 
Analogous for decay.

\item \label{ICcode} {\tt Code bite to compute} $\displaystyle \frac{d\Phi_{\rm IC \gamma}}{dE_\gamma}$. 

\item \label{BoostF} {\tt BoostF[z,minhalomass,cmodel]}: cosmological boost factor $B(z)$ as a function of redshift {\tt z} for a choice of $M_{\rm min}$ and $c(M)$.

\item \label{ETau} {\tt ETau[E,z',UVmodel]}: optical depth of the Universe $e^{\tau(E,z')}$ for a choice of UV background.

\item \label{EGgammaFluxAnn} {\tt EGgammaFluxAnn[primary,minhalomass,cmodel,UVmodel][mass,$\sigma$v,lE]}: differential flux $\displaystyle \frac{d\Phi_{{\rm EG}\gamma}}{dE_\gamma}$.\\ 
Analogous for decay.

\end{enumerate}

The {\sc Mathematica}$^{\tiny{\textregistered}}$ {\tt InterpolationFunction}s and the code bite provided in \ref{ICcode} have been produced with {\sc Mathematica}$^{\tiny{\textregistered}}$ version 7.0.0. The {\tt InterpolationFunction}s are expected to be compatible with version 2 and any later version. The code bite employs solution methods included in {\sc Mathematica}$^{\tiny{\textregistered}}$ version 4 and later.

\bigskip

Table~\ref{tab:choices} lists the discrete choices for the variables employed in the functions above, together with a reference to the corresponding discussion in the text (when available).  
\begin{table}
\begin{center}
\begin{tabular}{c|c|c}
Variable & Values or range & Refer to\\
\hline
{\tt primary} & $ \begin{array}{c}
{\tt eL,\ 
eR,\
\mu L,\ 
\mu R,\ 
\tau L,\ 
\tau R,}\\[1mm]
{\tt q, \ 
c, \ 
b, \  
t, \ 
\gamma,\ 
g,
WL,\ 
WT,\ 
ZL,\ 
ZT, }\\[1mm]
{\tt h_{115}, \
h_{135}, \
h_{170}, \
h_{200}, \
h_{300}, \
h_{400}, \
h_{500},} \\[1mm]
{\tt \nu e, \ 
\nu \mu, \ 
\nu \tau,
V \to e, \
V \to \mu, \
V \to \tau }
\end{array}$
  & Eq. (\ref{primarychannels})\\
\hline
{\tt final} & e, p, $\gamma$, d, $\nu$e, $\nu \mu$, $\nu \tau$ & Sec.~\ref{primary} \\
\hline
{\tt DMmass} & 
$ \begin{array}{c}
5 \ {\rm GeV} \to 100 \ {\rm TeV\ (annihilation)}  \\
10 \  {\rm GeV} \to 200 \ {\rm TeV\ (decay)} 
\end{array}$  & Sec.~\ref{fluxesresults} \\
\hline
{\tt halo} & {\tt NFW, Ein, EiB, Iso, Bur, Moo} & Fig.~\ref{fig:DMprofiles} \\
\hline
{\tt propag} & {\tt MIN, MED, MAX} &  Table \ref{tab:proparam} \\
\hline
{\tt $\sigma$v} or {\tt $\Gamma$} & any & \\
\hline
{\tt minhalomass} & {\tt 10}$^{{\tt -6}}$, {\tt 10}$^{{\tt -9}}$ &  Sec.~\ref{subsec:EGclusering} \\
\hline
{\tt cmodel} & {\tt maccio, powerlaw}  & Sec.~\ref{subsec:EGclusering} \\
\hline
{\tt UVmodel} & {\tt noUV, minUV, maxUV} &  Sec.~\ref{subsec:absorption} \\
\end{tabular}
\end{center}
\caption{\em \small {\bfseries Variables of the numerical functions and their admitted values}. 
\label{tab:choices}}
\end{table}

\subsection{Recipes}

The main recipes for computing DM indirect detection signals are:

\begin{enumerate}[label=\Roman*.]

\item {\bf Computing $\displaystyle \frac{d\Phi_{e^\pm}}{dE}$: the differential flux of $e^\pm$ at Earth}:\\[2mm]
eq.~(\ref{eq:positronsflux}) with \ref{b[E,r,z]}, \ref{dlNdlxEW} and \ref{ElectronHaloFunctEarthAnnI}. 
Or use \ref{ElectronFluxAnn}.

\item {\bf Computing $\displaystyle \frac{d\Phi_{e^\pm}}{dE}$ with approximated energy losses}:\\[2mm]
 eq.~(\ref{eq:positronsfluxnox}) with \ref{dlNdlxEW} and \ref{fitcoefficientspos}. 

\item {\bf Computing $\displaystyle \frac{d\Phi_{\bar p}}{dK}$: the differential flux of antiprotons at Earth}:\\[2mm] 
eq.~(\ref{eq:fluxpbar}) with \ref{dlNdlxEW} and \ref{fitcoefficientspbar} in eq.~(\ref{eq:fitantiprotons}).  
Or use \ref{ProtonFluxAnn}.

\item {\bf Computing $\displaystyle \frac{d\Phi_{\bar d}}{dK_d}$: the differential flux of antideuterons at Earth }:\\[2mm] 
eq.~(\ref{eq:fluxdbar}) with \ref{dlNdlxEW} and \ref{fitcoefficientsdbar} in eq.~(\ref{eq:fitantideuterons}).  
Or use \ref{DeuteronFluxAnn}.

\item {\bf Computing $\displaystyle \frac{d\Phi_{\gamma}}{dE_\gamma}$: the differential flux of prompt $\gamma$ rays}:\\[2mm] 
eq.~(\ref{gammafluxI}) with \ref{dlNdlxEW} and $\bar J$ from table~\ref{tab:Jfactors} or from eq.~(\ref{eq:formulaeJfactors}) with \ref{Jave}.  

\item {\bf Computing $\displaystyle \frac{d\Phi_{\rm IC \gamma}}{dE_\gamma}$: the differential flux of galactic ICS $\gamma$ rays}:\\[2mm]
eq.~(\ref{eq:summaryICI}) with eq.~(\ref{eq:summaryIC}), with \ref{dlNdlxEW} and \ref{IIC}. Or use directly \ref{ICcode}.

\item {\bf Computing $\displaystyle \nu \frac{dW_{\rm syn}}{d\nu \, d\Omega}$: the differential flux of galactic synchrotron radiation}:\\[2mm]
eq.~(\ref{synspectrumfinal}), with \ref{dlNdlxEW} and \ref{ElectronHaloFunctGalaxyAnnI}. 

\item {\bf Computing $\displaystyle \frac{d\Phi_{\rm EG \gamma}}{dE_\gamma}$: the differential flux of extragalactic $\gamma$ rays}:\\[2mm]
eq.~(\ref{eq:EGfluxtoday}) with eq.~(\ref{jEGprompt}) and eq.~(\ref{jEGICS}), and with \ref{BoostF}, \ref{ETau} and \ref{dlNdlxEW}. Or use directly \ref{EGgammaFluxAnn}.

\end{enumerate}

%\begin{figure}[t]
%\begin{center}
%\includegraphics[width=0.4\textwidth]{xxx}\qquad
%\includegraphics[width=0.4\textwidth]{xxx}
%\caption{\em\label{fig:xxx} Caption.}
%\end{center}
%\end{figure}

\paragraph{Acknowledgements}
We thank Gianfranco Bertone, C\'eline Combet, David Maurin, Fabrizio Nesti, Michele Papucci, Stefano Pozzorini, Pierre Salati, Peter Skands, Pasquale D. Serpico, Torbj\"orn Sj\"ostrand and Marco Taoso for useful discussions. We are particularly grateful to Caner \"Unal (Middle East Technical University and CERN Summer Student Program 2010) for his important feedback on the numerical results. We also thank the anonimous JCAP referee for his/her useful and thorough comments on the first version of this paper. 
This work was supported by the ESF Grant 8090, ESF Grant 8499, Estonian Ministry of Education and Research project SF0690030s09 and European Social Fund (Mobility Grant MJD52). We also thank the EU Marie Curie Research \& Training network ``UniverseNet" (MRTN-CT-2006-035863) for support.

\bigskip
\appendix

\footnotesize
\begin{multicols}{2}
  
\end{multicols}


\begin{thebibliography}{nn}

% REFS FOR SECTION 1

\bibitem{JungmanReview}
G.~Jungman, M.~Kamionkowski and K.~Griest,
  %``Supersymmetric dark matter,''
  Phys.\ Rept.\  {267} (1996) 195
  [arXiv:\hhref{hep-ph/9506380}].
  %%CITATION = PRPLC,267,195;%%

\bibitem{BertoneReview}
G.~Bertone, D.~Hooper and J.~Silk,
  %``Particle dark matter: Evidence, candidates and constraints,''
  Phys.\ Rept.\  {405} (2005) 279
  [arXiv:\hhref{hep-ph/0404175}].
  %%CITATION = PRPLC,405,279;%%

\bibitem{EinastoReview}
J.~Einasto,
  %``Dark Matter,''
  arXiv:\hhref{0901.0632} [astro-ph.CO].
  %%CITATION = ARXIV:0901.0632;%%

\bibitem{rotation}
F.~Zwicky, Helv. Phys. Acta 6 (1933) 110.

V.~Rubin and W.~Ford, 
%``Rotation of the Andromeda Nebula from a Spectroscopic Survey of Emission Regions'',
Astrophysical Journal, vol. 159, p.379

V.~Rubin, N.~Thonnard and W.~Ford,
%``Extended rotation curves of high-luminosity spiral galaxies. IV - Systematic dynamical properties, SA through SC'',
Astrophysical Journal, Part 2 - Letters to the Editor, vol. 225, Nov. 1, 1978, p. L107-L111.

V.~Rubin, W.~Ford and N.~Thonnard,
%``Rotational properties of 21 SC galaxies with a large range of luminosities and radii, from NGC 4605 /R = 4kpc/ to UGC 2885 /R = 122 kpc",
Astrophysical Journal, Part 1, vol. 238, June 1, 1980, p. 471-487
  %%CITATION = ASJOA,238,471;%%


\bibitem{lensing}
See for instance 
R.~Massey {\it et al.},
  %``Dark matter maps reveal cosmic scaffolding,''
  Nature {445} (2007) 286
  [arXiv:\hhref{astro-ph/0701594}].
  %%CITATION = NATUA,445,286;%%

\bibitem{bullet}
D.~Clowe, M.~Bradac, A.~H.~Gonzalez, M.~Markevitch, S.~W.~Randall, C.~Jones and D.~Zaritsky,
  %``A direct empirical proof of the existence of dark matter,''
  Astrophys.\ J.\  {648} (2006) L109
  [arXiv:\hhref{astro-ph/0608407}].
  %%CITATION = ASJOA,648,L109;%%
 M.~Bradac {\it et al.},
  %``Strong and weak lensing united III: Measuring the mass distribution of the
  %merging galaxy cluster 1E0657-56,''
  Astrophys.\ J.\  {652} (2006) 937
  [arXiv:\hhref{astro-ph/0608408}].
  %%CITATION = ASJOA,652,937;%%

\bibitem{cosmoDM}
E.~Komatsu {\it et al.},
  %``Seven-Year Wilkinson Microwave Anisotropy Probe (WMAP) Observations:
  %Cosmological Interpretation,''
  arXiv:\hhref{1001.4538} [astro-ph.CO].
  %%CITATION = ARXIV:1001.4538;%%
%These numbers summarize 
%various recent global analyses of cosmological data
%within the $\Lambda$CDM model
%that found compatible values and uncertainties:
%\hepart[astro-ph/0603449]{D. N. Spergel {\it et al.} [WMAP collaboration]},
%\art[astro-ph/0607086]{M. Cirelli and A. Strumia}{JCAP}{0612}{013}{2006}, %%CITATION = ARXIV:ASTRO-PH/0607086;%%
%\art[astro-ph/0608632]{M. Tegmark et al.}{Phys. Rev.}{D74}{123507}{2006}.

\bibitem{Gunn}
J.~E.~Gunn, B.~W.~Lee, I.~Lerche, D.~N.~Schramm and G.~Steigman,
  %``Some astrophysical consequences of the existence of a heavy stable  neutral
  %lepton,''
  Astrophys.\ J.\  {\bf 223} (1978) 1015.
  %%CITATION = ASJOA,223,1015;%%

\bibitem{Stecker}  
F.~W.~Stecker,
  %``The Cosmic Gamma-Ray Background From The Annihilation Of Primordial Stable
  %Neutral Heavy Leptons,''
  Astrophys.\ J.\  {\bf 223} (1978) 1032.
  %%CITATION = ASJOA,223,1032;%%
  
\bibitem{Zeldovich}  
  Y.~B.~Zeldovich, A.~A.~Klypin, M.~Y.~Khlopov and V.~M.~Chechetkin,
  %``Astrophysical Constraints On The Mass Of Heavy Stable Neutral Leptons,''
  Sov.\ J.\ Nucl.\ Phys.\  {\bf 31} (1980) 664
  [Yad.\ Fiz.\  {\bf 31} (1980) 1286].
  %%CITATION = YAFIA,31,1286;%%

\bibitem{Ellis} 
J.~R.~Ellis, R.~A.~Flores, K.~Freese, S.~Ritz, D.~Seckel and J.~Silk,
  %``COSMIC RAY CONSTRAINTS ON THE ANNIHILATIONS OF RELIC PARTICLES IN THE
  %GALACTIC HALO,''
  Phys.\ Lett.\  B {\bf 214} (1988) 403.
  %%CITATION = PHLTA,B214,403;%%

\bibitem{SilkSrednicki}
  J.~Silk and M.~Srednicki,
  %``Cosmic-ray antiprotons as a probe of a photino-dominated universe,''
  Phys.\ Rev.\ Lett.\  {\bf 53} (1984) 624.
  %%CITATION = PRLTA,53,624;%%

\bibitem{SteckerRudazWalsh}  
F.~W.~Stecker, S.~Rudaz and T.~F.~Walsh,
  %``Galactic Anti-Protons From Photinos,''
  Phys.\ Rev.\ Lett.\  {\bf 55} (1985) 2622.
  %%CITATION = PRLTA,55,2622;%% 

\bibitem{SteckerRudaz}   
S.~Rudaz and F.~W.~Stecker,
  %``COSMIC RAY ANTI-PROTONS, POSITRONS AND GAMMA-RAYS FROM HALO DARK MATTER
  %ANNIHILATION,''
  Astrophys.\ J.\  {\bf 325} (1988) 16.
  %%CITATION = ASJOA,325,16;%%  

\bibitem{SteckerTylka}  
F.~W.~Stecker and A.~J.~Tylka,
  %``THE COSMIC RAY ANTI-PROTON SPECTRUM FROM DARK MATTER ANNIHILATION AND ITS
  %ASTROPHYSICAL IMPLICATIONS: A NEW LOOK,''
  Astrophys.\ J.\  {\bf 336} (1989) L51.
  %%CITATION = ASJOA,336,L51;%%

\bibitem{TurnerWilczek}
M.~S.~Turner and F.~Wilczek,
  %``Positron Line Radiation from Halo WIMP Annihilations as a Dark Matter
  %Signature,''
  Phys.\ Rev.\  D {\bf 42} (1990) 1001.
  %%CITATION = PHRVA,D42,1001;%%
  
\bibitem{pioneerDbar}
  F.~Donato, N.~Fornengo and P.~Salati,
  %``Antideuterons as a signature of supersymmetric dark matter,''
  Phys.\ Rev.\  D {\bf 62} (2000) 043003
  [arXiv:\hhref{hep-ph/9904481}].
  %%CITATION = PHRVA,D62,043003;%%

\bibitem{Dbar2}
  H.~Baer and S.~Profumo,
  %``Low energy antideuterons: Shedding light on dark matter,''
  JCAP {\bf 0512} (2005) 008
  [arXiv:\hhref{astro-ph/0510722}].
  %%CITATION = JCAPA,0512,008;%%

\bibitem{followupDbar}
  F.~Donato, N.~Fornengo and D.~Maurin,
  %``Antideuteron fluxes from dark matter annihilation in diffusion models,''
  Phys.\ Rev.\  D {\bf 78} (2008) 043506
  [arXiv:\hhref{0803.2640} [hep-ph]].
  %%CITATION = PHRVA,D78,043506;%%

\bibitem{BerezinskyGurevich}
V.~S.~Berezinsky, A.~V.~Gurevich and K.~P.~Zybin,
  %``Distribution of dark matter in the galaxy and the lower limits for the
  %masses of supersymmetric particles,''
  Phys.\ Lett.\  B {\bf 294} (1992) 221.
  %%CITATION = PHLTA,B294,221;%%
  
\bibitem{BerezinskyBottino}  
  V.~Berezinsky, A.~Bottino and G.~Mignola,
  %``High-energy gamma radiation from the galactic center due to neutralino
  %annihilation,''
  Phys.\ Lett.\  B {\bf 325} (1994) 136
  [arXiv:\hhref{hep-ph/9402215}].
  %%CITATION = PHLTA,B325,136;%%

\bibitem{Gondolo}
P.~Gondolo,
  %``Either neutralino dark matter or cuspy dark halos,''
  Phys.\ Lett.\  B {\bf 494} (2000) 181
  [arXiv:\hhref{hep-ph/0002226}].
  %%CITATION = PHLTA,B494,181;%%

\bibitem{BertoneSiglSilk}
G.~Bertone, G.~Sigl and J.~Silk,
  %``Astrophysical limits on massive dark matter,''
  Mon.\ Not.\ Roy.\ Astron.\ Soc.\  {\bf 326} (2001) 799
  [arXiv:\hhref{astro-ph/0101134}].
  %%CITATION = MNRAA,326,799;%%
  
\bibitem{AloisioBlasiOlinto}  
  R.~Aloisio, P.~Blasi and A.~V.~Olinto,
  %``Neutralino annihilation at the galactic center revisited,''
  JCAP {\bf 0405} (2004) 007
  [arXiv:\hhref{astro-ph/0402588}].
  %%CITATION = JCAPA,0405,007;%%

\bibitem{BergstomEdsjoUllio}
L.~Bergstrom, J.~Edsjo and P.~Ullio,
  %``Spectral gamma-ray signatures of cosmological dark matter  annihilations,''
  Phys.\ Rev.\ Lett.\  {\bf 87} (2001) 251301
  [arXiv:\hhref{astro-ph/0105048}].
  %%CITATION = PRLTA,87,251301;%%

\bibitem{BaltzWai}
E.~A.~Baltz and L.~Wai,
  %``Diffuse inverse Compton and synchrotron emission from dark matter
  %annihilations in galactic satellites,''
  Phys.\ Rev.\  D {\bf 70} (2004) 023512
  [arXiv:\hhref{astro-ph/0403528}].
  %%CITATION = PHRVA,D70,023512;%%
  
\bibitem{Cholis}
  I.~Cholis, G.~Dobler, D.~P.~Finkbeiner, L.~Goodenough and N.~Weiner,
  %``The Case for a 700+ GeV WIMP: Cosmic Ray Spectra from ATIC and PAMELA,''
  Phys.\ Rev.\  D {\bf 80} (2009) 123518
  [arXiv:\hhref{0811.3641} [astro-ph]].
  %%CITATION = PHRVA,D80,123518;%%
  
\bibitem{Zhang}
  J.~Zhang, X.~J.~Bi, J.~Liu, S.~M.~Liu, P.~F.~Yin, Q.~Yuan and S.~H.~Zhu,
  %``Discriminating different scenarios to account for the cosmic e + /- excess
  %by synchrotron and inverse Compton radiation,''
  Phys.\ Rev.\  D {\bf 80} (2009) 023007
  [arXiv:\hhref{0812.0522} [astro-ph]].
  %%CITATION = PHRVA,D80,023007;%%

  
\bibitem{website}
 \myurl{www.marcocirelli.net/PPPC4DMID.html}{www.marcocirelli.net/PPPC4DMID.html}

\bibitem{galprop}
 \myurl{galprop.stanford.edu}{galprop.stanford.edu}.
 See A.~E.~Vladimirov {\it et al.},
  %``GALPROP WebRun: an internet-based service for calculating cosmic ray
  %propagation and associated diffuse emissions,''
  arXiv:\hhref{1008.3642} [astro-ph.HE] 
  %%CITATION = ARXIV:1008.3642;%%
for the latest release.

\bibitem{darksusy}
P. Gondolo, J. Edsj\"o, P. Ullio, L. Bergstr\"om, M. Schelke, E.A. Baltz, T. Bringmann and G. Duda, \myurl{www.physto.se/~edsjo/darksusy/}{http://www.darksusy.org.}.
P.~Gondolo, J.~Edsjo, P.~Ullio, L.~Bergstrom, M.~Schelke and E.~A.~Baltz,
  %``DarkSUSY: Computing supersymmetric dark matter properties numerically,''
  JCAP {\bf 0407} (2004) 008
  [arXiv:\hhref{astro-ph/0406204}].
  %%CITATION = JCAPA,0407,008;%%

\bibitem{micromegas}
 \myurl{lapth.in2p3.fr/micromegas/}{lapth.in2p3.fr/micromegas}.
See G.~Belanger, F.~Boudjema, P.~Brun, A.~Pukhov, S.~Rosier-Lees, P.~Salati and A.~Semenov,
  %``Indirect search for dark matter with micrOMEGAs2.4,''
  arXiv:\hhref{1004.1092} [hep-ph]
  %%CITATION = ARXIV:1004.1092;%%
for the latest release.

\bibitem{isatools}
H.~Baer, A.~Belyaev, T.~Krupovnickas and J.~O'Farrill,
  %``Indirect, direct and collider detection of neutralino dark matter,''
  JCAP {\bf 0408} (2004) 005
  [arXiv:\hhref{hep-ph/0405210}].
  %%CITATION = JCAPA,0408,005;%%

\bibitem{wimpsim}
J. Edsj\"o, WimpSim Neutrino Monte Carlo, \myurl{www.physto.se/~edsjo/wimpsim/}{http://www.physto.se/$\sim$edsjo/wimpsim/}.
 M.~Blennow, J.~Edsjo and T.~Ohlsson,
  %``Neutrinos from WIMP Annihilations Using a Full Three-Flavor Monte Carlo,''
  JCAP {\bf 0801} (2008) 021
  [arXiv:\hhref{0709.3898} [hep-ph]].
  %%CITATION = JCAPA,0801,021;%%



%%%%%%%%%%%%%%%%%%%%%%%%%%%%%%%%%%%%%%%%%%%%%%%%%%%%

% REFS FOR SECTION 2

\bibitem{Navarro:1995iw}
  J.~F.~Navarro, C.~S.~Frenk and S.~D.~M.~White,
  %``The Structure of Cs Dark Matter Halos,''
  Astrophys.\ J.\  {462} (1996) 563
  [arXiv:\hhref{astro-ph/9508025}].
  %%CITATION = ASJOA,462,563;%%

\bibitem{Graham:2005xx}
  A.~W.~Graham, D.~Merritt, B.~Moore, J.~Diemand and B.~Terzic,
  %``Empirical models for Dark Matter Halos. I. Nonparametric Construction of
  %Density Profiles and Comparison with Parametric Models,''
  Astron.\ J.\  {132} (2006) 2685
  [arXiv:\hhref{astro-ph/0509417}].
  %%CITATION = ANJOA,132,2685;%%

\bibitem{Navarro:2008kc}
  J.~F.~Navarro {\it et al.},
  %``The Diversity and Similarity of Cold Dark Matter Halos,''
  arXiv:\hhref{0810.1522} [astro-ph].
  %%CITATION = ARXIV:0810.1522;%%

\bibitem{Begeman}
K.~G.~Begeman, A.~H.~Broeils, R.~H.~Sanders, MNRAS 249, 523 (1991).

\bibitem{Bahcall:1980fb}
  J.~N.~Bahcall and R.~M.~Soneira,
  %``The Universe At Faint Magnetidues. 2. Models For The Predicted Star
  %Counts,''
  Astrophys.\ J.\ Suppl.\  {44}, 73 (1980).
  %%CITATION = APJSA,44,73;%%

\bibitem{Burkert}
A.~Burkert,
  %``The Structure of dark matter halos in dwarf galaxies,''
  IAU Symp.\  {171} (1996) 175
  [Astrophys.\ J.\  {447} (1995) L25]
  [arXiv:\hhref{astro-ph/9504041}].
  %%CITATION = ASJOA,447,L25;%%
  
See also:  
P.~Salucci and A.~Burkert,
  %``Dark Matter Scaling Relations,''
  arXiv:\hhref{astro-ph/0004397}.
  %%CITATION = ASTRO-PH/0004397;%%
G.~Gentile, P.~Salucci, U.~Klein, D.~Vergani and P.~Kalberla,
  %``The cored distribution of dark matter in spiral galaxies,''
  Mon.\ Not.\ Roy.\ Astron.\ Soc.\  {351} (2004) 903
  [arXiv:\hhref{astro-ph/0403154}] and 
  %%CITATION = MNRAA,351,903;%%
P.~Salucci, A.~Lapi, C.~Tonini, G.~Gentile, I.~Yegorova and U.~Klein,
  %``The universal rotation curve of spiral galaxies. II: The dark matter
  %distribution out to the virial radius,''
  Mon.\ Not.\ Roy.\ Astron.\ Soc.\  {378} (2007) 41
  [arXiv:\hhref{astro-ph/0703115}].
  %%CITATION = MNRAA,378,41;%%
  
  \bibitem{Moore04}
J.~Diemand, B.~Moore and J.~Stadel,
  %``Convergence and scatter of cluster density profiles,''
  Mon.\ Not.\ Roy.\ Astron.\ Soc.\  {353} (2004) 624
  [arXiv:\hhref{astro-ph/0402267}].
  %%CITATION = MNRAA,353,624;%%
  
  \bibitem{DMandBaryons}
See e.g.  O.~Y.~Gnedin, A.~V.~Kravtsov, A.~A.~Klypin {\it et al.},
  %``Response of dark matter halos to condensation of baryons: Cosmological simulations and improved adiabatic contraction model,''
  Astrophys.\ J.\  {616 } (2004)  16-26.
  [\hhref{astro-ph/0406247}];
  F.~Prada, A.~Klypin, J.~Flix Molina {\it et al.},
  %``Dark Matter Annihilation in the Milky Way Galaxy: Effects of Baryonic Compression,''
  Phys.\ Rev.\ Lett.\  {93 } (2004)  241301.
  [\hhref{astro-ph/0401512}].

\bibitem{Tissera}
P.~B.~Tissera, S.~D.~M.~White, S.~Pedrosa and C.~Scannapieco,
  %``Dark matter response to galaxy formation,''
  arXiv:\hhref{0911.2316}.
  %%CITATION = ARXIV:0911.2316;%%

\bibitem{Jing:2002np}
  Y.~P.~Jing and Y.~Suto,
  %``Triaxial Modeling of Halo Density Profiles with High-resolution N-body
  %Simulations,''
  Astrophys.\ J.\  {\bf 574} (2002) 538
  [arXiv:\hhref{astro-ph/0202064}].
  %%CITATION = ASJOA,574,538;%%
  
\bibitem{Law:2009yq}
  D.~R.~Law, S.~R.~Majewski and K.~V.~Johnston,
  %``Evidence for a Triaxial Milky Way Dark Matter Halo from the Sagittarius
  %Stellar Tidal Stream,''
  Astrophys.\ J.\  {\bf 703} (2009) L67
  [arXiv:\hhref{0908.3187} [astro-ph.GA]].
  %%CITATION = ASJOA,703,L67;%%

\bibitem{Gnedin:2005pt}
  O.~Y.~Gnedin, A.~Gould, J.~Miralda-Escude and A.~R.~Zentner,
  %``Probing the Shape of the Galactic Halo with Hyper-Velocity Stars,''
  Astrophys.\ J.\  {\bf 634} (2005) 344
  [arXiv:\hhref{astro-ph/0506739}].
  %%CITATION = ASJOA,634,344;%%  
  
\bibitem{Calcaneo}
C.~Calcaneo-Roldan and B.~Moore,
  %``The surface brightness of dark matter: Unique signatures of neutralino
  %annihilation in the galactic halo,''
  Phys.\ Rev.\  D {\bf 62}, 123005 (2000)
  [arXiv:\hhref{astro-ph/0010056}].
  %%CITATION = PHRVA,D62,123005;%%


\bibitem{rSun}
S.~Gillessen, F.~Eisenhauer, S.~Trippe, T.~Alexander, R.~Genzel, F.~Martins and T.~Ott,
  %``Monitoring stellar orbits around the Massive Black Hole in the Galactic
  %Center,''
  Astrophys.\ J.\  {692} (2009) 1075
  [arXiv:\hhref{0810.4674} [astro-ph]].
  %%CITATION = ASJOA,692,1075;%%

\bibitem{rSun2}
J.~Bovy, D.~W.~Hogg and H.~W.~Rix,
  %``Galactic masers and the Milky Way circular velocity,''
  Astrophys.\ J.\  {704} (2009) 1704
  [arXiv:\hhref{0907.5423} [astro-ph.GA]].
  %%CITATION = ASJOA,704,1704;%%
A.~M.~Ghez {\it et al.},
  %``Measuring Distance and Properties of the Milky Way's Central Supermassive
  %Black Hole with Stellar Orbits,''
  Astrophys.\ J.\  {689} (2008) 1044
  [arXiv:\hhref{0808.2870} [astro-ph]].
  %%CITATION = ASJOA,689,1044;%%
  
\bibitem{Kerr}
F.~J.~Kerr and D.~Lynden-Bell,
  %``Review of galactic constants,''
  Mon.\ Not.\ Roy.\ Astron.\ Soc.\  {221} (1986) 1023.
  %%CITATION = MNRAA,221,1023;%%  
  See also M. Shen and Z. Zhu, 
 %``A Kinematical Calibration of the Galactocentric Distance"
  Chin. Astron. Astrophys. 7, 120 (2007)
 for a recent compilation.
    
\bibitem{PDG}
C. Amsler et al. (\myurl{pdg.lbl.gov/}{Particle Data Group}), Physics Letters B667, 1 (2008) and 2009 partial update for the 2010 edition.

\bibitem{CatenaUllio}
R.~Catena and P.~Ullio,
  %``A novel determination of the local dark matter density,''
  arXiv:\hhref{0907.0018} [astro-ph.CO].
  %%CITATION = ARXIV:0907.0018;%%

\bibitem{deBoer}
M.~Weber and W.~de Boer,
  %``Determination of the Local Dark Matter Density in our Galaxy,''
  arXiv:\hhref{0910.4272} [astro-ph.CO].
  %%CITATION = ARXIV:0910.4272;%%

\bibitem{Nesti}
P.~Salucci, F.~Nesti, G.~Gentile and C.~F.~Martins,
  %``The dark matter density at the Sun's location,''
  arXiv:\hhref{1003.3101}.
  %%CITATION = ARXIV:1003.3101;%%

\bibitem{deBoer2}  
  W.~de Boer and M.~Weber,
  %``The Dark Matter Density in the Solar Neighborhood reconsidered,''
  arXiv:\hhref{1011.6323} [astro-ph.CO].
  %%CITATION = ARXIV:1011.6323;%%

\bibitem{SDSSMWmass}
X.~X.~Xue {\it et al.}  [SDSS Collaboration],
  %``The Milky Way's Circular Velocity Curve to 60 kpc and an Estimate of the
  %Dark Matter Halo Mass from Kinematics of ~2400 SDSS Blue Horizontal Branch
  %Stars,''
  Astrophys.\ J.\  {684} (2008) 1143
  [arXiv:\hhref{0801.1232} [astro-ph]].
  %%CITATION = ASJOA,684,1143;%%
  
\bibitem{SakamotoMWmass}
T.~Sakamoto, M.~Chiba and T.~C.~Beers,
  %``The Mass of the Milky Way: Limits from a Newly Assembled Set of Halo
  %Objects,''
  Astron.\ Astrophys.\  {397} (2003) 899
  [arXiv:\hhref{astro-ph/0210508}].
  %%CITATION = AAEJA,397,899;%%
  
  \bibitem{PrzybillaMWmass}
  N.~Przybilla, A.~Tillich, U.~Heber and R.~D.~Scholz,
  %``Weighing the Galactic dark matter halo: a lower mass limit from the fastest
  %halo star known,''
  arXiv:\hhref{1005.5026}.
  %%CITATION = ARXIV:1005.5026;%%

\bibitem{Ghigna:1999sn}
S.~Ghigna, B.~Moore, F.~Governato, G.~Lake, T.~R.~Quinn and J.~Stadel,
  %``Density profiles and substructure of dark matter halos: converging results
  %at ultra-high numerical resolution,''
  Astrophys.\ J.\  {\bf 544} (2000) 616
  [arXiv:\hhref{astro-ph/9910166}].
  %%CITATION = ASJOA,544,616;%%

\bibitem{boost}
  J.~Silk and A.~Stebbins,
  %``Clumpy cold dark matter,''
  Astrophys.\ J.\  {\bf 411} (1993) 439.
  %%CITATION = ASJOA,411,439;%%
  L.~Bergstrom, J.~Edsjo, P.~Gondolo and P.~Ullio,
  %``Clumpy neutralino dark matter,''
  Phys.\ Rev.\  D {\bf 59} (1999) 043506
  [arXiv:\hhref{astro-ph/9806072}].
  %%CITATION = PHRVA,D59,043506;%%
  V.~Berezinsky, V.~Dokuchaev and Y.~Eroshenko,
  %``Small-scale clumps in the galactic halo and dark matter annihilation,''
  Phys.\ Rev.\  D {\bf 68} (2003) 103003
  [arXiv:\hhref{astro-ph/0301551}].
  %%CITATION = PHRVA,D68,103003;%%

  \bibitem{Lavalle}
  J.~Lavalle, J.~Pochon, P.~Salati and R.~Taillet,
  %``Clumpiness of Dark Matter and Positron Annihilation Signal: Computing the
  %odds of the Galactic Lottery,''
  Astron.\ Astrophys.\  {\bf 462} (2007) 827
  [arXiv:\hhref{astro-ph/0603796}].
  %%CITATION = AAEJA,462,827;%%
  J.~Lavalle, Q.~Yuan, D.~Maurin and X.~J.~Bi,
  %``Full Calculation of Clumpiness Boost factors for Antimatter Cosmic Rays in
  %the light of \LambdaCDM N-body simulation results,''
  Astron.\ Astrophys.\  {\bf 479} (2008) 427
  [arXiv:\hhref{0709.3634} [astro-ph]].
  %%CITATION = AAEJA,479,427;%%
  
  \bibitem{Sommerfeld}
A. Sommerfeld, ``\"Uber die Beugung und Bremsung der Elektronen'', Ann. Phys. 403, 257 (1931).
J.~Hisano, S.~Matsumoto and M.~M.~Nojiri,
  %``Explosive dark matter annihilation,''
  Phys.\ Rev.\ Lett.\  {92} (2004) 031303
  [arXiv: \hhref{hep-ph/0307216}].
  %%CITATION = PRLTA,92,031303;%%
J.~Hisano, S.~Matsumoto, M.~M.~Nojiri and O.~Saito,
  %``Non-perturbative effect on dark matter annihilation and gamma ray
  %signature from galactic center,''
  Phys.\ Rev.\  D {71} (2005) 063528
  [arXiv: \hhref{hep-ph/0412403}].
  %%CITATION = PHRVA,D71,063528;%%
See also previous work in 
  K.~Belotsky, D.~Fargion, M.~Khlopov and R.~V.~Konoplich,
  %``May heavy neutrinos solve underground and cosmic ray puzzles?,''
  Phys.\ Atom.\ Nucl.\  {71} (2008) 147
  [arXiv:\hhref{hep-ph/0411093}] and references therein. 
  %%CITATION = PANUE,71,147;%%

\bibitem{MDMastro}
M.~Cirelli, A.~Strumia, M.~Tamburini,
  %``Cosmology and Astrophysics of Minimal Dark Matter,''
  Nucl.\ Phys.\  B {\bf 787} (2007) 152
  [arXiv:\hhref{0706.4071} [hep-ph]].
  %%CITATION = NUPHA,B787,152;%%

\bibitem{CKRS}
M.~Cirelli, M.~Kadastik, M.~Raidal and A.~Strumia,
  %``Model-independent implications of the e+, e-, anti-proton cosmic ray
  %spectra on properties of Dark Matter,''
  Nucl.\ Phys.\  B {813} (2009) 1
  [arXiv:\hhref{0809.2409}].
  %%CITATION = NUPHA,B813,1;%%

\bibitem{AHFSW}
N.~Arkani-Hamed, D.~P.~Finkbeiner, T.~R.~Slatyer, N.~Weiner,
  %``A Theory of Dark Matter,''
  Phys.\ Rev.\  {D79 } (2009)  015014.
  [arXiv:\hhref{0810.0713}].

\bibitem{Sommerfeld2}
 M.~Lattanzi and J.~I.~Silk,
  %``Can the WIMP annihilation boost factor be boosted by the Sommerfeld
  %enhancement?,''
  arXiv:\hhref{0812.0360} [astro-ph].
  %%CITATION = ARXIV:0812.0360;%%
  J.~D.~March-Russell and S.~M.~West,
  %``WIMPonium and Boost Factors for Indirect Dark Matter Detection,''
  arXiv:\hhref{0812.0559} [astro-ph].
  %%CITATION = ARXIV:0812.0559;%%
L.~Pieri, M.~Lattanzi and J.~Silk,
  %``Constraining the Sommerfeld enhancement with Cherenkov telescope
  %observations of dwarf galaxies,''
  arXiv:\hhref{0902.4330} [astro-ph.HE].
  %%CITATION = ARXIV:0902.4330;%%
  R.~Iengo,
  %``Sommerfeld enhancement: general results from field theory diagrams,''
  arXiv:\hhref{0902.0688} [hep-ph].
  %%CITATION = ARXIV:0902.0688;%%

\bibitem{Kuhlen:2009kx}
  M.~Kuhlen, P.~Madau and J.~Silk,
  %``Exploring Dark Matter with Milky Way substructure,''
  Science {\bf 325} (2009) 970
  [arXiv:\hhref{0907.0005} [astro-ph.GA]].
  %%CITATION = SCIEA,325,970;%%
  
  \bibitem{Hutsi:2010ai}
  G.~Hutsi, A.~Hektor, M.~Raidal,
  %``Implications of the Fermi-LAT diffuse gamma-ray measurements on annihilating or decaying Dark Matter,''
  JCAP {\bf 1007} (2010) 008
  [arXiv:\hhref{1004.2036}].

%%%%%%%%%%%%%%%%%%%%%%%%%%%%%%%%%%%%%%%%%%%%%

% REFS FOR SECTION 3

\bibitem{higgsupdate04072012}
The ATLAS collaboration, arXiv:\hhref{1207.7214} [hep-ex].
The CMS collaboration, arXiv:\hhref{1207.7235} [hep-ex].
The ATLAS collaboration, 
%Observation of an Excess of Events in the Search for the Standard Model Higgs boson with the ATLAS detector at the LHC,
\myurl{http://cdsweb.cern.ch/record/1460439?ln=en}{ATLAS-CONF-2012-093}.
The CMS collaboration,
% Observation of a new boson with a mass near 125 GeV,
\myurl{http://cdsweb.cern.ch/record/1460438/}{CMS-PAS-HIG-12-020}.
See J.~Incandela's and F.~Gianotti's \myurl{indico.cern.ch/conferenceDisplay.py?confId=197461}{seminars} on behalf of the CMS and ATLAS collaborations at CERN on 4 July 2012.

\bibitem{PR}
M.~Pospelov, A.~Ritz,
  %``Astrophysical Signatures of Secluded Dark Matter,''
  Phys.\ Lett.\  {B671 } (2009)  391-397.
  [arXiv:\hhref{0810.1502}].
  
\bibitem{pythia8}
T. Sjostrand, S. Mrenna and P.Z. Skands, Comput. Phys. Commun. 178 (2008) 852.

\bibitem{herwig6}
G. Corcella et al., JHEP 0101 (2001) 010.
  
\bibitem{CCRSSU}
P.~Ciafaloni, D.~Comelli, A.~Riotto, F.~Sala, A.~Strumia and A.~Urbano,
  %``Weak Corrections are Relevant for Dark Matter Indirect Detection,''
  arXiv:\hhref{1009.0224}.
  %%CITATION = ARXIV:1009.0224;%%  

\bibitem{EWbrem}
In 
  M.~Kachelriess and P.~D.~Serpico,
  %``Model-independent dark matter annihilation bound from the diffuse $\gamma$
  %ray flux,''
  Phys.\ Rev.\  D {\bf 76} (2007) 063516
  [arXiv:\hhref{0707.0209} [hep-ph]]
  %%CITATION = PHRVA,D76,063516;%%
  the authors first point out the potential relevance of the effect.
See also:
  N.~F.~Bell, J.~B.~Dent, T.~D.~Jacques and T.~J.~Weiler,
  %``Electroweak Bremsstrahlung in Dark Matter Annihilation,''
  Phys.\ Rev.\  D {\bf 78} (2008) 083540
  [arXiv:\hhref{0805.3423} [hep-ph]].
  %%CITATION = PHRVA,D78,083540;%%
More recent analyses and a more complete appraisal of the phenomenological importance of the effect have been performed in:
  M.~Kachelriess, P.~D.~Serpico and M.~A.~Solberg,
  %``On the role of electroweak bremsstrahlung for indirect dark matter
  %signatures,''
  Phys.\ Rev.\  D {\bf 80} (2009) 123533
  [arXiv:\hhref{0911.0001} [hep-ph]];
  %%CITATION = PHRVA,D80,123533;%%
  P.~Ciafaloni and A.~Urbano,
  %``TeV scale Dark Matter and electroweak radiative corrections,''
  Phys.\ Rev.\  D {\bf 82} (2010) 043512
  [arXiv:\hhref{1001.3950} [hep-ph]];
  %%CITATION = PHRVA,D82,043512;%%
N.~F.~Bell, J.~B.~Dent, T.~D.~Jacques and T.~J.~Weiler,
  %``W/Z Bremsstrahlung as the Dominant Annihilation Channel for Dark Matter,''
  arXiv:\hhref{1009.2584} [hep-ph].
  %%CITATION = ARXIV:1009.2584;%%

%\bibitem{ewpt} J.~Alcaraz et al., LEP Collaborations and LEP Electroweak Working Group,
%arXiv:\hhref{0712.0929} [hep-ex].

\bibitem{daniela}
G. Corcella and D. Rebuzzi, in 
Proceedings of the Workshop on Monte Carlo's, Physics and Simulations at
the LHC PART II', Frascati 2006,
  arXiv:\hhref{0902.0180}.

\bibitem{hdecay}
A. Djouadi, J. Kalinowski and M. Spira, Comput. Phys. Commun. 108 (1998) 56.

\bibitem{isasugra}
 H. Baer, F.E. Paige, S.D. Protopopescu and X. Tata, \hhref{hep-ph/0312045}.

\bibitem{tauola}
S. Jadach and Z. Was, Comput. Phys. Commun. 76 (1993) 361.

\bibitem{IB}
T.~Bringmann, L.~Bergstrom, J.~Edsjo,
  %``New Gamma-Ray Contributions to Supersymmetric Dark Matter Annihilation,''
  JHEP {0801 } (2008)  049.
  [arXiv:\hhref{0710.3169}].

\bibitem{3body}
X.~-l.~Chen, M.~Kamionkowski,
  %``Three body annihilation of neutralinos below two-body thresholds,''
  JHEP {9807 } (1998)  001.
  [\hhref{hep-ph/9805383}].
C.~E.~Yaguna,
  %``Large contributions to dark matter annihilation from three-body final states,''
  Phys.\ Rev.\  {D81 } (2010)  075024.
  [arXiv:\hhref{1003.2730}].
  L.~L.~Honorez, C.~E.~Yaguna,
  %``The inert doublet model of dark matter revisited,''
  JHEP {1009}, 046 (2010).
  [arXiv:\hhref{1003.3125}].
 
\bibitem{marweb}
G. Marchesini and B.R. Webber, Nucl. Phys. B 238 (1984) 1;  
ibid. B 310 (1988) 461.

\bibitem{pythia}
T. Sj\"ostrand, S. Mrenna and P. Skands, JHEP 0605 (2006) 036.

\bibitem{pythiakt}
T. Sj\"ostrand and P. Skands, Eur. Phys. J. C 39 (2005) 129.

\bibitem{cdf} K. Ackerstaff et al., Phys. Rev. D 50 (1994) 5562.

\bibitem{bcd}
A. Banfi, G. Corcella and M. Dasgupta, JHEP 0703 (2007) 050.

\bibitem{ng} M. Dasgupta and G.P. Salam, Phys. Lett. B 512 (2001) 323.

\bibitem{cmw} 
S. Catani, G. Marchesini and B.R. Webber, Nucl. Phys. B 349 (1991) 635.

\bibitem{cluster}
B.R. Webber, Nucl. Phys. B 238 (1984) 492.

\bibitem{string}
B. Andersson, G. Gustafson, G. Ingelman, T. Sj\"ostrand,
Phys. Rept. 97 (1983) 31.

\bibitem{volker}
G. Corcella and V. Drollinger, Nucl. Phys. B 730 (2005) 82.

\bibitem{cdf1}
A. Affolder et al., Phys Rev. D 65 (2002) 092002. 

\bibitem{herpp}
M. Bahr et al., Eur. Phys. J C 58 (2008) 639.

\bibitem{hamilton} K. Hamilton and P. Richardson, JHEP 0607 (2006) 010.

\bibitem{alpgen} M.L. Mangano, M. Moretti, F. Piccinini, R. Pittau, A. Polosa, JHEP 0307 (2003) 001.

\bibitem{madgraph} J. Alwall, P. Demin, S. de Visscher, R. Frederix, M. Herquet, F. Maltoni, T. Plehn, D.L. Rainwater and T. Stelzer, JHEP 0709 (2007) 028.

\bibitem{BG}
\href{http://www.balticgrid.org}{www.balticgrid.org}.

  \bibitem{Cembranos:2010dm}
  J.~A.~R.~Cembranos, A.~de la Cruz-Dombriz, A.~Dobado, R.~A.~Lineros and A.~L.~Maroto,
  %``Photon spectra from WIMP annihilation,''
  arXiv:\hhref{1009.4936}.
  %%CITATION = ARXIV:1009.4936;%%

\bibitem{Dbar}
M.~Kadastik, M.~Raidal and A.~Strumia,
  %``Enhanced anti-deuteron Dark Matter signal and the implications of PAMELA,''
  Phys.\ Lett.\  B {683} (2010) 248
  [arXiv:\hhref{0908.1578}].
  %%CITATION = PHLTA,B683,248;%%

\bibitem{DMnu}
M.~Cirelli, N.~Fornengo, T.~Montaruli, I.~A.~Sokalski, A.~Strumia and F.~Vissani,
  %``Spectra of neutrinos from dark matter annihilations,''
  Nucl.\ Phys.\  B {\bf 727} (2005) 99
  [Erratum-ibid.\  B {\bf 790} (2008) 338]
  [arXiv:\hhref{hep-ph/0506298}].
  %%CITATION = NUPHA,B727,99;%%
  See \myurl{www.marcocirelli.net/DMnu.html}{www.marcocirelli.net/DMnu.html} for the numerical results from that paper.

%%%%%%%%%%%%%%%%%%%%%%%%%%%%%%%%%%%%%%%%%

% REFS FOR SECTION 4

\bibitem{SalatiCargese}
For a lucid and pedagogical review of the CR propagation formalism relevant for DM, see: 
P.~Salati, ``Indirect and direct dark matter detection'', Proceedings of the 2007 Carg\`ese Summer School: Cosmology and Particle Physics Beyond the Standard Models, PoS(cargese)009, \myurl{pos.sissa.it/cgi-bin/reader/conf.cgi?confid=49}{pos.sissa.it/cgi-bin/reader/conf.cgi?confid=49}

\bibitem{Delahaye:2008ua}
  T.~Delahaye, F.~Donato, N.~Fornengo, J.~Lavalle, R.~Lineros, P.~Salati and R.~Taillet,
  %``Galactic secondary positron flux at the Earth,''
  Astron.\ Astrophys.\  {\bf 501} (2009) 821
  [arXiv:\hhref{0809.5268} [astro-ph]].
  %%CITATION = AAEJA,501,821;%%

\bibitem{DiffusionCylinder}
The now standard ``two-zone diffusion model" introduced in 
V.L.~Ginzburg, Ya.M.~Khazan and V.S.~Ptuskin, 
Astrophysics and Space Science 68, 295-314 (1980).  
W.R.~Webber, M.A.~Lee and M.~Gupta, Astrophysical Journal 390, 96-104 (1992).

\bibitem{perelstein}
M.~Perelstein and B.~Shakya,
  %``Remarks on calculation of positron flux from galactic dark matter,''
  Phys.\ Rev.\  D {\bf 82} (2010) 043505
  [arXiv:\hhref{1002.4588} [astro-ph.HE]].
  %%CITATION = PHRVA,D82,043505;%%

\bibitem{Blum}
  K.~Blum,
  %``Cosmic ray propagation time scales: lessons from radioactive nuclei and
  %positron data,''
  arXiv:\hhref{1010.2836} [astro-ph.HE].
  %%CITATION = ARXIV:1010.2836;%%

\bibitem{fermibubbles}
  M.~Su, T.~R.~Slatyer and D.~P.~Finkbeiner,
  %``Giant Gamma-ray Bubbles from Fermi-LAT: AGN Activity or Bipolar Galactic
  %Wind?,''
  Astrophys.\ J.\  {\bf 724} (2010) 1044
  [arXiv:\hhref{1005.5480} [astro-ph.HE]].
  %%CITATION = ASJOA,724,1044;%%
See also \myurl{www.nasa.gov/mission_pages/GLAST/news/new-structure.html}{www.nasa.gov/mission$\_$pages/GLAST/news/ new-structure.html}


 \bibitem{CP}
 M.~Cirelli and P.~Panci,
  %``Inverse Compton constraints on the Dark Matter e+e- excesses,''
  Nucl.\ Phys.\  B {\bf 821} (2009) 399
  [arXiv:\hhref{0904.3830} [astro-ph.CO]].
  %%CITATION = NUPHA,B821,399;%%

\bibitem{Meade}
  P.~Meade, M.~Papucci, A.~Strumia and T.~Volansky,
  %``Dark Matter Interpretations of the Electron/Positron Excesses after
  %FERMI,''
  Nucl.\ Phys.\  B {831} (2010) 178
  [arXiv:\hhref{0905.0480}].

\bibitem{StrongMoskalenkoReimer98}
 A.~W.~Strong, I.~V.~Moskalenko and O.~Reimer,
  %``Diffuse continuum gamma rays from the Galaxy,''
  Astrophys.\ J.\  {537} (2000) 763
  [Erratum-ibid.\  {541} (2000) 1109]
  [arXiv:\hhref{astro-ph/9811296}].
  %%CITATION = ASJOA,537,763;%%
  
  \bibitem{perelstein2}
  M.~Perelstein and B.~Shakya,
  %``Antiprotons from Dark Matter: Effects of a Position-Dependent Diffusion
  %Coefficient,''
  arXiv:\hhref{1012.3772} [astro-ph.HE].
  %%CITATION = ARXIV:1012.3772;%%

\bibitem{FornengoDec2007}
T.~Delahaye, R.~Lineros, F.~Donato, N.~Fornengo and P.~Salati,
  %``Positrons from dark matter annihilation in the galactic halo: theoretical
  %uncertainties,''
  Phys.\ Rev.\  D 77 (2008) 063527
  [arXiv:\hhref{0712.2312}].
  %%CITATION = PHRVA,D77,063527;%%

\bibitem{DonatoPRD69}
F.~Donato, N.~Fornengo, D.~Maurin and P.~Salati,
  %``Antiprotons in cosmic rays from neutralino annihilation,''
  Phys.\ Rev.\  D {69} (2004) 063501
  [arXiv:\hhref{astro-ph/0306207}].
  %%CITATION = PHRVA,D69,063501;%%
   
  \bibitem{Maurin2010}
A.~Putze, L.~Derome and D.~Maurin, 
%``A Markov Chain Monte Carlo technique to sample transport and source %parameters of Galactic cosmic rays: II. Results for the diffusion model %combining B/C and radioactive nuclei,'' 
Astron.\ Astrophys.\ {516} (2010) A66 [arXiv:\hhref{1001.0551} [astro-ph.HE]]. 
%%CITATION = AAEJA,516,A66;%%

\bibitem{Dragon2010}
G.~Di Bernardo, C.~Evoli, D.~Gaggero, D.~Grasso and L.~Maccione, 
%``Unified interpretation of cosmic-ray nuclei and antiproton recent measurements,'' 
Astropart.\ Phys.\ {34} (2010) 274 [arXiv:\hhref{0909.4548} [astro-ph.HE]]. 
%%CITATION = APHYE,34,274;%%

\bibitem{Trotta2010}
R.~Trotta, G.~Johannesson, I.~V.~Moskalenko, T.~A.~Porter, R.~R.~de Austri and A.~W.~Strong, 
%``Constraints on cosmic-ray propagation models from a global Bayesian %analysis,'' 
arXiv:\hhref{1011.0037} [astro-ph.HE]. 
%%CITATION = ARXIV:1011.0037;%%

\bibitem{Zupan}
I.~Z.~Rothstein, T.~Schwetz and J.~Zupan,
  %``Phenomenology of Dark Matter annihilation into a long-lived intermediate state,''
  JCAP {0907} (2009) 018
  [arXiv:\hhref{0903.3116} [astro-ph.HE]], Appendix B.
  %%CITATION = JCAPA,0907,018;%%
  
\bibitem{smoothing}
A.~Barrau, P.~Salati, G.~Servant, F.~Donato, J.~Grain, D.~Maurin and R.~Taillet,
  %``Kaluza-Klein Dark Matter and Galactic Antiprotons,''
  Phys.\ Rev.\  D {\bf 72} (2005) 063507
  [arXiv:\hhref{astro-ph/0506389}].
  %%CITATION = PHRVA,D72,063507;%%

\bibitem{crosssection}
L.~C.~Tan and L.~K.~Ng,
  %``Calculation Of The Equilibrium Anti-Proton Spectrum,''
  J.\ Phys.\ G {9} (1983) 227.
  %%CITATION = JPHGB,G9,227;%%

\bibitem{HisanoAntiparticles}
J.~Hisano, S.~Matsumoto, O.~Saito and M.~Senami,
  %``Heavy Wino-like neutralino dark matter annihilation into antiparticles,''
  Phys.\ Rev.\  D {73} (2006) 055004
  [arXiv:\hhref{hep-ph/0511118}].
  %%CITATION = PHRVA,D73,055004;%%

\bibitem{DonatoApJ563}
F.~Donato, D.~Maurin, P.~Salati, A.~Barrau, G.~Boudoul and R.~Taillet,
  %``Antiprotons from spallation of cosmic rays on interstellar matter,''
  Astrophys.\ J.\  {563} (2001) 172
  [arXiv:\hhref{astro-ph/0103150}].
  %%CITATION = ASJOA,563,172;%%

  \bibitem{MaurinApJ555}
D.~Maurin, F.~Donato, R.~Taillet and P.~Salati,
  %``Cosmic Rays below Z=30 in a diffusion model: new constraints on propagation
  %parameters,''
  Astrophys.\ J.\  {555} (2001) 585
  [arXiv:\hhref{astro-ph/0101231}].
  %%CITATION = ASJOA,555,585;%%

\bibitem{methodPbar}
P.~Chardonnet, G.~Mignola, P.~Salati and R.~Taillet,
  %``Galactic diffusion and the antiproton signal of supersymmetric dark
  %matter,''
  Phys.\ Lett.\  B {384} (1996) 161
  [arXiv:\hhref{astro-ph/9606174}].
  %%CITATION = PHLTA,B384,161;%%
  
A.~Bottino, F.~Donato, N.~Fornengo and P.~Salati,
  %``Which fraction of the measured cosmic ray antiprotons might be due to
  %neutralino annihilation in the galactic halo?,''
  Phys.\ Rev.\  D {58} (1998) 123503
  [arXiv:\hhref{astro-ph/9804137}].
  %%CITATION = PHRVA,D58,123503;%%

    See also: 
L.~Bergstrom, J.~Edsjo and P.~Ullio,
  %``Cosmic antiprotons as a probe for supersymmetric dark matter?,''
  Astrophys.\ J.\  {526} (1999) 215
  [arXiv:\hhref{astro-ph/9902012}].
  %%CITATION = ASJOA,526,215;%%


\bibitem{TailletRRDA}  
D.~Maurin, R.~Taillet, F.~Donato, P.~Salati, A.~Barrau and G.~Boudoul,
  %``Galactic cosmic ray nuclei as a tool for astroparticle physics,''
  arXiv:\hhref{astro-ph/0212111}.
  %%CITATION = ASTRO-PH/0212111;%%
  
\bibitem{GA}
L.J. Gleeson and W.I. Axford, ApJ 149, L115 (1967) and 
L.J. Gleeson and W.I. Axford, ApJ 154, 1011 (1968).

\bibitem{DonatoDbar}
  F.~Donato, N.~Fornengo and D.~Maurin,
  %``Antideuteron fluxes from dark matter annihilation in diffusion models,''
  Phys.\ Rev.\  D {\bf 78} (2008) 043506
  [arXiv:\hhref{0803.2640} [hep-ph]].
  %%CITATION = PHRVA,D78,043506;%%
  
  \bibitem{DMantideuterium}
  C.~B.~Braeuninger and M.~Cirelli,
  %``Anti-deuterons from heavy Dark Matter,''
  Phys.\ Lett.\  B {\bf 678} (2009) 20
  [arXiv:\hhref{0904.1165} [hep-ph]].
  %%CITATION = PHLTA,B678,20;%%

%%%%%%%%%%%%%%%%%%%%%%%%%%%%%%%%%%%%%%%%%

% REFS FOR SECTION 6
  
 \bibitem{rybicki}
 G. B. Rybicki and A. P. Lightman, `Radiative Processes in Astrophysics', Wiley-Interscience, New York, 1979.
 
 \bibitem{BCST}
 G.~Bertone, M.~Cirelli, A.~Strumia and M.~Taoso,
  %``Gamma-ray and radio tests of the e+e- excess from DM annihilations,''
  JCAP {\bf 0903} (2009) 009
  [arXiv:\hhref{0811.3744} [astro-ph]].
  %%CITATION = JCAPA,0903,009;%%
 
%%%%%%%%%%%%%%%%%%%%%%%%%%%%%%%%%%%%%%%%%

% REFS FOR SECTION 7


\bibitem{Huetsi:2009ex}
G.~H{\"u}tsi, A.~Hektor, and M.~Raidal,
  %``Constraints on leptonically annihilating Dark Matter from reionization and extragalactic gamma background,''
  Astron.\ Astrophys.\ {\bf 505} (2009) 999
  [arXiv:\hhref{0906.4550}].
  
  \bibitem{CPS}
  M.~Cirelli, P.~Panci and P.~D.~Serpico,
  %``Diffuse gamma ray constraints on annihilating or decaying Dark Matter after
  %Fermi,''
  Nucl.\ Phys.\  B {\bf 840} (2010) 284
  [arXiv:\hhref{0912.0663} [astro-ph.CO]].
  %%CITATION = NUPHA,B840,284;%%

\bibitem{2002PhR...372....1C}
  A.~Cooray, R.~K.~Sheth,
  %``Halo models of large scale structure,''
  Phys.\ Rept.\  {\bf 372 } (2002) 1-129.
  [arXiv:\hhref{astro-ph/0206508}].

\bibitem{PressSchechter}
W.~H.~Press and P.~Schechter,
  %``Formation of galaxies and clusters of galaxies by selfsimilar gravitational
  %condensation,''
  Astrophys.\ J.\  {\bf 187} (1974) 425.
  %%CITATION = ASJOA,187,425;%%

\bibitem{1999MNRAS.308..119S}
  R.~K.~Sheth, G.~Tormen,
  %``Large scale bias and the peak background split,''
  Mon.\ Not.\ Roy.\ Astron.\ Soc.\  {\bf 308 } (1999) 119
  [arXiv:\hhref{astro-ph/9901122}].


\bibitem{1997ApJ...490..493N}
  J.~F.~Navarro, C.~S.~Frenk, S.~D.~M.~White,
  %``A Universal density profile from hierarchical clustering,''
  Astrophys.\ J.\  {\bf 490 } (1997) 493-508
  [arXiv:\hhref{astro-ph/9611107}].

\bibitem{2008MNRAS.391.1940M}
  A.~V.~Macci{\`o}, A.~A.~Dutton, and F.~C.~van~den~Bosch,
  %``Concentration, spin and shape of dark matter haloes as a function of the cosmological model: WMAP1, WMAP3 and WMAP5 results,''
  MNRAS {\bf 391} (2008) 1940-1954
  [arXiv:\hhref{0805.1926}].

\bibitem{Neto:2007vq}
A.~F.~Neto {\it et al.},
  %``The statistics of LCDM Halo Concentrations,''
  arXiv:\hhref{0706.2919} [astro-ph].
  %%CITATION = ARXIV:0706.2919;%%

\bibitem{Martinez:2009jh}
  G.~D.~Martinez, J.~S.~Bullock, M.~Kaplinghat, L.~E.~Strigari and R.~Trotta,
  %``Indirect Dark Matter Detection from Dwarf Satellites: Joint Expectations
  %from Astrophysics and Supersymmetry,''
  JCAP {\bf 0906} (2009) 014
  [arXiv:\hhref{0902.4715} [astro-ph.HE]].
  %%CITATION = JCAPA,0906,014;%%

\bibitem{Bringmann:2009vf}
  T.~Bringmann,
  %``Particle Models and the Small-Scale Structure of Dark Matter,''
  New J.\ Phys.\  {\bf 11} (2009) 105027
  [arXiv:\hhref{0903.0189}].

%\bibitem{Inoue:2009kd}
%  S.~Inoue, R.~Salvaterra, T.~R.~Choudhury {\it et al.},
%  %``Probing intergalactic radiation fields during cosmic reionization through gamma-ray absorption,''
%  [arXiv:\hhref{0906.2495}].

\bibitem{Aliu:2008ay}
  E.~Aliu {\it et al.} [ MAGIC Collaboration ],
  %``Very-High-Energy Gamma Rays from a Distant Quasar: How Transparent Is the Universe?,''
  Science {\bf 320} (2008) 1752.
  [arXiv:\hhref{0807.2822}].
  
  \bibitem{Dominguez:2010bv}
A.~Dominguez {\it et al.},
  %``Extragalactic Background Light Inferred from AEGIS Galaxy SED-type
  %Fractions,''
  arXiv:\hhref{1007.1459} [astro-ph.CO].
  %%CITATION = ARXIV:1007.1459;%%
  
  \bibitem{Franceschini}
  A.~Franceschini, G.~Rodighiero and M.~Vaccari,
  %``The extragalactic optical-infrared background radiations, their time
  %evolution and the cosmic photon-photon opacity,''
  arXiv:\hhref{0805.1841} [astro-ph].
  %%CITATION = ARXIV:0805.1841;%%
  
%\bibitem{Stecker:2005qs}
%  F.~W.~Stecker, M.~A.~Malkan, S.~T.~Scully,
%  %``Intergalactic photon spectra from the far ir to the uv lyman limit for 0 < Z < 6 and the optical depth of the universe to high energy gamma-rays,''
%  Astrophys.\ J.\  {\bf 648} (2006) 774-783.
%  [arXiv:\hhref{astro-ph/0510449}].

\bibitem{Zdziarski:1989aa}
  A.~A.~Zdziarski and R.~Svensson,
  %``Absorption of X-rays and gamma rays at cosmological distances,''
  Astrophys.\ J. {\bf 344} (1989) 551--566.
  
\bibitem{1988ApJ...335..786Z}
  A.~A.~Zdziarski,
  %``Saturated pair-photon cascades on isotropic background photons,''
  {\em Astrophys. J.} {\bf 335} (1988) 786--802.

\bibitem{Fan:2006dp}
 X.~H.~Fan, C.~L.~Carilli, B.~G.~Keating,
  %``Observational constraints on cosmic reionization,''
  Ann.\ Rev.\ Astron.\ Astrophys.\ {\bf 44} (2006) 415-462
  [arXiv:\hhref{astro-ph/0602375}].

\end{thebibliography}
\end{document}